\documentclass[%
 amsmath,amssymb,
 aps,
 prfluids,
floatfix,
]{revtex4-2}

\usepackage{upgreek}
\usepackage{natbib}
\usepackage{ulem}
\usepackage{enumitem}
\usepackage{hyperref}
\hypersetup{
colorlinks=true, 
citecolor=magenta,
breaklinks=true, 
urlcolor= blue, 
linkcolor= blue, 
bookmarksopen=true, 
}

\usepackage{graphicx}
\usepackage{dcolumn}
\usepackage{bm}

\usepackage{xcolor}

\newcommand{\MA}{\textcolor{blue}}

\begin{document}


\title{Breaking of a floating particle raft by water waves}

%
%
%
%

\author{Louis Saddier} 
\affiliation{Paris Saclay University, France}  \altaffiliation[Present address: ]{ENSL, UCBL, CNRS, Laboratoire de physique, F-69342 Lyon, France} 
\author{Ambre Palotai}
\affiliation{\'Ecole Normale Sup\'erieure, Paris Sciences \& Lettres University, France}
 \author{Mathéo Aksil} 
 \affiliation{\'Ecole Normale Sup\'erieure, Paris Sciences \& Lettres University, France}
  \author{Michel Tsamados} 
  \affiliation{Center for Polar observation, University College London, London, UK}
\author{Michael Berhanu } 
\affiliation{MSC, Université Paris Cit\'e, CNRS (UMR 7057), 75013 Paris, France}
\email{michael.berhanu@u-paris.fr}


\date{\today}

\begin{abstract}
When particles of a few tens of microns are spread on the surface of water, they aggregate under the action of capillary forces and form a thin floating membrane, a particle raft. In a tank with a raft made of graphite powder, we generate in the laboratory gravity surface waves, whose wavelength {about $17$ cm} is very large compared to the thickness of the raft {of order $10 \, \mu$m}. For a sufficiently strong wave amplitude, the raft breaks up progressively by developing cracks and producing fragments whose sizes decrease on a time scale long compared to the period of the wave. We characterize the breaking mechanisms. Then, we investigate the area distribution of the fragments produced during the fragmentation process. 
The visual appearance of the fragments distributed in size and surrounded by open water bears a {notable} resemblance to the floes produced by the fracturing of sea ice by waves in the polar oceans. Fragmentation concepts and morphological tools built for sea ice floes can be applied to our macroscopic analog, on which the entire dynamic evolution is accessible. {However, the mechanic of the two systems differ, as our particle raft breaks due to the viscous stresses, whereas the sea-ice fractures due to its bending by the waves.}
\end{abstract}

\maketitle


\section{Introduction}

\begin{figure}
 \begin{center}
    \includegraphics[width=0.7    \columnwidth]{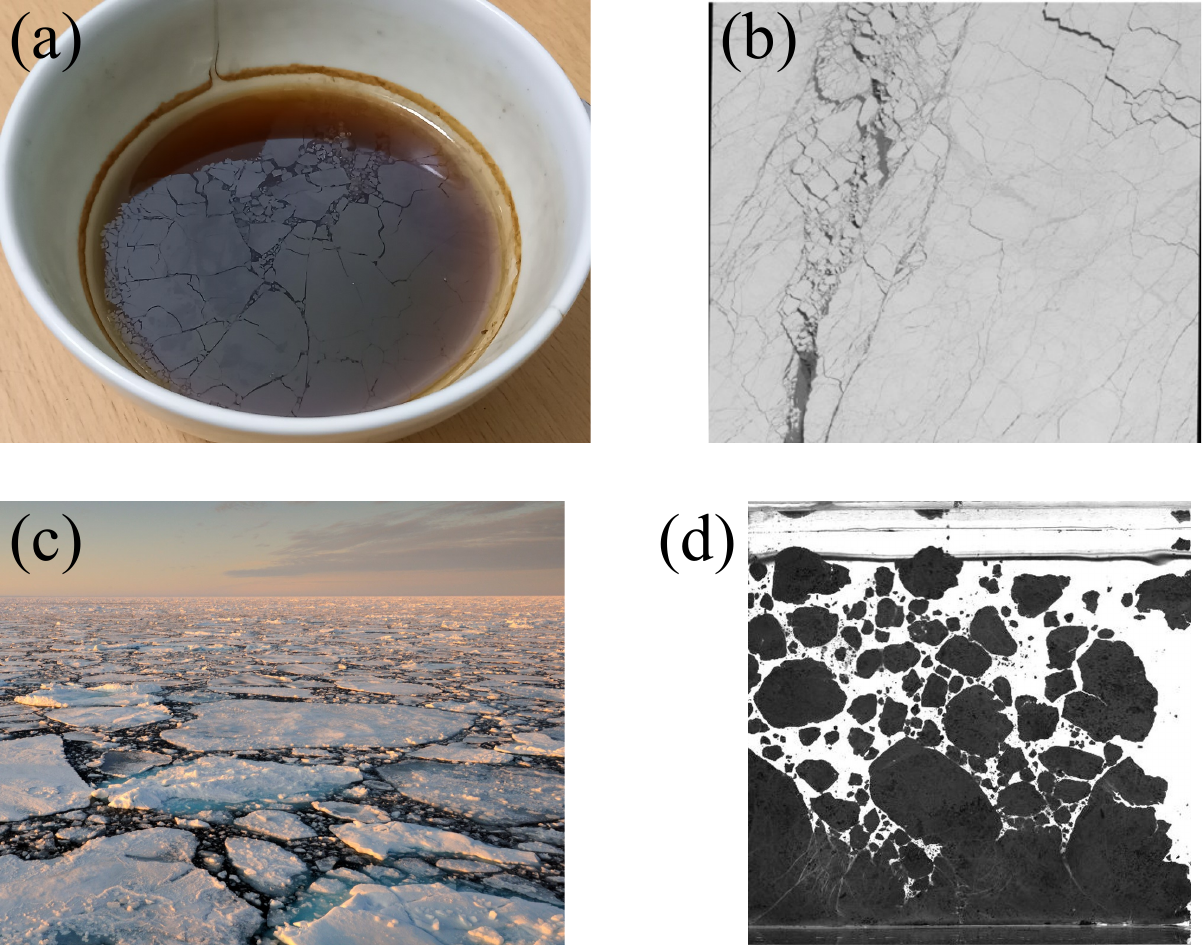}  
 \caption{(a) fragmented tea skin in a bowl of approximate diameter 14 cm. A black tea infusion is let to rest and is then slightly fractured by inducing a flow with a spoon. 
 (b) Arctic sea ice cover seen from the satellite SPOT, April 6th, 1996, (80$^\circ$ 11'N, 108$^\circ$ 33'W). image size: 60 km~\cite{Lehoucq2015}. 
(c) picture of ice floes near Svalbard (Norway), (Credits Andy Rousse). (d) one example of a fractured particle raft made of graphite powder, that we use in our experiments view from the top. The height of the image is 20 cm. 
 }
 \label{Images1}
 \end{center}
\end{figure}

Every tea drinker can observe the formation of a floating scum on the surface of their favorite beverage within a few minutes~\cite{Spiro1993,Enzweiler1994}. This solid but delicate ``skin of tea" can be easily broken by a gentle blow or stirring, resulting in a network of fractures across the tea skin (Fig.~\ref{Images1} (a)). Visually, this image bears striking resemblance to the crack patterns observed on the surface of sea ice in polar regions (Fig.~\ref{Images1} (b)), where the characteristic cracks span  several kilometers. The interaction between fractured sea ice and incoming oceanic gravity surface waves gives rise to the formation of polygonal fragments known as ice floes (Fig.~\ref{Images1} (c)). Similar polygonal fragments can also be observed  in the tea skin when stirred more vigorously. How can we explain the similarity between these two examples? From a mechanical perspective, both cases involve a thin, elastic floating membrane that is deformed by the movement of the liquid beneath it. When the deformation exceeds a certain threshold, the membrane ruptures and cracks form. The repetition of successive fractures then initiates a fragmentation process, dividing the membrane into smaller fragments. The size distribution, the shape of the fragments and the fragmentation dynamics are far from trivial, even in this two-dimensional system.

We propose conducting a laboratory-scale model experiment to investigate the fracture and  fragmentation processes of an elastic floating membrane subjected to surface { gravity water waves of typical frequency and wavelength, $3\,$ Hz and $17$ cm, respectively}. While it is not possible to directly scale the mechanical properties of a floating membrane to those of sea ice, we anticipate observing general features in the fragmentation of a two-dimensional (2D) thin elastic membrane. However, the chemical composition of tea scum is complex, involving calcium carbonate crystals within an organic matrix that contains polyphenolic components~\cite{Spiro1993,Enzweiler1994,Tanizawa2007}. Consequently, ensuring reproducibility in the creation of the membrane and its mechanical properties becomes challenging. We opted to perform laboratory experiments using an alternative analog system, replacing the tea scum with a thin floating particle raft~\cite{protiere2023particle,lagarde2019capillary}, composed of micrometric particles. In this study, we demonstrate that a graphite powder composed of rough particles with a diameter of approximately 10 $\mu$m of diameter yields a nearly homogeneous and solid-like two-dimensional membrane. Despite being denser than water, the particles remain at the free surface due to capillary forces and agglomerate via capillary attractive interactions, {forming the particle raft. For particle sizes of order the millimeter, the attraction mechanism is commonly known} as the ``Cheerios effect"~\cite{chan1981interaction,Vella2005cheerios,vassileva2005capillary,dalbe2011aggregation} {and is related to the interface deformation due to the particle weight. For particles smaller than $5\, \mu$m the gravity effect is negligible~\cite{kralchevsky2000capillary} and the capillary attractive interactions are either caused by the irregularity of the contact line~\cite{stamou2000long,kralchevsky2001particles,fournier2002anisotropic,botto2012capillary} due to shape or chemical heterogeneities. The mechanical properties of these particle rafts} can be described as an elastic floating membrane~\cite{Vella2004} with a Young's modulus, $E$, on the order of $\gamma /d \approx 10^4$ Pa, where $\gamma$ represents the air-water surface tension and $d$ is the particle diameter. In such particle rafts, previous studies have reported the formation of propagating fractures generated by localized addition of surfactants, leading to a Marangoni flow~\cite{Vella2006,bandi2011shock,peco2017influence}. Furthermore, the fracture and fragmentation of granular rafts have been investigated under mechanical excitations, particularly during shearing or stretching of a capillary aggregate by an underlying flow~\cite{vassileva2006restructuring,vassileva2007fragmentation,huang2012wet,kim2019failure,xiao2020strain,to2023rifts}. In particular, Vassileva et al.~\cite{vassileva2007fragmentation} distinguish between fragmentation, where a fragment divides in two parts, and erosion processes, where a single particle is removed from the edge. By balancing viscous drag and capillary attraction within the raft, they proposed  critical threshold shear rates, which vary slightly depending on the raft size.  {The erosion and the cohesion of a particle raft also have been probed during its motion~\cite{lagarde2020probing}. For particle radius below the millimeter, the cohesion force in the raft is strongly underestimated by the  ``Cheerios effect". The tilting of the contact line and its pinning variations seem to enhance the raft cohesion.} The description of the fracture of particle raft as brittle or ductile, motivated some recent studies. Xiao et al.~\cite{xiao2020strain} discovered that by reducing the particle size the material becomes less brittle as rearrangements occur more easily. On the other hand, 
To et al.~\cite{to2023rifts} suggested that for a fast enough stretching velocity the fracture becomes brittle. 

Another mechanism for breaking a granular raft involves subjecting it to water surface waves.  In that case, which has not been studied to our knowledge, a sufficient deformation of the raft may cause its fracture and fragmentation. This situation is {to some extent similar to the} fracturing of the sea ice at its edges in contact with the polar ocean, the so-called Marginal Ice Zone (MIZ). 
In the MIZ, oceanic gravity surface waves are indeed the main mechanism of fragmentation of sea ice~\cite{squire1995ocean,dumont2011wave}. Sea ice, which covers extensive areas in polar oceans (thousands of kilometers) is a brittle material that fractures at relatively small strains and corresponding stresses compared to its Young's modulus. The bulk of sea ice experiences internal and external stresses at large scales, leading to multiple kilometer-length cracks with a wide range of scales (see Fig.~\ref{Images1}, (b)). This complex fracturing behavior necessitates the use of a complex rheology for describing large-scale motions~\cite{feltham2008sea,herman2022granular}. In the MIZ, swells consisting of  water surface waves with wavelengths on the order of 10 to 100 m deform and break the sea ice, generating multiple fragments called floes. These floes interact with incoming waves by scattering them and increasing their damping rate~\cite{squire2020ocean}. However, due to wind, water currents, and wave momentum, the floes drift away from the sea ice edge. The features of the MIZ, resulting from ice breaking by the waves, grinding by the collisions between the floes, freezing or melting, and drift, exhibit high variability and evolve over time~\cite{Horvat2022}. {Recent numerical simulations show that the interactions between the floes themselves and with the flow generate turbulence at the ice-ocean surface~\cite{Brenner2023}}. Consequently, there have been numerous studies focusing on the statistical analysis of floes, particularly the floe-size distribution, both theoretically~\cite{Herman2010,Gherardi2015,Horvat2017,montiel2022theoretical} and in the field~\cite{Dumas2021}. Recently, experimental investigations using the Large Ice Model Basin in Hamburg have been conducted, where a water flume of $72\,$m\,$\times$\,$10$\,m is used to generate gravity surface waves on a artificial floating ice sheet, allowing for the characterization of ice floe size and shape distributions from surface images~\cite{Herman2018}. However, these measurements are static in nature, and the dynamic fragmentation process has not been explored. Here, we {adopt} a similar experimental protocol on a smaller scale, on the order of the ten of centimeter, to study the fragmentation of a particle raft subjected to surface waves and potentially draw analogies with the fracturing of sea ice and the production of ice floes. As described in the following sections, the fracture patterns observed in our particle rafts exhibit a visual resemblance to the sea ice (see Fig.~\ref{Images1}, (d)). {Nevertheless, the analogy remains essentially visual, because as we will show later, the breaking mechanisms differ strongly between our particle raft and the sea-ice. Yet, our experiment constitutes a model system to study fragmentation of a two-dimensional material, which constitutes a challenging domain relevant to sea-ice.}

This article begins by presenting the properties of the particle raft created using graphite powder and describing the experimental setup. Subsequently, we provide qualitative visual observations of the breaking and fragmentation of particle rafts induced by gravity surface waves under various configurations. To interpret these observations, we model the mechanical response of the raft as a thin elastic plate. We then investigate the onset of raft breaking, marked by the appearance of initial fractures. Finally, we briefly examine the fragmentation process by tracking the fragments of the raft, which analogously refer to as ``graphite floes". The size of these graphite floes decreases as a result of fragmentation and erosion in response to the surface currents generated by the incoming waves. The distribution of floe area is also characterized.

\section{Experimental methods}
\subsection{Graphite particle raft}
\label{raft}

The quantitative experiments were conducted using graphite particles (Graphite powder, synthetic, conducting grade, -325 mesh, 99.9995 \% from Alfa Aesar). The 325 mesh grade ensures that these particles are smaller than $44\,\mu$m but the size distribution is polydisperse. The wetting angle of graphite depends strongly on the conditions and the mineralogy of the sample, as reported in the literature~\cite{Garcia2008,Kozbial2018}. Under common conditions at $20^\circ$\,C, the contact angle is approximately $\theta \approx 85^\circ$, indicating that the material is mildly hydrophilic. The bulk density of graphite is $2260$\,kg m$^{-3}$, making the particles denser than water, whose density is $\rho=998$ kg m$^{-3}$ at ambient temperature. However, when the graphite powder is gently sprinkled onto the surface, capillary forces allow the particles to remain at the water-air interface. By stirring the free-surface with a low-amplitude, fast motion, the particles aggregate due capillary interactions.
{Schematically, the attractive force between two grains arises from the minimization of surface energy at the water-air interface. For identical particles, the minimal surface area is achieved when the particles are in contact. For particles of order of $10\, \mu$m, small in front of the capillary length $L_c=\sqrt{\gamma/(\rho\,g)}$, the ``Cheerios" effect~\cite{kralchevsky2000capillary,Vella2005cheerios} due to flotation has a negligible influence. The anisotropy of the contact line associated with the non-spherical particle shape and/or wetting heterogeneities induce quadrupolar attractive forces~\cite{stamou2000long,kralchevsky2001particles,fournier2002anisotropic,botto2012capillary}. After aggregation, the particles form a nearly homogeneous floating layer, referred to as a particle raft~\cite{lagarde2019capillary,protiere2023particle}. However, inside the raft, for two particles in close contact, the estimation of the cohesion force due to anistropic capillary interactions is challenging~\cite{lehle2008ellipsoidal,botto2012capillary}, due to near field effects and to the contact line geometry, even for simple particle shapes. 
In the particle rafts we use, the situation is even more complex, because the interactions involve numerous particles, the powder is polydisperse and the grains are rough.}

 \begin{figure}
 \begin{center}
 \includegraphics[width=.7\columnwidth]{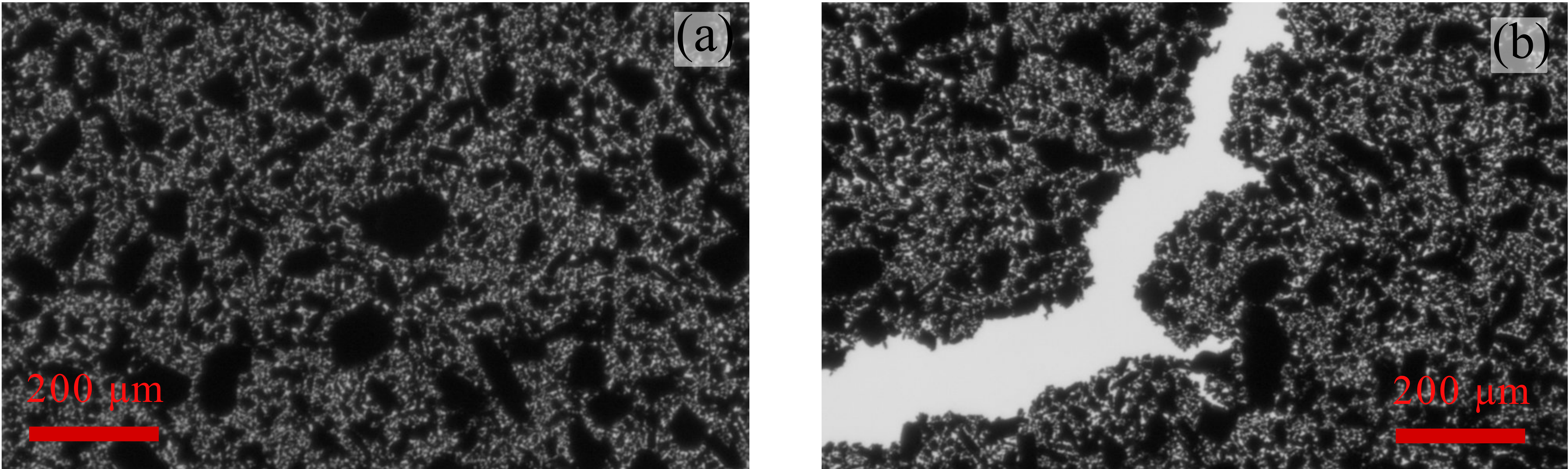} 
  \ \caption{Images of the particle raft taken with a Leica Z16 APO microscope, {with a resolution of $0.6\, \mu$m per pixel}. The black elements correspond to the graphite grains while the bright parts correspond to the liquid substrate. We can see that the powder is polydisperse and that the grains are not circular. On the image (b), a fracture is present on the raft.}
 \label{Macroscope1}
 \end{center}
\end{figure}

 \begin{figure}
 \begin{center}
 \includegraphics[width=.7\columnwidth]{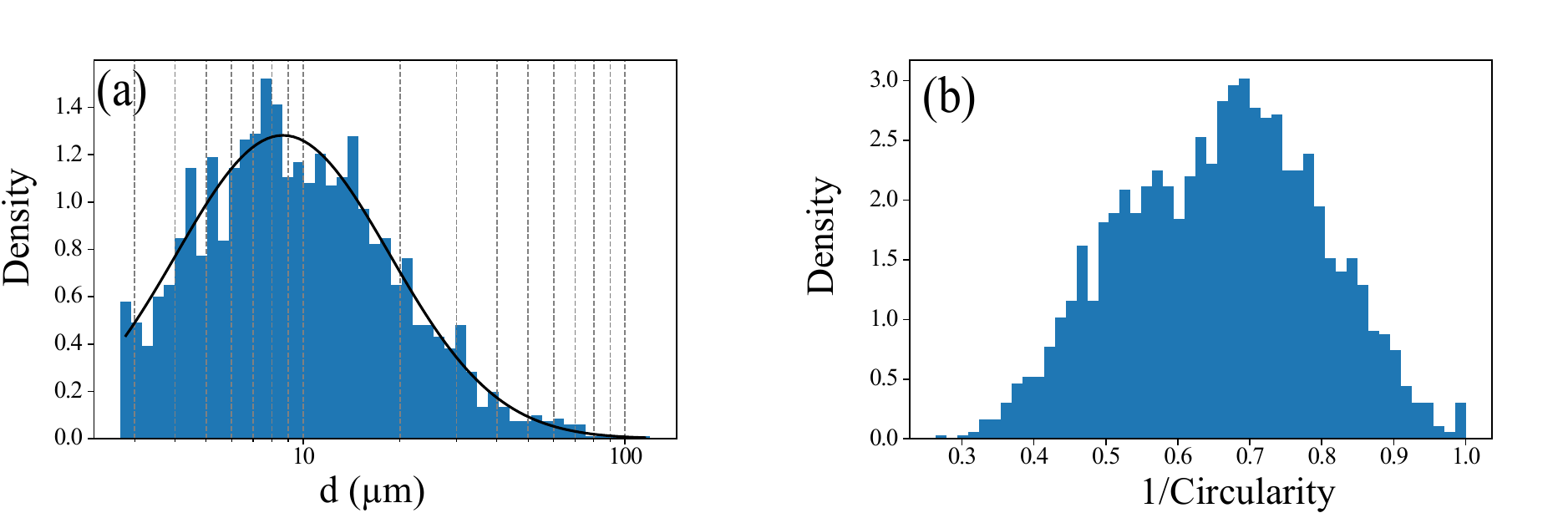} 
  \ \caption{(a) Distribution of the effective diameter of the grains composing the raft from image (a) in Figure~\ref{Macroscope1}. The data are fitted with a lognormal distribution of parameter $\mu= 2.2$ and $\sigma^2 = 0.58$. The maximum observed size is $120$ $\mu$m $> 44$ $\mu$m, this would mean that the ``\textit{watershed treatment}" (\textit{ImageJ)} does not distinguish all of the grains in contact. Moreover, {due to the resolution of the image, an arbitrary minimal cutoff diameter of 3 $\mu$m has been chosen in order well separate the particle}. {(b)} Distribution of the {inverse circularity ${1/\kappa}=(4\pi\,\mathcal{S})/(\mathcal{P}^2)$} of the grains composing the particle raft, where $\mathcal{S}$ and $\mathcal{P}$ are the surface area and the perimeter of the grains, respectively. The closer the circularity is to 1, the more circular the grains are.}
 \label{Macroscope2}
 \end{center}
\end{figure}

The thickness of the graphite layer is approximately controlled by adding a known mass of graphite powder over a known area with a precision of $0.1\,$g. For this study, we selected a mass of $11.1$ g m$^{-2}$, which corresponds to a thickness of  $4.9$\,$\mu$m for fully filled layer. In a compact 2D arrangement of spherical particles, the minimal thickness would be $6.6$\,$\mu$m. The small-scale structure of our particle raft can be visualized using a Leica Z16 APO microscope as shown in Fig.~\ref{Macroscope1}. We observe that the graphite powder used is indeed polydisperse with a large variety of sizes and shapes. Additionally, the floating layer does not completely cover the water free-surface. By employing the ``\textit{Watershed}" algorithm (\textit{ImageJ}), we evaluated the effective diameter of the grains as the square-root of the observed surface area $d=\sqrt{\mathcal{A}}$, along with the circularity parameter {${1/\kappa=}(4\pi\,\mathcal{A})/(\mathcal{P}^2)$}, where $\mathcal{A}$ and $\mathcal{P}$ represents the perimeter. The particle diameter distribution in this powder is fitted using a log-normal distribution $p(d) \propto \dfrac{1}{d}\, \mathrm{exp} \left( - \dfrac{(\mathrm{log\, d-\mu)^2}}{2\sigma^2} \right) $ as expected for a powder resulting from the fragmentation of a solid~\cite{villermaux2007fragmentation}. The results are plotted in Fig.~\ref{Macroscope2}. Although particles are not detected below 3 $\mu$m, we find that the most probable diameter is about 10 $\mu$m, which is also the typical thickness of the raft, henceforth assimilated to a particle monolayer.

By depositing a graphite powder consisting of 10 $\mu$m particles on the surface of water, a thin and loose monolayer is thus formed, creating a very thin solid floating membrane. The mechanical behavior of this membrane can be described using the standard tools of elasticity~\cite{Vella2004}. Directly measuring the elastic parameters is challenging due to the membrane's thinness. However, the appearance of wrinkles when the raft is compressed~\cite{Vella2004} provides an estimation of the Young's modulus, with a value of approximately $E \approx 0.8 \times 10^4$ Pa (see Appendix \ref{Emod}). Water surface waves can propagate in presence of this floating raft, causing deformation on its surface. Similar to sea ice, when the thickness is small enough compared to the wavelength, gravity becomes the dominant restoring force, and the mass contribution of the raft is negligible compared to that of water (refer to Section \ref{SurfaceWavespropagation}).  Consequently, we consider the surface waves deforming the interface to be akin to classical gravity surface waves, regardless of the presence or absence of a floating particle raft that may fracture or fragment. In this study, we will examine the response of the raft to the surface deformation without feedback on wave propagation.

As will be demonstrated later, the fracture of the particle raft by surface waves exhibits visual similarity to the breaking of sea ice by waves originating from the open ocean. Both scenarios involved the rupture of a thin, quasi-two dimensional, elastic floating membrane by surface waves with wavelengths much larger than the membrane thickness. However, there are several significant differences to note. Sea ice is indeed an ice layer formed through the freezing of polar ocean waters, typically having a thickness of 1 m and a Young's Modulus, $E$, of approximately 10 GPa~\cite{Schwarz1977,Mellor1986,Timco2010}. Its mechanical properties depend significantly on factors such as thickness, water salinity, porosity and age. Therefore, the sea ice is considerably more rigid and orders of magnitude thicker compared to the graphite membrane under investigation.

In addition to the scale difference, a fundamental distinction between a particle raft and sea ice lies in their cohesive origin. In the case of ice, the water molecules are chemically bonded through strong hydrogen bounds. Conversely, the bonds {due to menisci} between particles in the particle raft are significantly less energetic and easier to break. Furthermore, the capillary attraction between the floating particles that constitute the particle raft allows for some reorganization of the membrane at small scale. Fractures in the raft can ``heal" if the separation between particles is not too large, rendering them not always irreversible. A similar process may occur in sea ice, when ice floes can thermodynamically adhere to each other through the freezing of sea surface between the floes~\cite{Heorton2018}.


\subsection{Experimental setup}
\label{setup}

 \begin{figure}[h!]
 \begin{center}

 \includegraphics[width=.8\columnwidth]{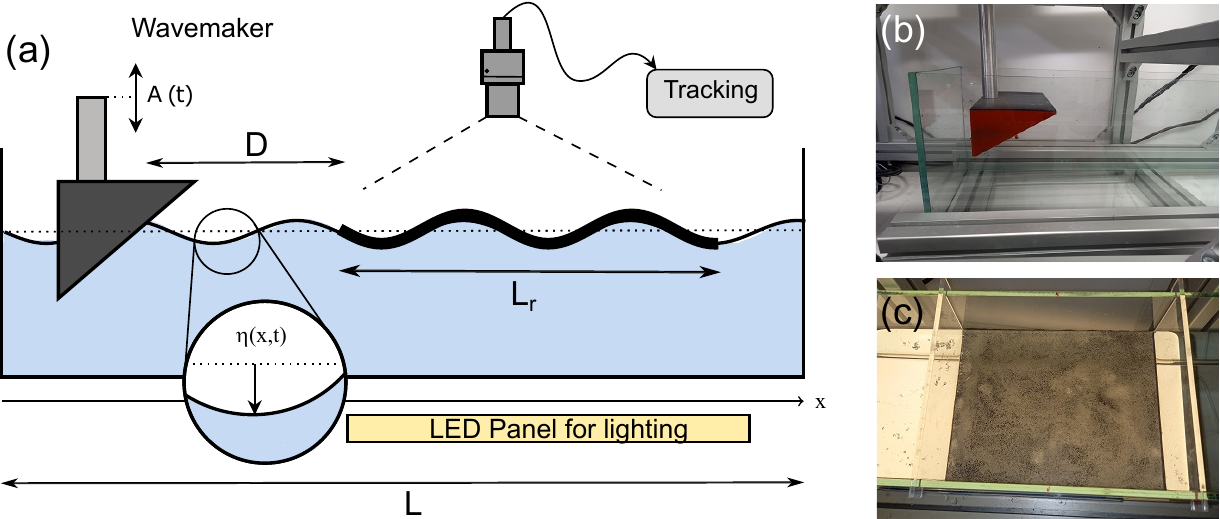}
   \caption{(a) Side view of the experimental setup. Plane waves are generated in a glass tank of length $L$ directed along the $x$ axis and width $W$ in which a graphite particle raft has been created at the surface over a length $L_r$ and all the tank width $W$. The distance between the wavemaker and the raft is noted $D$. Two tanks are used: the first is $L=300$ cm long and $W=7.5$ cm wide and is filled with tap water at a level of 12 cm, the second is $L=120$ cm long and $W=20$ cm wide and is filled at a level of 13 cm. Due to the wave propagation, the free-surface $\eta(x,t)$ is deformed. The tank is lightened from below and the raft is imaged with one or two top view cameras. As the fragments are dark on a light background, they can be easily detected and tracked. (b) Picture of the prismatic wavemaker {($10 \times 5.6 \times 18$ cm$^3$ and angle $29^\circ$)} used in the second tank. The wavemaker is located $10.5$ cm away from the left wall and is moved vertically using an electromagnetic vibrator with a prescribed motion $A(t)$. (c) Picture of the particle raft between two plastic separators, which will be removed before application of the waves.}
 \label{ExSetup}
 \end{center}
\end{figure}

The experiment consists in creating a graphite particle raft on a section of a water tank, followed by the application of gravity waves on one side of the tank. The setup is illustrated on the schematic in Fig.~\ref{ExSetup}. Two transparent glass tanks directed along the $x$ axis were used. The first tank has dimension of 3 meter in length and 7.5 centimeters in width, filled with 12 cm of tap water. A prismatic wavemaker ($10 \times 5.6 \times 7$ cm$^3$ and angle $29^\circ$) is placed on one side of the tank to generate monochromatic gravity surface waves parallel to the tank length on the frequency range $[2.5,4]$ Hz, with an amplitude less than $3$ mm. In these conditions, the wave steepness, which quantifies the deformation of the free surface,  denoted as $ k\,a $ is smaller than $0.1$, where $k$ represents the wavenumber and $a$ denotes the wave amplitude. Therefore, we consider weak deformations and neglect nonlinear effects in the wave propagation throughout this article. The wavemaker undergoes vertical oscillatory motion using an electromagnetic vibrator (TIRA TV 51144), and its prismatic shape ensures the displacement of a water volume to generate waves. The vertical motion of the wavemaker is monitored using an accelerometer. In most experiments, an inclined polystyrene wall is placed at the end of the tank to act as an absorbing boundary for the waves, resembling a ``beach" scenario. In this case, we study progressive waves. The amplitude of these waves is measured by visualizing the vertical position of the contact line as a function of $x$ through a lateral camera (not shown). The experiments do not indicate a significant change in wave amplitude in the presence or absence of the raft on the water surface, consistent with the discussion in Appendix~\ref{SurfaceWavespropagation}. The raft also does not appear to significantly alter the wave attenuation in the range of frequencies tested.

The wave generation setup is nearly identical in the second tank, which has dimensions of 1.20 m in length and 20 cm in width, with the exception of a wider wavemaker {($10 \times 5.6 \times 18$ cm$^3$ and angle $29^\circ$)}. For the experiments  conducted in this tank, the end wall is always vertical and acts as a reflecting boundary for the waves. Therefore, the forcing frequency is chosen to be 3 Hz to excite a standing waves mode. At this frequency and a water depth of $h=13$ cm, according to the dispersion relation of surface waves (see Appendix~\ref{SurfaceWavespropagation}), the wavelength reads $\lambda=(2\pi)/k =17.3$ cm and the waves are in a deep-water regime since $k\,h \gg 1$. Between the wavemaker and the end wall of the tank, the forced standing wave mode corresponds to 6 wavelengths. Additionally, several runs were performed with random waves, where the instantaneous frequency ranged between 2.5 and 3.5 Hz. To achieve this, a phase noise was generated numerically, bandpass filtered to maintain a constant wave amplitude, and then applied to the electromagnetic vibrator, which vertically displaces the wavemaker. Under these conditions, the overall structure of the wave field is preserved, while the heterogeneity of the wave field amplitude resulting from standing wave modes is reduced. In this second tank, the wave amplitude is measured using a capacitance water height gauge~\cite{mcgoldrick1971sensitive} at the location of the raft, which provides the local wave amplitude. \\

To reduce dissipation at the contact line, we conducted additional experiments using a fine nylon fabric with a mesh of 250 $\mu$m applied to the vertical side walls. This hydrophilic fabric exhibits a phenomenon known as hemiwicking~\cite{kim2016dynamics}, where water is captured on the mesh, replacing the pinning of the contact line with smooth sliding. As a result, the wave dissipation caused by contact line motion is reduced, and the wave field becomes homogeneous perpendicular to its propagation direction~\cite{monsalve2022space}. In our experiments, we did not observe a strong difference in the propagation of the surface waves, but the lateral boundary conditions for the particular raft are strongly modified. \\

The raft and the formed fragments are imaged from above using two cameras, while a LED panel illuminates the transparent tank from below. The first camera is a scientific 16 bits CMOS PCO Edge camera with a resolution of 2560 $\times$ 2160 pixels equiped with a 35 mm focal length lens. The second camera is an 8 bits CMOS Basler Ace camera (2048 $\times$ 2048 pixels) with a 50 mm focal length lens. To ensure sufficient spatial resolution, only a portion of the tank is captured by the cameras. Due to the strong light absorption of the graphite raft and fragments, they appear as dark regions against a bright background representing the free water surface. After thresholding and binarization of the images using clustering algorithms, the position and dimensions of each fragment can be measured in each image. \\

During a typical experiment, the graphite particle raft is initially created  following the protocol described in section~\ref{raft}, spanning the entire width $W$ and length $L_r$ at a distance $D$ from the wavemaker, using two plastic separators. Once the separators are carefully removed, image acquisition is launched, and the wavemaker is activated to generate the gravity surface waves that arrive with a normal incidence on the raft. The evolution of the raft and its fragmentation is observed over time with the top cameras. The frame rate is often set equal to the frequency of the waves to facilitate visualization of the breaking process and the gradual drift of the fragments by removing their oscillatory motions associated with the waves.

\section{Experimental Observations}
\label{Observations}
\subsection{Progressive waves}

 \begin{figure}[h!]
 \begin{center}

 \includegraphics[width=1\columnwidth]{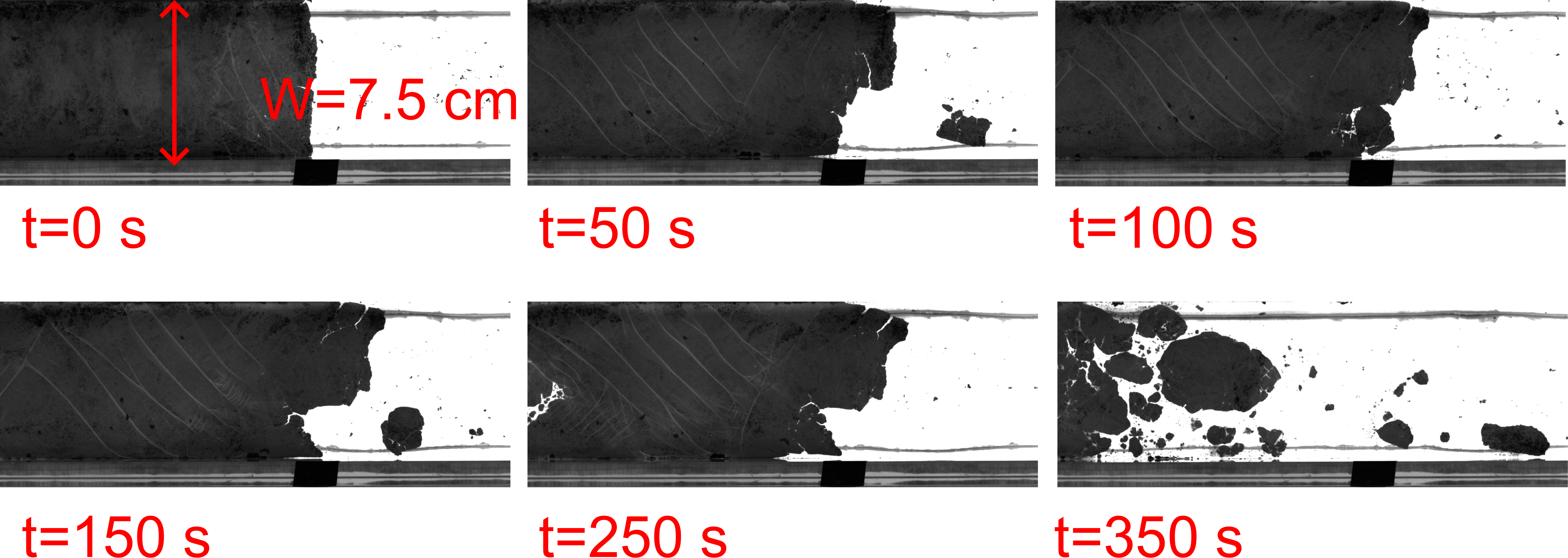}

   \caption{Fragmentation of the graphite particle raft seen from above for selected time steps in the tank 1 for progressive waves {(PCO Edge camera, resolution 0.19 mm per pixel)}. Regular sine waves of $f=3.0$ Hz and $a=2.20$ mm, coming from the right. Tank width, $W=7.5$ cm, raft length $L_r=30$ cm, distance between the raft and the wavemaker $D=70$ cm. Fragments are detached and are carried away from the raft due to a drift flow occuring on a time scale long compared to the wave period. (See Movie S1 in SI).}
 \label{MosProgWaves}
 \end{center}
\end{figure}

The fragmentation of the raft under the influence of progressive waves in tank 1 (3 meters long) is examined. A top view of the fragmentation is presented in Fig.~\ref{MosProgWaves} and Movie S1. After a few cycles of waves, cracks begin to appear at the edges of the raft when the wave amplitude is sufficiently large. These cracks can be oriented either perpendicular or parallel to the direction of wave propagation. They propagate gradually through the raft and may form branching structures. Simultaneously, oblique cracks with an angle of $45^\circ$ emerge within the bulk of the raft, spaced approximately $8$ mm apart. In the case of progressive waves, the wave field is generally homogeneous, and the cracks do not appear at specific locations. Over a time period longer  than the wave period, the fractures divide the raft into fragments that detach from each other and appear to be carried by an underlying flow, causing the dispersion of the fragments. It is also observed that the oblique cracks allow for significant mechanical rearrangements, leading to a complex rheological behaviour at larger scales. Finally, it is worth noting that once the raft has been weakened by numerous internal cracks, the disintegration of the raft into fragments can occur relatively rapidly, such as between $t=250$ and $t=350$ s, while raft remains quite similar between $t=150$ and $t=250$ s.

 \begin{figure}[h!]
 \begin{center}

 \includegraphics[width=1\columnwidth]{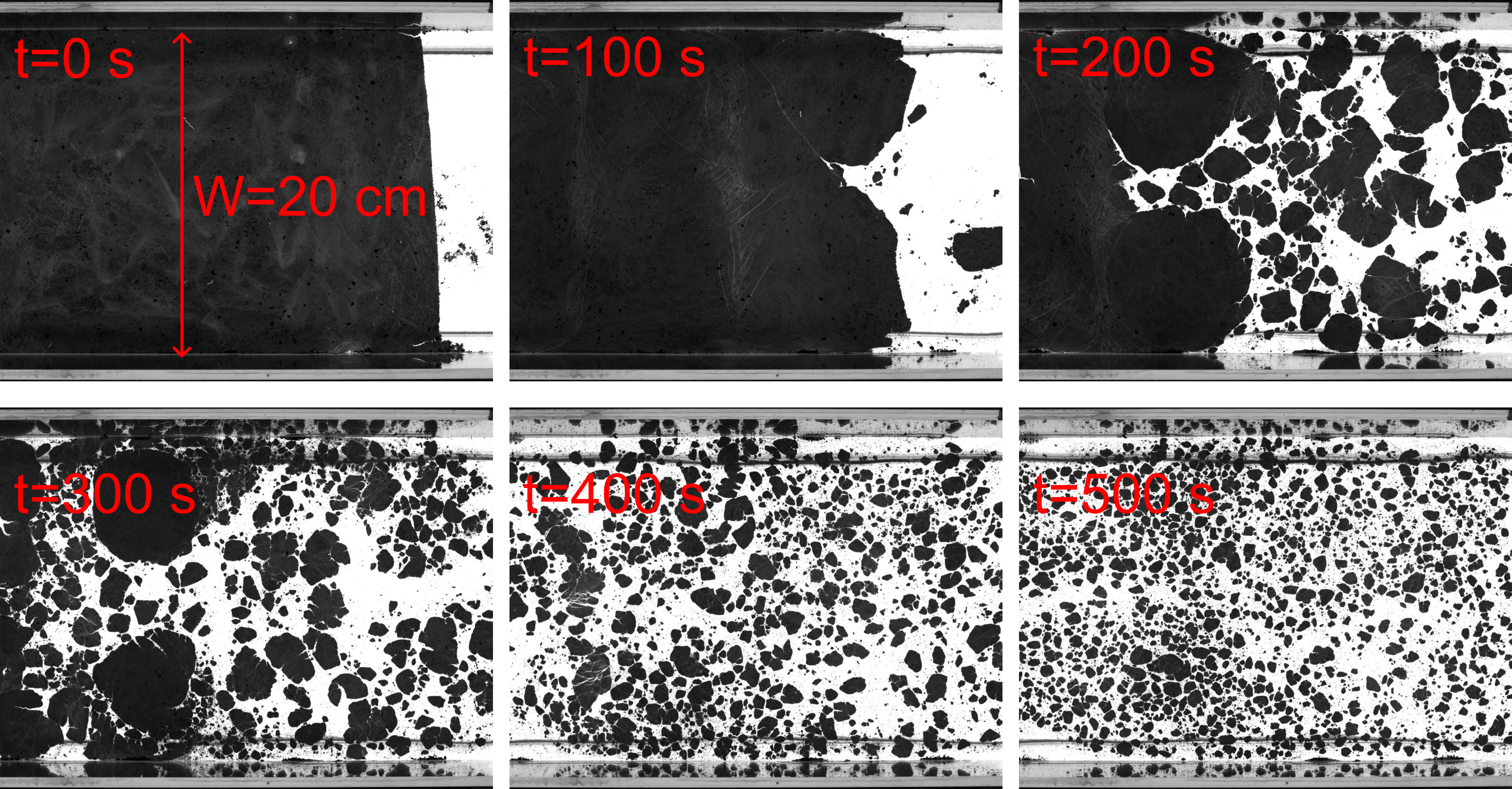}

   \caption{Fragmentation of a graphite particle raft for standing waves ($f=3.0$ Hz and $a=1.48$ mm) in the tank 2 {(PCO Edge camera, resolution 0.22 mm per pixel)}. The wavemaker is located on the right. Initially the raft length is $L_r=30$ cm and its distance to the wavemaker is $D=52.8$ cm. At $t=100$ s, we observe at the position of an antinode, numerous small scale cracks, which will develop further. Then, polygonal fragments are produced and move due to the presence of a drift flow, which induce a net displacement on timescales larger than the wave period. We note, that this flow induce circular trajectories, which participate to the fragmentation. Fragments become more and more smaller with time. (See Movie S2 in SI).}
 \label{MosStandingWaves}
 \end{center}
\end{figure}

\subsection{Standing waves}
\label{ObservationsStandingWaves}
\subsubsection*{Regular waves}

Next, we consider the experiments conducted in the second tank, which has different dimensions ($120$ cm long and $20$ cm wide). To achieve a standing wave mode, we choose a wave forcing frequency of $f=3$ Hz, corresponding to a wavelength of $\lambda={17.3}$ cm. This setup results in a standing wave mode between the end of the wave tank and the inclined wall of the wavemaker, with a length about $104.5\,$ cm, close to $6$ times the wavelength $\lambda$. The fragmentation process for a wave amplitude of $1.48$ mm is illustrated in Fig.~\ref{MosStandingWaves} and Movie S2. Again, when the wave amplitude is sufficient, the raft progressively breaks over a time scale longer than the wave period ($T=0.33$ s), generating fragments that are carried by an underlying drift flow. Initially, in addition to the horizontal and vertical fractures appearing at the edges of the raft, small cracks are predominantly observed within the raft, specifically at the positions of the antinodes where the wave amplitude is maximum. These cracks contribute to the fragility of the raft and aid in the fragmentation process. Notably, rotational motions are observed, resulting in large-scale fractures and the formation of some of the largest fragments. These motions indicate the presence of a long-term drift flow associated with the waves. Subsequently, the fragments are further transported by the surface flow, and recirculation cells with a width equal to half of the tank become apparent on the water surface. The fragments gradually decrease in size due to successive breaking events and the loss of graphite grains from the edges of these floating particle rafts.


\subsubsection*{Regular waves and sliding walls}

\begin{figure}[h!]
 \begin{center}
 \includegraphics[width=1\columnwidth]{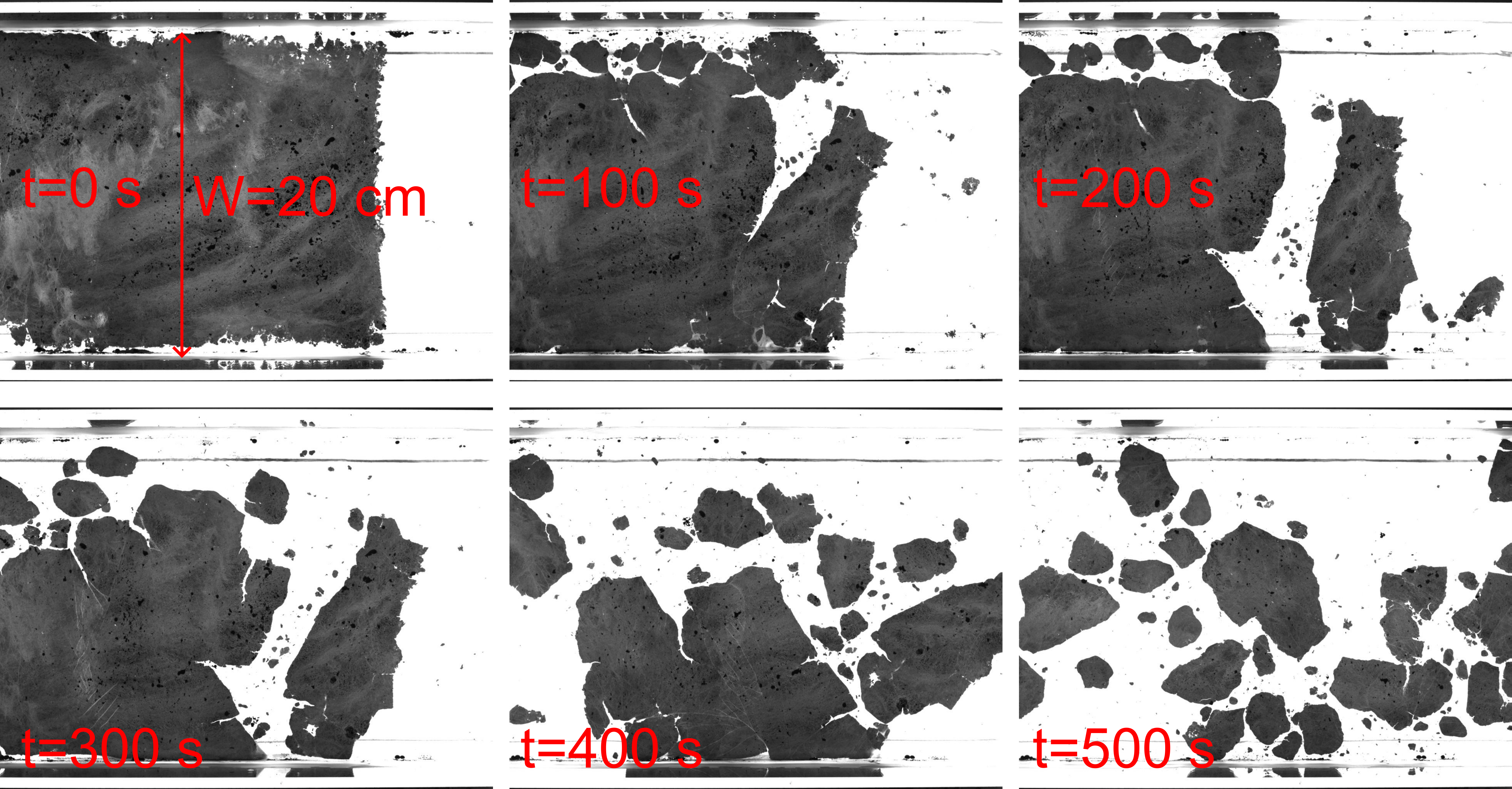}
   \caption{Fragmentation of a graphite particle raft for standing waves ($f=3.0$ Hz and $a=1.43$ mm) in the tank 2 in presence of a nylon fabric ensuring sliding of the contact line {(PCO Edge camera, resolution 0.22 mm per pixel)}. The initial length of the raft is $L_r=30$ cm. The wavemaker is located on the right at a distance $D=52.8$ cm from the raft edge. We note that the raft does not adhere tightly with the glass walls due to the change of wetting conditions. The raft is progressively broken and fragmented, generating polygonal floes which are carried away by a surface flow. (See Movie S3 in SI).}
 \label{MosStandingWavesTissu}
 \end{center}
\end{figure}

In the previous experiment, the lateral glass walls were left uncovered, and the dynamics of the air/water contact played a significant role. The partial pinning and hysteresis of the contact line resulted in increased wave dissipation and change to the structure of the wave field~\cite{monsalve2022space}, which deviated from perfect 
one-dimensionality. Furthermore, in the presence of a particle raft, the boundary conditions were likely influenced by the dynamics of the contact line at the edges. To test this effect, we conducted an experiment where a fine nylon fabric with a mesh size of $250$ $\mu$m was applied to the internal vertical glass walls. Under these conditions, the air/water contact line exhibited a smooth sliding motion~\cite{monsalve2022space}. We assume that a similar sliding motion occurs when the water surface is covered by graphite particles. The fragmentation process in the presence of regular standing waves and sliding walls is depicted in Fig.~\ref{MosStandingWavesTissu} and Movie S3. The process is qualitatively similar; however, we observe that the lateral edges of the raft do not appear to be firmly attached to the vertical walls. Consequently, the raft has more freedom to displace along the $x$ axis, and for a same wave amplitude level, the fragmentation appears to be less intense. Once again, over longer time scales, the fragments follow a surface flow, which now appears to consist of a large recirculation cell spanning the entire width of the tank. While this particular boundary condition has not been systematically tested, this experiment demonstrates the crucial importance of the lateral boundary conditions in the fragmentation mechanism of the particle rafts, and potentially other floating solid membranes.

\subsubsection*{Random waves}

 \begin{figure}[h!]
 \begin{center}
 \includegraphics[width=1\columnwidth]{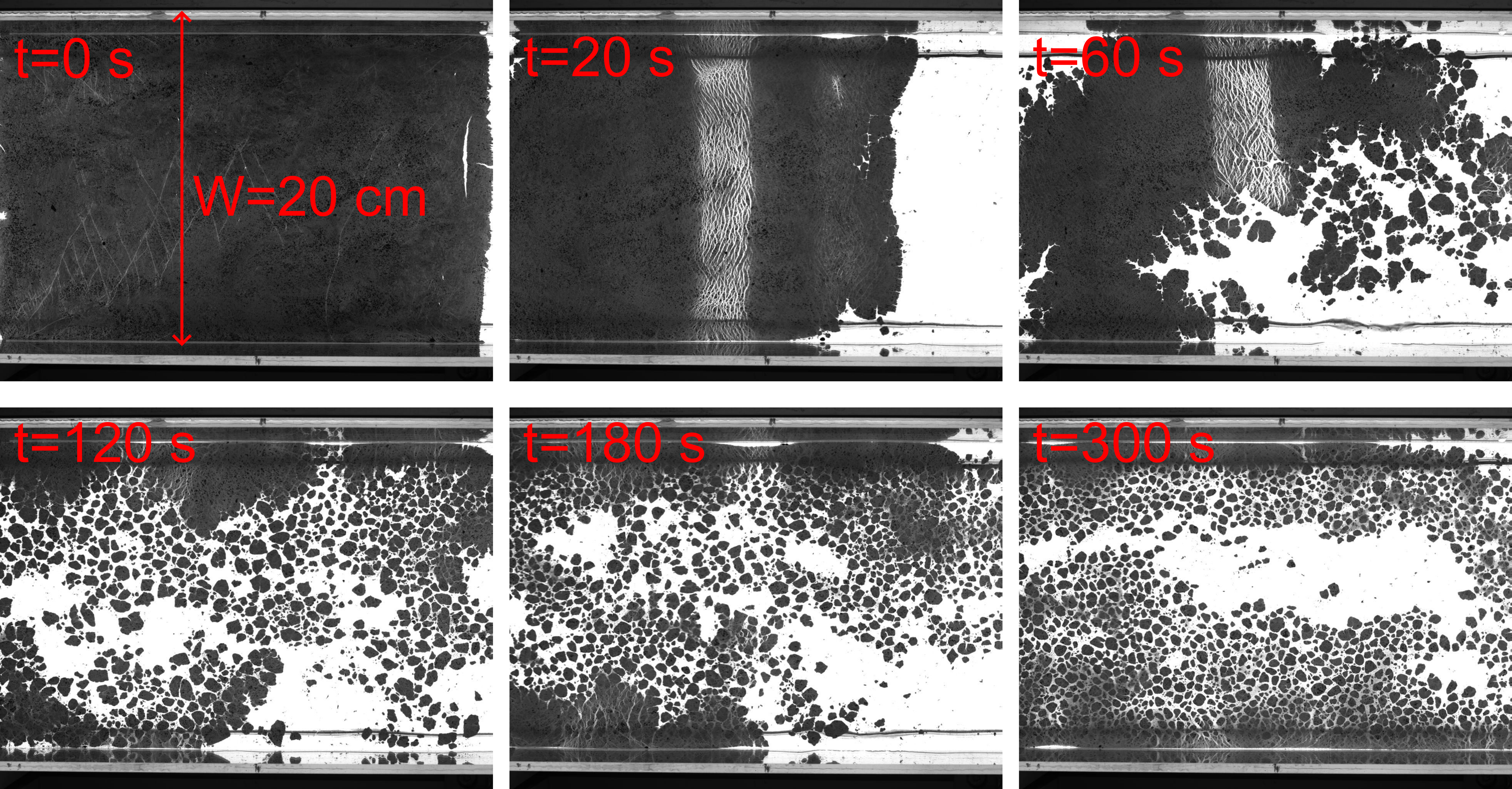}
   \caption{Fragmentation of a graphite particle raft for random waves ($f \in [2.5,3.5]$ Hz and $a=1.45$ mm) in the tank 2 {(PCO Edge camera, resolution 0.22 mm per pixel, frame rate 3 Hz)}. The initial length of the raft is $L_r=30$ cm. The wavemaker is located on the right at a distance $D=52.8$ cm from the raft edge. In the first steps, at the positions of local maxima, we observe transient array of oblique fracture network. Then, the raft is progressively broken and fragmented. We note that with random waves, the raft is broken more rapidly into small fragments and that the drift of the fragments due to the surface flow is reduced. (See Movie S4 in SI).}
 \label{MosRandom}
 \end{center}
\end{figure}


In order to minimize the cumulative wave-induced drift and achieve a more homogeneous excitation of the raft, we conducted additional experiments in tank 2 with uncovered glass walls and random waves. We generated phase noise with an instantaneous frequency ranging from 2.5 Hz to 3.5 Hz while maintaining a constant amplitude. The correlation time of the wave forcing was approximately one second. As a result, the positions of the nodes and antinodes varied throughout the duration of the experiment. It is expected that eigenmodes corresponding to constructive interferences would emerge over longer periods due to reflections on the walls. 

An example of the fragmentation process is shown in Fig.~\ref{MosRandom} and Movie S4. We observed that the fragmentation process occured more rapidly compared to the experiment shown in Fig.~\ref{MosStandingWaves}. At both $t=30$ s and $t=60$ s into the experiment, an array of small-scale cracks appeared at positions of maximum wave amplitude on the raft. Similarly, over longer durations, the raft disintegrated into smaller fragments.

\newpage
\subsection{Wave induced drift flow}

The previous experiments have clearly demonstrated that, over time scales longer than the wave period, the produced fragments undergo a net motion, despite the oscillatory velocity field associated with the waves (propagative or standing). The applied forcing consists of gravity surface waves propagating along the $x$ axis, which reveals the presence of a drift flow at the water surface. For progressive waves, the most well-known example of such a flow is the Stokes drift~\cite{vandenBremer2018}, where particles following the Lagrangian velocity field associated with the wave experience a net motion in the direction of wave propagation. According to this nonlinear effect of kinematics origin, the particle velocity at the free surface denoted as $u_{SD}$, can be approximated as $u_{SD}=c_\phi\,(a \, k)^2$, at the leading order, where $c_{\phi}$ is the phase velocity and $a \,k$ is the wave steepness (a dimensionless nonlinear parameter). For a forcing frequency of 3 Hz and millimeter amplitudes, $u_{SD}$ is on the order of a few millimeters per second. However, this drift is directed only along the x-axis, causing the fragments to move away from the wave generation area. Other explanations are thus needed. In fact, in wave experiments at the meter scale or below, although not well documented,  wave induced streaming effects are significant and generate large recirculation flows. We interpret this flow generation by a non-linear effect injecting vorticity on a time scale of a few periods, similar to previous observations in Faraday wave experiments~\cite{vonKameke2011,Francois2013,perinet2017streaming} or progressive waves near wavemaker~\cite{punzmann2014generation}.
In our experiments, at the end of the fragmentation process, we use the very small floes formed as floating tracers to measure the surface flow using Particle Image Velocimetry (PIV) algorithms (see Appendix~\ref{PIV}). We find that the surface mean flow is an order of magnitude smaller than the orbital flow. While this secondary flow is likely negligible in the fracture process compared to the orbital flow, it plays a crucial role in the drift of the floating fragments due to its cumulative effect. However, the precise geometry of the mean surface flow is not reproducible and appears to depend heavily on the experiment geometry and environmental conditions. Further dedicated studies would be valuable to quantify these streaming effects, but they are beyond the scope of this study.

\section{Breaking of a floating particle raft}
Motivated by the aforementioned observations, in this section, we aim to describe the initial breaking and failure of a floating particle raft subjected to waves with a wavelength much larger than the raft's thickness. As previously seen, the raft does not significantly modify the wave propagation, and in the first approximation, the deformation of the free-surface due to the wave is applied to the raft{. In this section, we make another important assumption: we consider the particle rafts as a thin homogeneous elastic plate with a three-dimensional Young's Modulus, although the raft is a monolayer of granular particles. This approach has been used to model the buckling and the wrinkling of particle raft for a large range of particle size and shape~\cite{Vella2004,protiere2023particle}, as well as to describe the propagation of hydro-elastic waves on particle-laden interfaces~\cite{Planchette2012}. In our experiments, the} appearance of straight cracks in the fragments suggests that the raft can be considered as a brittle material, undergoing irreversible breaking after linear elastic deformation. For this quasi-2D plate under weak deformation, the in-plane stress resulting from the deformation can induce either Opening fracture mode (Mode I) when the tensile stress is normal to the crack or Sliding fracture mode (Mode II - also know as Shearing Mode) when the shear stress acts parallel to the crack. The tearing mode (Mode III) is not considered since stresses perpendicular to the raft are not relevant.

\subsection{Breaking by bending}
\label{Bending}

In order to model the breaking process and make predictions about the fracture geometry, we consider now the raft as an elastic thin plate that deforms according to the vertical displacement of the free surface associated with a progressive wave of amplitude $a$:
\begin{equation}
\xi(x,t)=a\,\cos (k\,x- \omega t) \, .
\label{Def}
\end{equation}
Although experimentally challenging to achieve, we assume that the raft, with a thickness $e$, has homogeneous and isotropic mechanical properties. For a thin plate, two deformations modes can be observed: stretching and bending. {When the layer is compressed, the tensile deformation corresponds to a displacement aligned with the initial plane of the layer.} Conversely, in a bending deformation, the plate will bend and deviate from its initial plane. Let us define $\sigma$ as the local stress in the plate and $\epsilon$ as the corresponding strain, which are related through Hooke's law: $\sigma \sim E \, \epsilon$.

As the edges of the raft are not constrained to stay at the same $x$ location during the wave propagation and are free to move along the $x$ direction, we neglect the contribution of stretching along the $x$ axis and we try now to study with further details the breaking by bending.

\begin{figure}[h!]
\centering
\includegraphics[width=.5\textwidth]{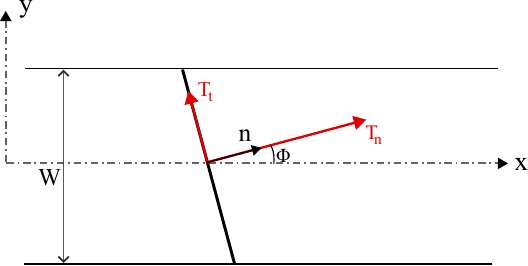}
\caption{Schematics showing a top view of the granular raft at the first breaking onset. The black oblique segment represents a fracture.}
\label{schemafrac}
\end{figure}

We consider the case where the elastic raft follow the deformation of the free-surface due to the wave propagation. As the raft is not constrained along the $x$ axis, its length is supposed constant and we neglect first the stretching. However, when the raft adopts a sinusoidal shape, it becomes subjected to a flexural strain, which leads to breaking by bending. Note that for sea ice breaking by oceanic waves, bending effects are assumed to be the main mechanism of fracture~\cite{dumont2011wave,voermans2020experimental}.
In this bending dominated scenario and according to the thin plate theory~\cite{LandauElasticity}, the components of the stress tensor read:

\begin{equation}
  \left\{
      \begin{aligned}
     {\sigma_{xx}} & =  - \dfrac{ E \, z}{1-\nu^2}\,\dfrac{\partial^2 \xi}{\partial x^2}=\dfrac{ E \, z}{1-\nu^2}\,a\,k^2\,\cos (k\,x - \omega t) \\
       {\sigma_{yy}} & =  - \dfrac{ E \, z}{1-\nu^2}\,\nu \,\dfrac{\partial^2 \xi}{\partial x^2}=\nu \, {\sigma_{xx}} \\
     \sigma_{xy}& =  - \dfrac{ E \, z}{1+\nu}\,\dfrac{\partial^2 \xi}{\partial x \, \partial y}=0
      \end{aligned}
    \right. \, ,
\end{equation}
with $E$ the Young's modulus of the raft and $\nu$ its Poisson coefficient. $z$ is the vertical component along the thickness of the raft, which is $0$ in the middle plane of the raft. As we aim to evaluate the maximal stress, we replace $z$ by $e/2$, because the stress is maximal at the bottom and top boundaries of the raft.\\
We seek to determine where the fractures will occur and which mode of fracture is selected between opening and shearing.
Consider a fracture whose normal $\mathbf{n}$ makes an angle $\phi$ with the $x$ axis (see Fig.~\ref{schemafrac}). The stress on this fracture reads~\cite{OswaldBook}{, where $ {\mathbf{e}_x}$ and $ {\mathbf{e}_y}$ are the base unit vectors}:
\begin{equation}
\mathbf{T}= {  \mathbf{\sigma} \cdot \mathbf{n}  }={\sigma_{xx}} \, ( \cos \phi \,\mathbf{e}_x + \nu  \sin \phi \, \mathbf{e}_y )
\label{StressT}
\end{equation}
Then, one can deduce the normal component to the fracture:
\begin{equation}
T_n =  \mathbf{T} \cdot \mathbf{n} = {\sigma_{xx}} \, [1-(1-\nu)\, \sin^2 \phi] \, ,
\end{equation}
and the tangential component:
\begin{equation}
\mathbf{T_t}=  \mathbf{T} - T_n \, \mathbf{n}= {\sigma_{xx}} \,(1-\nu)\, \sin \phi \, \cos \phi \, (\, \sin \phi   \,\mathbf{e}_x + \cos \phi  \, \mathbf{e}_y ) \,
\end{equation}
whose norm is:
\begin{equation}
T_t =\frac{1}{4} \, (1-\nu) \, {\sigma_{xx}} |\sin (2 \phi)| 
\end{equation}
The opening fracture modes are caused by the normal stress, which are maximal for $\phi=0$ ($\nu <1/2$):
\begin{equation}
T_ {n,\mathrm{max}}={\sigma_{xx}}=\frac{1}{2}\,\dfrac{E\, e}{1-\nu^2} \, a \, k^2 | \cos (k\,x-\omega\,t) | \,.
\label{TnBendingOp}
\end{equation}
Therefore, opening fractures due to bending must be perpendicular to the propagation direction of the waves. This reasoning holds also for standing waves and in that case, the fracture occurs on the antinodes, where the deformation is maximum.
In contrast, shearing fractures originate from the tangential stress, which is maximal for $\phi= \pm 45^\circ$:
\begin{equation}
T_ {t,\mathrm{max}}=\frac{1}{4}\,\dfrac{E\, e}{1+\nu} \, a \, k^2 | \cos (k\,x-\omega\,t) | \,.
\end{equation}
Shearing fractures are thus oblique as commonly observed for the compression or traction of homogeneous isotropric brittle materials in three dimensions. For standing waves, the fractures must occur again on the waves maxima, \textit{i.e.} on the antinodes.
Under the conditions of our experiment, we use a graphite particle raft with a Young's modulus $E  \sim 10^4$ Pa, a Poisson ratio of $\nu \approx 1/3$ and a thickness $e \sim 10$ $\mu$m. The frequency $f$, the amplitude $a$, are of order of $3$ Hz and $1$ mm, respectively. We find thus, the orders of magnitude of the maximal stresses, $T_ {n,\mathrm{max}} \approx 70 \times  10^{-3} $ Pa and $T_ {t,\mathrm{max}} \approx 1/3\,T_ {n,\mathrm{max}}$. These values are very weak compared to the Young's Modulus of the raft even if a brittle material breaks for stresses small compared to its Young's Modulus. Therefore, for the low frequencies used, we conclude that our particle rafts do not break by a bending mechanism like is the case for sea ice. To observe such a mechanism in a laboratory scale experiment, higher raft thickness and wave frequency are needed.

\subsection{Breaking caused by the viscous stresses exerted by the waves}
\label{ViscousStresses}
\begin{figure}[h!]
\centering

\includegraphics[width=1\textwidth]{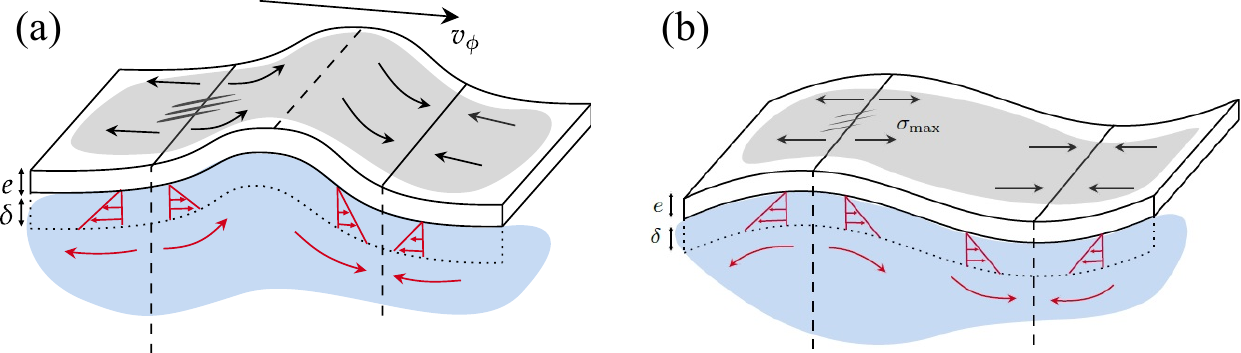}
\caption{Schematics of breaking of a particle raft of thickness $e$ by viscous stresses in presence (a) of progressive waves on the left and (b) of standing waves on the right. For progressive waves at a given time, the near surface velocity field due to wave propagation plotted in red vanishes at the zeros of the wave elevation. For the zeros behind the maxima, the  diverging flow can create a fracture, but these locations propagate with the wave. In contrast, the velocity field in red associated to standing waves displays at the time of maximum amplitude at the antinodes a diverging flow at the maxima and a converging flow at the minima. The diverging flow can initiate a fracture due to the resulting viscous stress on the direction of the velocity field associated to the waves in red. Its magnitude decreases to zero when approaching the raft on a length $\delta$ defining a viscous boundary layer. However, the viscous stress acts on the bottom face of the raft, whereas the rupture occurs on a surface of thickness $e$. The maximal viscous stress $\sigma_{max,xx}$ is thus obtained by an integration of $\sigma_{xz}$.} 
\label{schemaviscous}
\end{figure}

In the previous sections, we considered the elastic stresses on the raft directly related to its deformation. However, the linear theory of water waves demonstrates that wave propagation is associated with a potential flow in the liquid phase. Moreover, the displacements of the fragments indicate that they follow a velocity field at the free surface, which is due to a nonlinear streaming effect (see Appendix~\ref{PIV}) and contributes to the breaking of the raft and the drift of the fragments. Therefore, we need to quantify the interactions between a floating particle raft and the flow in the underlying liquid phase. Since the thickness of the rafts used is on the order of a few microns, we neglect the inertial effects resulting from a change in the position of the center of mass of the raft in the presence of waves. Furthermore, for the wavelengths used, the wave propagation is not significantly affected by the raft, and thus the waves are not reflected by the raft. Hence, we also neglect the radiation pressure. The main contribution arises from the viscous stresses on the bottom side of the raft, which behaves as a solid wall.
If the raft is held in place by the lateral walls and remains motionless, the fluid velocity must be zero, despite the orbital flow related to wave propagation. Below the raft, near the wall, the velocity gradient accommodates this constraint within a viscous boundary layer of thickness $\delta$, {as depicted Fig.~\ref{schemaviscous} (a)}. This generates a viscous stress $\sigma_{v,xz} = \eta\, \dfrac{\partial u_x}{\partial z}$ on the raft, where $\eta$ is the dynamic viscosity of water and $u_x$ the component of the velocity field assumed to be directed along the $x$ axis. For a free fragment, this stress forces it to follow the flow. In the case of an oscillatory flow at an angular frequency $\omega$, such as the one associated with a water wave, the characteristic length of the boundary layer is $\delta \sim \sqrt{\nu_w/\omega}$ where $\nu_w=\eta/\rho$ being the kinematic viscosity of water. In the conditions of our experiments, with $f\approx 3$ Hz and $\nu_w \approx 10^{-6}$ m$^2$ s$^{-1}$ , $\delta \approx 0.2 $ mm, which is small compared to the wavelength $\lambda$ but large in front of the raft thickness.
We first consider a motionless raft before breaking, subjected to linear progressive waves of amplitude $a$. As the surface velocity field is $ u(x,z=\xi,t) = a\,\omega\, \cos (k\, x - \omega\, t)$, for small deformations $\xi$ of the raft, the viscous stress below the raft reads:
\begin{equation}
\sigma_{v,xz} \approx \eta\, \dfrac{u(x,z=\xi,t)}{\delta} \approx (\eta \rho)^{1/2} \, \omega^{3/2}\, a \, \cos (k\,x - \omega \, t) \,.
\label{viscousprog}
\end{equation}
Nevertheless, the raft breaks along a horizontal fracture on a lateral surface of thickness $e$ and the relevant stress is $\sigma_{xx}$ if we suppose for now an initial fracture perpendicular to the waves.  This stress can be obtained by a force balance by considering the raft as a solid and considering the cumulative contribution of each slice of width $ \mathrm{d}x$. The horizontal force in magnitude is thus equal $\mathrm{d}F=\sigma_{x,z} \, \mathrm{d}x \, \mathrm{d}y = e \, \mathrm{d}\sigma_{xx}   \, \mathrm{d}y $.
Then, by integration of  $\mathrm{d}\sigma_{xx} $ and using the dispersion relation of water surface wave $\omega^2=g\,k$, we obtain:
  \begin{equation}
\sigma_{v,xx}(x,t) \approx \dfrac{\sqrt{\nu/\omega}}{e} \, \rho \, g \, a  \, \sin(k\,x - \omega \,t) \, .
\end{equation}
{The integration constant is $0$ by symmetry argument: the stress exerted at $x$ must be the opposite at $x + \lambda/2$.} We note that $\sigma_{v,xx}(x,t)$ is thus maximal {in the wave slope minima}. Using the same reasoning as in the section~\ref{Bending} (by replacing $\sigma_xx$ by $\sigma_{v,xx}$ in Eq.~\ref{StressT}), we find that shear fracture is favored for a fracture at a $45^\circ$ angle. While for the opening fracture, it is favored when the fracture is orthogonal to the wave propagation.

$\sigma_{v,xx}(x,t)$ increases linearly with the wave amplitude $a$. On the other hand, it decreases when the angular frequency $\omega$ increases. This may seem counter-intuitive because the velocity at the water surface increases with $\omega$. In fact, it should also be noted that the lower the angular frequency, the longer the wavelength of the waves and thus the more viscosity stresses will be able to exert on large areas under the raft. Thus, the sum of the forces becomes larger. We also see that $\sigma_{v,xx}$ is maximum on the maxima of the waves. In our experiment for $f \sim 3$ Hz and $a \sim 1$ mm, the maximum stresses on the raft at these points are :
$$ || \sigma_{v,xx} ||_{max} \sim 230 \, \mathrm{Pa} \, .$$
This value is {two orders} below than the estimated Young's modulus $E$ of the raft that is about $10^4$ Pa, but seems sufficient to break it, as we know that for brittle materials with defects at small scale, the experimental breaking stress threshold is small compared to the Young's modulus. \\

For standing waves, the situation differs due to the presence of stagnation points for the surface fluid velocity, where the viscous shear is maximal as depicted in Fig.~\ref{schemaviscous} (b).

We aim now to estimate the maximal axial stress directed along the $x$ axis at a standing wave antinode. First, for small wave steepness, the linear theory of water waves predicts that for a surface deformation 
\begin{equation}
\xi(x,t)=a\,\cos(k\,x)\,  \cos(\omega\,t)
\end{equation}
 the surface velocity field reads:
\begin{equation}
u_x(x,t)=a\,\omega\,\sin(k\,x)\,\sin(\omega\,t)
\end{equation}
The local viscous stress $\sigma_v(x,t)$ is estimated using the thickness of the viscous boundary layer $\delta$:
\begin{equation}
\sigma_{v,xz}(x,t)=\eta\,\dfrac{\partial u}{\partial z} \approx \eta \, \dfrac{u(x,z=\xi,t)}{\delta} \approx \dfrac{\eta\,a\,\omega}{\delta} \, \sin(k\,x)\,\sin(\omega\,t)
\end{equation}
  Again, the viscous stress acts on the bottom face of the raft, whereas the rupture occurs on the lateral surface of the raft of thickness $e$, with a stress $\sigma_{xx}$. 
By the same reasoning than for propagative waves, we obtain for standing waves:
  \begin{equation}
\sigma_{v,xx}(x,t) \approx -\dfrac{\sqrt{\nu/\omega}}{e} \, \rho \, g \, a  \, \cos(k\,x)\,\sin(\omega\,t)
\end{equation}
The areas of maximal stresses are now located on the wave antinodes and the corresponding magnitude is also of order $|| \sigma_{v,xx} ||_{max} \sim 10^2$ Pa.
Like for progressive waves, this is {two orders} of magnitude less than the Young's modulus $E$. Under our experimental conditions, the viscous stresses are therefore the most likely candidate to explain the breaking of the particles raft.

\section{Experimental characterization at the onset of breaking}

\subsection{Fracture patterns}
\label{Fracturepatterns}
\begin{figure}[h!]
\centering
\includegraphics[width=.78\textwidth]{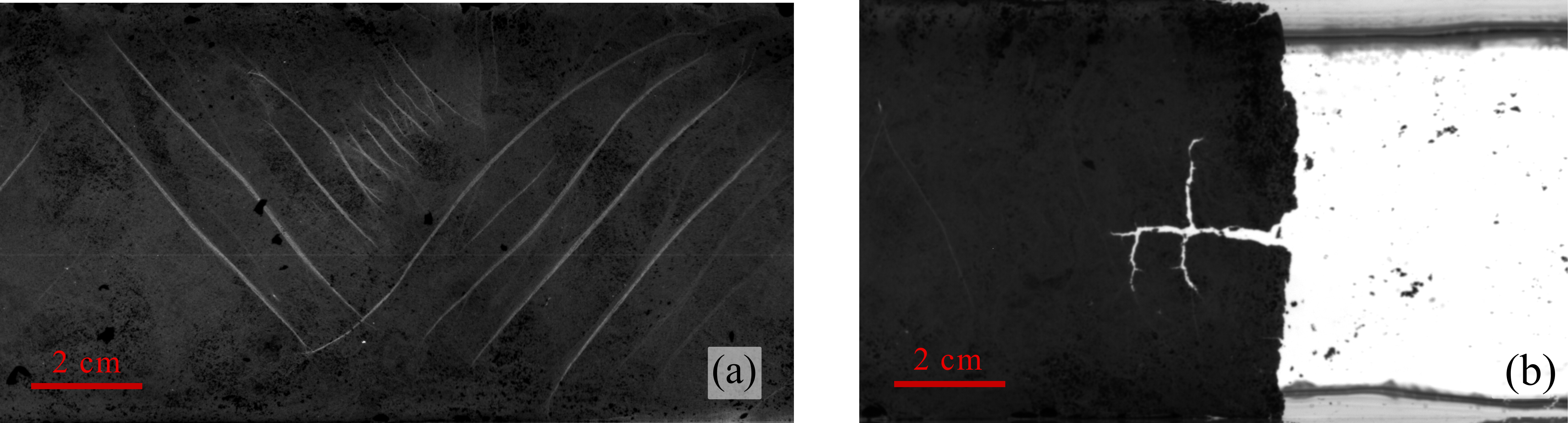}
\caption{(a)  Example of oblique fractures made by shearing in the bulk of the raft before breaking. They can be oriented at either $+45$ or $-45^\circ$. A characteristic distance between fractures about $8$ mm seems to emerge, but without a clear periodicity. Parameters of the incoming surface waves. $a = 2.1 \pm 0.1$ mm, $\lambda = {17.3}$ cm, $f = 3$ Hz, tank 1, $W=7.5$ cm. (b) Branching fracture at the edge of the raft at the first steps of the breaking process. The  rupture occurs by an opening fracture mode. Parameters of the incoming surface waves. $a = 2.2 \pm 0.1$ mm, $\lambda = {17.3}$ cm, $f = 3$ Hz. tank 1, $W=7.5$ cm.}
\label{FractModes}
\end{figure}

The experiments presented in the section~\ref{Observations} show at the beginning of the fragmentation process specific fractures patterns, which may infer about the physical mechanisms at play. Typically, for progressive waves we observe as shown in Fig.~\ref{FractModes} the following:
\begin{itemize}
\item oblique fractures that propagate rapidly. They result from shear mode fracture. They are rather localized in the bulk of the particle raft. Note that the fast propagation of cracks is not resolved experimentally in time.
\item horizontal and vertical fractures on the edges of the raft which open and close periodically. Their formation is rather slow and their structure can be ramified or branched.
\end{itemize}

\subsubsection{Shear fracture modes}

\begin{figure}[h!]
\centering
\includegraphics[width=.78\textwidth]{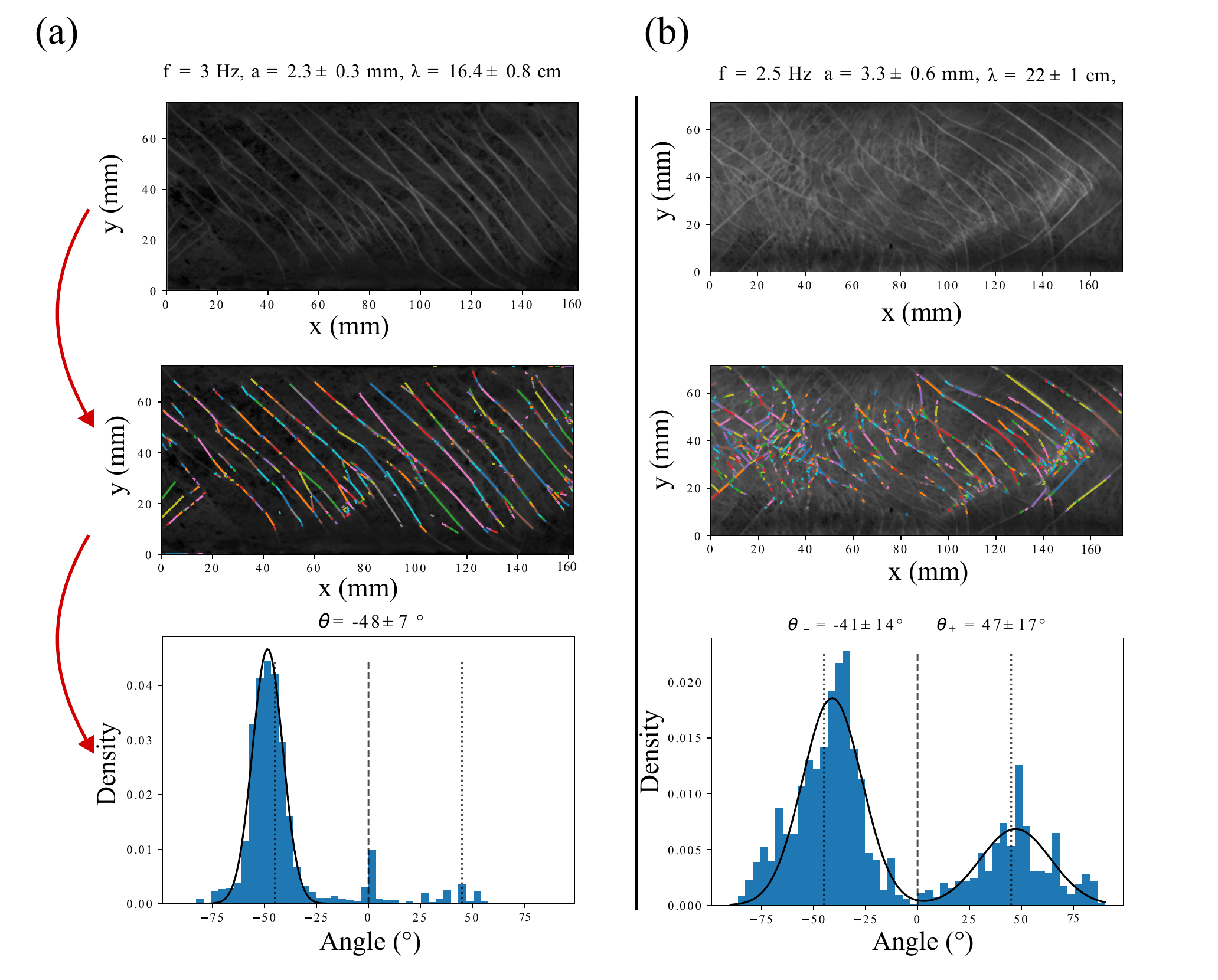}
\caption{Fracture detection and angle distribution. A Gaussian fit is drawn on the histograms. Two histograms from two different runs are presented. On the left a): ordered fractures, they are photographed 325 s after the appearance of the first fractures. Right b): complex patterns photographed 400 s after the appearance of the first fractures. Wave parameters, (a) $f=3$ Hz and $a=2.3$ mm, (b), $f=2.5$ Hz and $a=3.3$ mm. First tank. Similar observations are obtained for other runs with progressive waves.}
\label{Obliquedetect}
\end{figure}

As seen previously, shear fractures can arise from bending stresses or viscous stresses, although we assume the latter process to be dominant in our experiments. In both cases, the theoretical angle for the fracture is $ \pm 45^\circ$. To experimentally validate this result, image processing is employed to detect the fractures on the surface of the particle raft. The fractures are approximated by straight lines, and the angle they form with the $x$-axis is calculated for each fracture. The histogram of the angles is then plotted, with the length of the lines taken as a weighting factor (see Figure~\ref{Obliquedetect}). It should be noted that multiple lines can be used to represent a single fracture. Therefore, the image processing considers the possibility of fractures changing orientation or connecting to each other to form a ramified structure. Several regimes are observed in the experimental results. When the wave amplitude is too low, no fractures are distinguishable. As the amplitude is increased, fractures start to appear, and they exhibit a well-organized arrangement with a minimal overlap (Fig.~\ref{Obliquedetect} (a) ). The direction of the fracture can also alternate between $ + 45$ and $-45^\circ$ (Fig.~\ref{FractModes} (a) ). Finally, at high amplitudes, the fracturing becomes spontaneous, and fractures in both directions can appear in the same locations. The resulting fracture patterns are much more complex (Fig.~\ref{Obliquedetect} (b) ). Throughout the experiment, the density of oblique fractures increases over time, leading to the formation of increasingly intricate fracture patterns before fragmentation occurs. {The initial distance between cracks at low amplitude, which is about $8$ mm for progressive waves is not really understood, but we can note that this length is larger than the wrinkling length about $0.6$ mm (See Appendix~\ref{Emod}) and smaller than the wavelength $\lambda=17.3$ cm.}

By analyzing the histograms of the angles at selected times, we find the presence of the predicted $\pm 45^\circ$ values as predicted by the theory. {This observation validates the relevance of considering the particle raft as a homogeneous elastic material, despite the strong assumptions made.} For higher wave amplitudes, the angle distributions exhibit a wider spread around these theoretical values. In the regime where fractures are well organized, the wavelength of the resulting patterns appears to be regular. It is on the order of centimeter and smaller than the wavelength of the waves (ranging from $10$ to $25$ cm). It is noteworthy, that fractures often initiate near pre-existing fractures that formed during the raft making. This suggests that there may be a minimum distance required for the formation of new fractures, which in turn determines the wavelength of the patterns. However, obtaining precise quantitative results on this is challenging, due to the frequent reorganization of the patterns, the superposition or replacement of old patterns, and the complicating effects of raft wear. Nevertheless, qualitatively, it appears that patterns at lower amplitude have larger lengthscales. Over time, other patterns appear between the previous ones because of the mechanical wear of the membrane, thus reducing the wavelength.

\begin{figure}[h!]
\centering
\includegraphics[width=.7\textwidth]{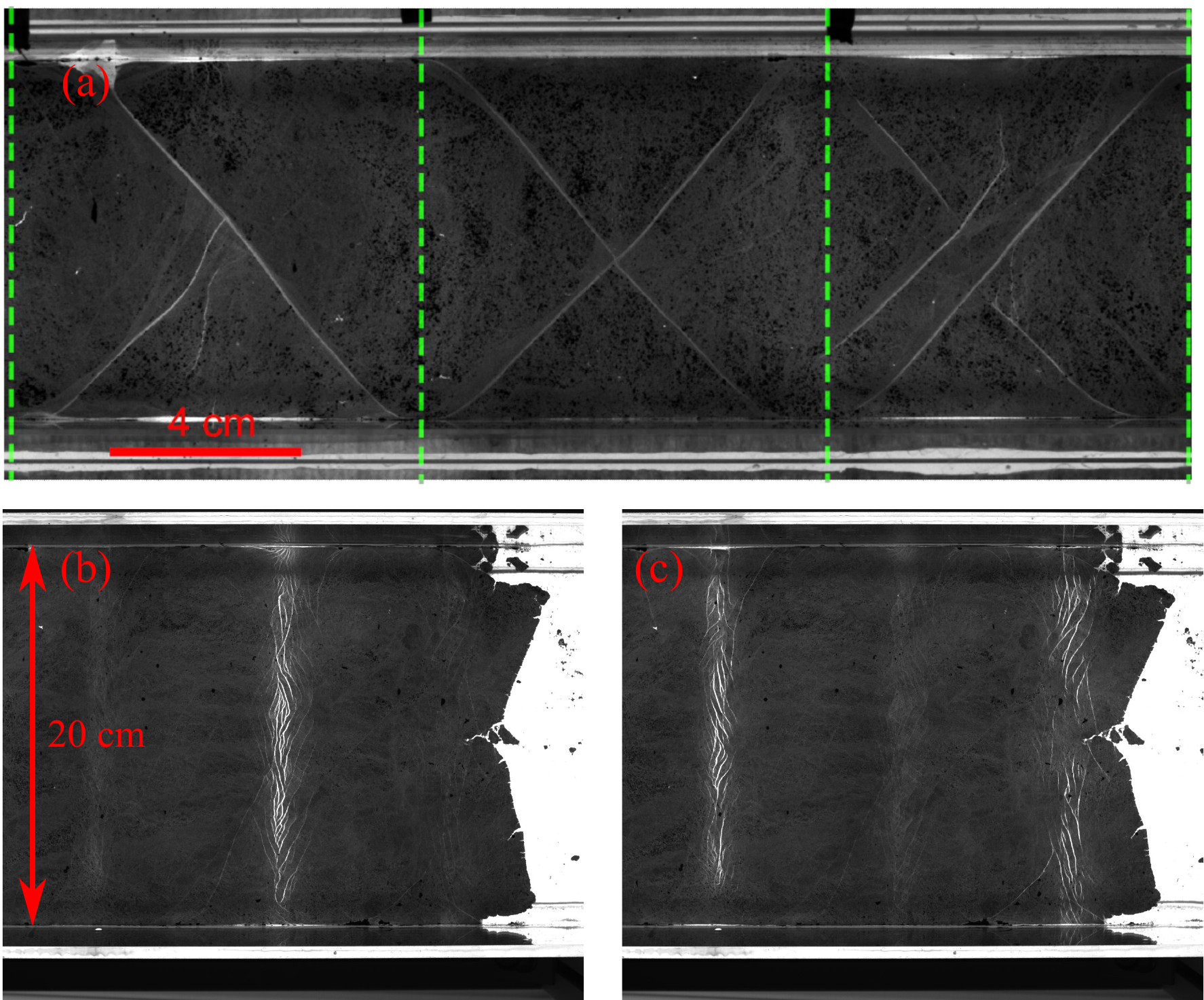}
\caption{Oblique fracture patterns in presence of standing waves. (a) Tank 1 equipped with a movable wall to set up a standing wave pattern. The wave amplitude is close to the breaking threshold. The dashed green lines correspond to the nodes where the amplitude of the wave is close to zero. $f=3.15$ Hz, $\lambda=17$ cm and $a=1.2$ mm (at the wavemaker position). The oblique fractures are located between the nodes. Fracture patterns for standing waves in the Tank 2, for $t=49.1$ s (b) and (c) half a period later,  ($f=3$ Hz and $a=1.48$ mm). When the wave is maximal, the oblique fractures appear transiently on the crests, where the viscous stresses shear the raft. After half a period the positions of the fractures correspond to the antinodes where the surface deformation is the largest. We observe a complex pattern of small cracks roughly oriented at $\pm 45^\circ$. These cracks induce a mechanical wear of the raft.}
\label{ObliquedetectStanding}
\end{figure}

In the case of standing waves at the beginning of an experiment, as depicted in Fig.~\ref{ObliquedetectStanding} (b-c), oblique fractures preferentially form on the antinodes where the wave amplitude is maximal, rather than uniformly distributed across the raft. This observation aligns with our model describing the influence of viscous stresses (Section~\ref{ViscousStresses}).  When the wave amplitude reaches a sufficient magnitude, oblique fracture patterns localize onto narrow stripes measuring a few centimeters (approximately $2.5$ cm for $\lambda={17.3}$ cm) precisely at the instants of maximum wave amplitude and surface deformation, where shearing by the orbital flow predominantly occurs.

Moreover, we observe that before the application of the waves, once the {plastic separators (see Fig.~\ref{ExSetup} (c))} are removed in the preparation of the raft, some oblique fractures can appear due to the sudden decompression, but without strong consequences for the following experiment{, because these fractures do not widen before the wave solicitation and the raft preserves its integrity}. Most of the shear fractures observed in our experiments have an oblique direction compared to the wave solicitation.

\subsubsection{Opening fracture modes}

\begin{figure}[h!]
\centering
\includegraphics[width=.5\textwidth]{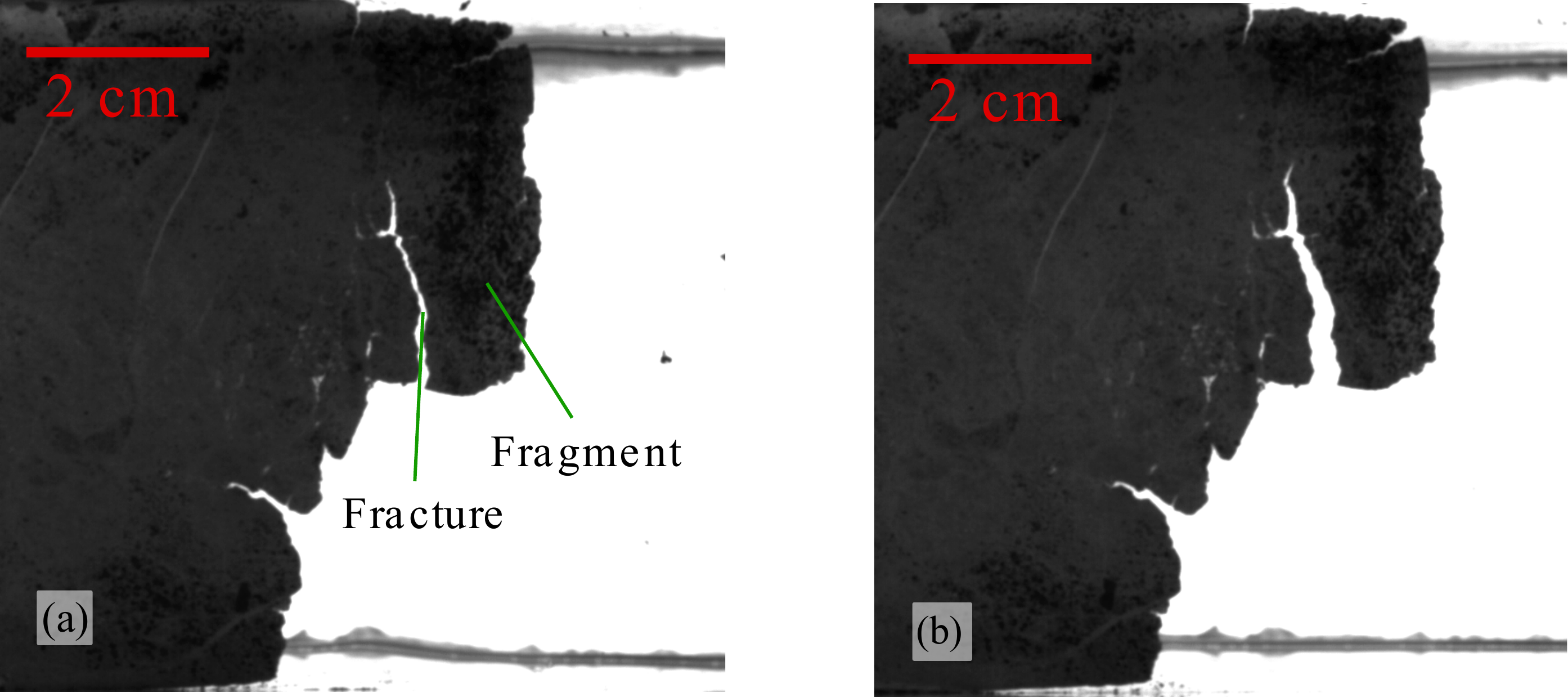}
\caption{Example of fracture by opening where the associated fragment is detached from the front of the raft in the tank 1 with progressive waves. (a), Passage of a wave crest. (b), Passage of a wave {trough}. Wave parameters, $a=2.2$ mm and $f = 3$ Hz.} 
\label{OpeningFractures}
\end{figure}

Opening fractures are predominantly observed at the ends of the raft, particularly on the side closest to the wavemaker. They typically initiate from imperfections present along the raft edge. During the fracture formation process, we observe that the fractures open at the troughs of the waves and close at their crests. This observation rules out bending at the cause of fracturing, as bending would result in the opposite phenomenon, where the opening would occur at the troughs and crests according to Eq.~\ref{TnBendingOp}. Instead, opening fractures are attributed to viscous stresses arising from currents at the fluid surface.

Consider a fracture located at the front of the raft, perpendicular to the direction of wave propagation ($90^\circ$) as shown in Fig.~\ref{OpeningFractures}. A mobile fragment that remains attached to the raft is associated with this fracture. According to Eq.~\ref{viscousprog}, the stress exerted on the wave crests propels the fragment in the direction of wave propagation. In contrast, at the wave troughs, the fragments move in the opposite direction. Since the fracture is situated at the front of the raft, the wave crests bring the fragment back towards the raft, causing the fracture to close. Conversely, at the troughs, the fragment breaks away from the raft, resulting in the opening and widening of the fracture. It should be noted that these opening fractures progress gradually, with the width of the fracture increasing and its tip advancing with each wave cycle. In contrast, oblique fractures span the entire tank almost instantaneously and remain narrow.

While the viscous stress model explains the opening of fractures orthogonal to the wave propagation, it does not account for the opening of fractures oriented in the direction as the wave propagation. Our model does not consider stresses along the $y$ axis. The opening mechanism of fracture of this type cannot be explained 
solely by the orbital velocity field associated with the waves, which is directed along the $x$-axis. For fractures located at the front of the raft, the opening of fractures has been observed at both the troughs and the crests (but not simultaneously for the same fracture). Therefore, the streaming flow, which is weaker but exhibits a more complex geometry, likely contributes to this breaking process. 

\subsection{Breaking threshold}

Determining the breaking threshold, which corresponds to the minimum wave excitation required to initiate fractures in the raft, is crucial for characterizing the breaking phenomenon. In this section, we aim to estimate the threshold values above which we observe opening or shearing fractures. We estimate the Young's Modulus of the graphite particle raft to be $E \approx 10^4$ Pa (in Appendix~\ref{Emod}). As calculated in sections~\ref{Bending} and \ref{ViscousStresses}, for incoming waves with a frequency $f=3$ Hz and amplitude $a=1$ mm, the maximal stress due to bending is about $70 \times 10^{-3}$ Pa, whereas the stress due to viscous shear is of order of $230$ Pa. Therefore, we consider the contribution of viscous stresses as the primary mechanism for fracturing.

In general, determining the breaking threshold theoretically is quite challenging. The cohesion between the constituents of a material (such as atomic planes in a crystal or particles in a granular raft) does not solely explain the critical stress required for material rupture. Experimental rupture values are typically three orders of magnitude lower than that the theoretical predictions~\cite{OswaldBook}. Defects within the material are in fact of great importance. The mechanism of rupture would not be due to the block displacement of particles or atomic planes, but to the propagation of defects also called dislocations. In our case, to elaborate a model is difficult since the graphite powder used is {polydisperse} and the grains which compose it are not spherical. We therefore propose an experimental approach to determine the breaking threshold of the graphite raft. We plot the  experimental points in an $(a-\lambda)$ (amplitude$-$wavelength) diagram and classify them according to the observed phenomena.

\begin{figure}[h!]
\centering
\includegraphics[width=1\textwidth]{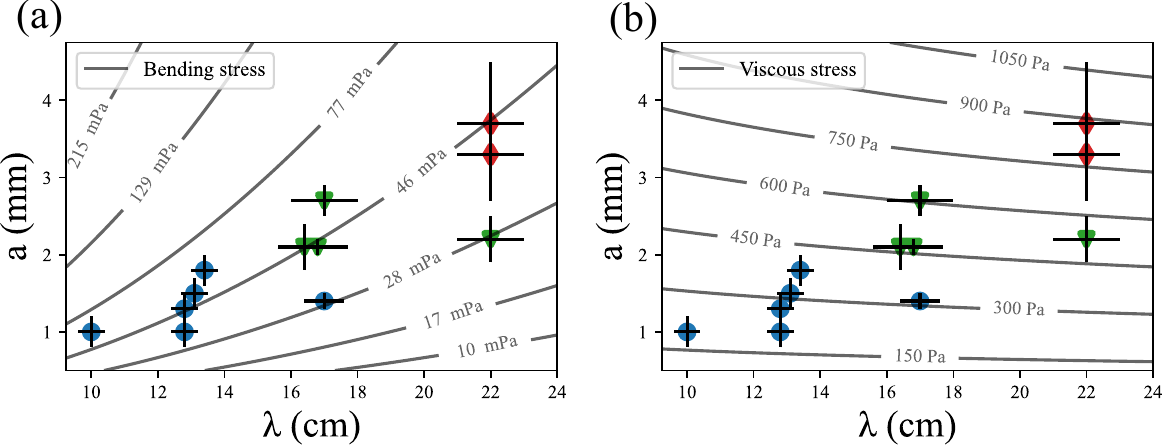}
\caption{Breaking threshold diagram. Experimental points. Blue dots indicate no fracturing, green triangles indicate well organized fracturing and red diamonds indicate disordered fracturing. 
(a) gray lines, {iso-values of the bending stresses which are proportional to the curvature $\mathcal{C}=ak^2$}. (b) iso-values of the viscous stresses. The second diagram separates better the experimental points with and without breaking.}
\label{Thresholdfract}
\end{figure}

Fig.~\ref{Thresholdfract} illustrates the experimental points obtained in tank 1 for progressive waves, excluding the presence of a nylon mesh on the walls. Wave attenuation restricts the availability of high-amplitude points for small wavelengths. The background of the figure depicts the physical phenomena responsible for the plate's stress: the maximum curvature of the raft representing bending on the left and the maximum theoretical viscous stress caused by surface currents on the right. By following, for instance, the iso-stress line $\sigma=46$ mPa, we observe situations with no fracturing (blue dots) as well as situations with intense fracturing (red diamonds). Fracture by bending seems again to be invalidated experimentally. In contrast, the viscous isostresses separate well the different regimes observed during fracturing and we can give an estimate of the threshold stress at approximately {$450$} Pa. Additionally, as the stresses increase, the severity of fracturing intensifies. Hence, the fracture model based on viscous stresses at the bottom side of the raft aligns well with the experimental measurements.
We exclude the stretching in this diagram. It may play a role in the pinning of the contact lines and induce perpendicular shearing fractures, as explained in Appendix~\ref{StretchingMechanism}. These cracks are, however, suppressed when we apply a nylon mesh on the lateral glass walls, allowing sliding of the contact line. We note also (not shown), that the experiments with the nylon mesh give a breaking threshold for viscous stresses slightly smaller about {$300$} Pa. Therefore, the boundary conditions must be taken into account to determine precisely the breaking thresholds. Finally, we note that the breaking thresholds for standing waves in the tank 2 occur for significantly lower wave amplitude ($0.5$ mm for $f=3$ Hz). As the maximal stresses are located on the wave antinodes, due to the cumulative solicitation and to the wearing of the raft, the breaking occurs more easily at these antinodes.

The fragmentation and the erosion of a particle raft have been investigated by Vassileva et al.~\cite{vassileva2006restructuring,vassileva2007fragmentation} for rafts made of spherical glass spheres of 65 and 115 $\mu$m in close packing. By balancing the viscous drag and the capillary attraction between the particles of the raft, they identified critical values of the shear rate for breaking or fragmentation, approximately $1.8 $ s$^{-1}$ and $0.8$  s$^{-1}$ for $65$ and $115$ $\mu$m particles. The critical values for erosion were slightly lower. The authors note that only the straining component of the shear contributes to the rupture of capillary bonds since rafts detached from the walls can rotate freely. In our experiments where {progressive} waves of frequency $3$ Hz and amplitude $2$ mm break the graphite particle raft, we can estimate the shear rates at the antinodes for the 2D surface flow to be about $a\,\omega /(1/2\,\lambda) = \omega\,(a \, k) /\pi \approx 0.4 $ {s$^{-1}$}, which seems lower than the shear rate found by Vassileva et al. but with a similar order of magnitude. In our experiments, we use graphite rafts made with a larger number of smaller and polydisperse particles, potentially rendering the raft more fragile due to irregular packing.

\section{Fragmentation of a particle raft}

After characterizing the onset of fracture, our focus shifts to the progressive fragmentation of the raft. It refers to the decomposition of the raft into multiple fragments or floes, similar to the separation and drift of sea ice caused by surface flow.
In the previous section, we have seen that different fracture mechanisms are associated with distinct crack locations. Therefore, we would expect the formation of initial fragments with sizes determined by the distance between positions of maximum stress, approximately $\lambda/2$. However, the analysis of experimental videos (see Movie S1, S2, S3, S4 and s5) reveals a much more complex situation. Oblique fractures weaken the raft and create zones where parts of the raft can slide easily, but without complete separation. We also observe rapid rearrangements of the fracture network, which will facilitate fragment generation. In contrast, opening fractures, occurring at the free ends of the raft, clearly generate the first fragments as the fracture width increases with the time of arrival of waves. After several tens of seconds, we observe the first detachment of floes influenced by the streaming flow. However, the precise scenario of initial fragmentation is highly variable between experiments. It depends, in particular, on the raft preparation, which always incorporates defects and heterogeneities, as well as on the streaming flow, whose geometry is not reproducible due to its sensitivity to initial conditions.

\begin{figure}[h]
\centering
\includegraphics[width=.85\textwidth]{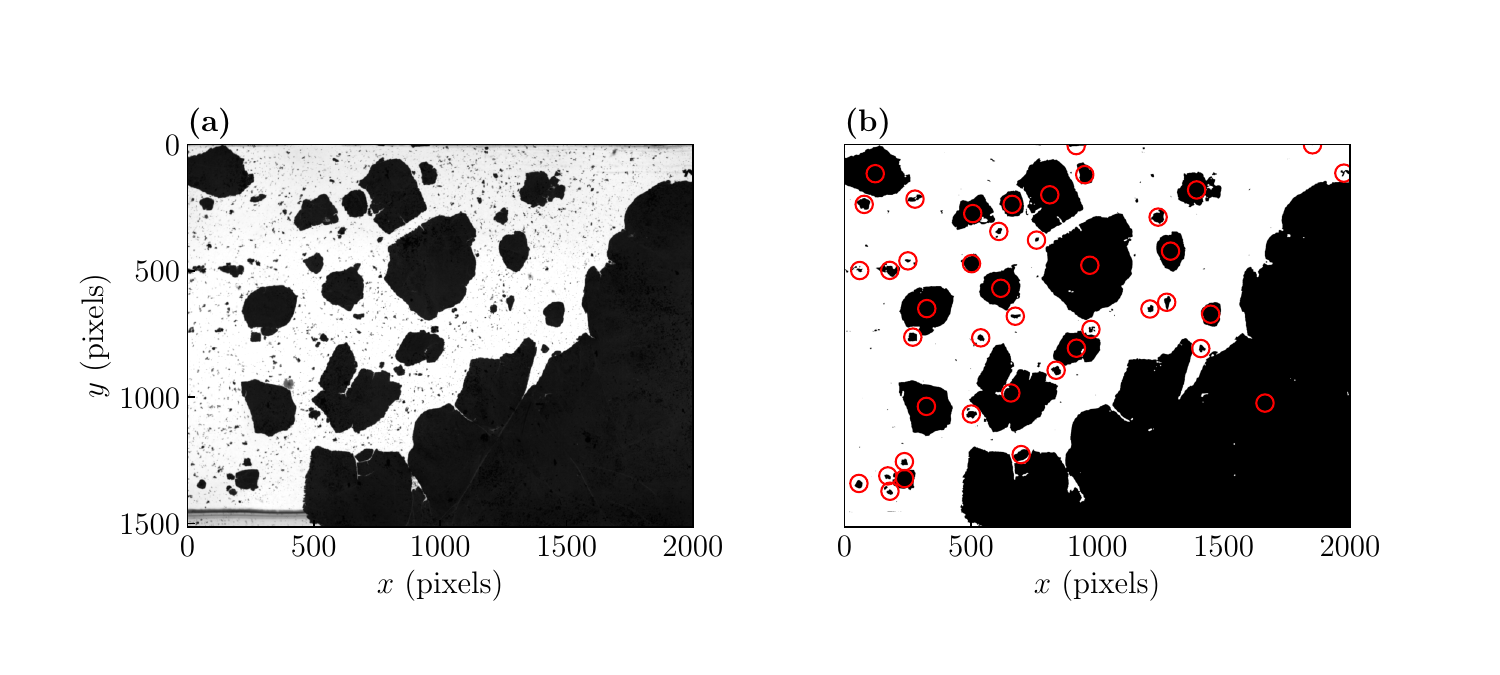}
\caption{Example of image processing and tracking of floes during the breakup of the particle raft using the package \textit{trackpy}~\cite{trackPy}. (a), raw image. (b), binarized images. The red circles correspond to the pixel barycenters of the detected clusters. A minimal cluster size of 200 pixels, i.e. approximately 1 mm has been chosen, in order to not count the small aggregates forming a dust between the fragments. We note also a large variation of cluster sizes.}  
\label{exemple_tracking}
\end{figure}

\begin{figure}[h]
\centering
\includegraphics[width=.6\textwidth]{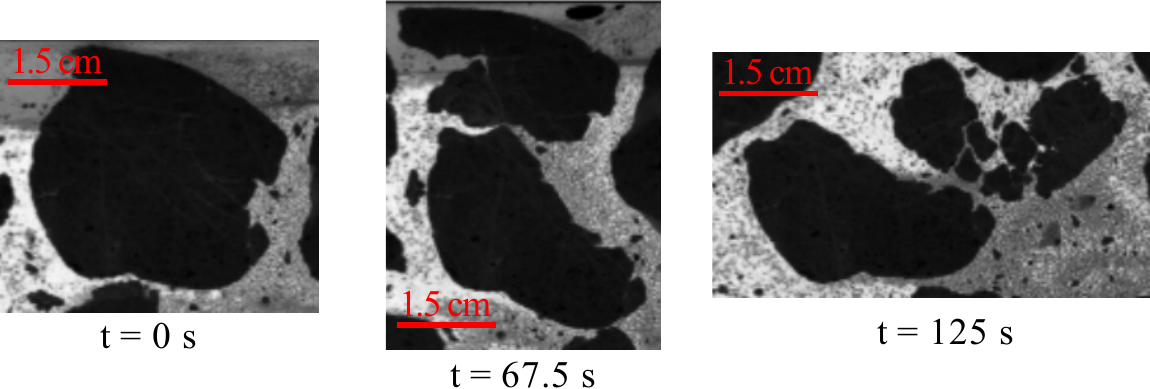}
\caption{Example of a fragmentation event of a graphite floe. It first splits in two then the fragments obtained are refragmented into smaller fragments. Parameters : $a=3.2$ mm, $f=2.5$ Hz and Tank 1.}  
\label{ExampleFragmentation}
\end{figure}

To quantitatively analyze the floes as individual clusters in the images and investigate their geometrical properties, we propose a tracking approach. Our goal is to characterize the sizes, positions, and shapes of the fragments during a raft fracture experiment. We therefore need to know several parameters describing the $N$ floes appearing in the image at time $t$, among which the area $A$ of the floe $i$ and its perimeter $p$.

As the graphite floes appear as dark objects on a light background, the tracking by image thresholding works well. We employ the \textit{trackpy} package of Python~\cite{trackPy} associated with the library for image processing \textit{scikit-image}~\cite{van2014scikit} , to detect large non spherical objects. After importing the images corresponding to an experiment, we start by cropping the area of interest and binarizing using a chosen threshold. The objective is to resolve each of the fragments and their edge while removing the dust made by individual floating graphite grains, corresponding to the smallest detected clusters, below approximately 1 mm. The acquisitions are carried out over a period about 10 minutes, and we film with a frame rate equal to the frequency of the waves (most often 3 Hz), to better visualize the drift of the floes. An example of image processing and tracking of floes is shown in Fig. 16.
 
In the study of the marginal ice zone, where floes are formed as a result of sea ice fragmentation, the distribution of floe sizes plays a crucial role as a characterization tool. Here, we propose to investigate the statistical properties of the floes, by studying the evolution and the relative distribution of the floe area $A$, which is geometrically related to the size distribution. After the breaking of the raft, the initial generation of floes undergoes two distinct phenomena. First, through fragmentation, larger floes can break into two smaller floes, as illustrated in Fig.~\ref{ExampleFragmentation}. The occurrence of multiple fragmentation events leads to an increased number of detected floes and a decrease in the average floe area. Second, through erosion, individual floes progressively lose graphite grains from their boundaries. For sole erosion, the average floe area should decrease while the number of detected floes remains constant. Nevertheless, both phenomena can occur simultaneously on the floe population and are thus challenging to isolate in the data. {Experiments shearing  a small particle raft made of about 100 $\mu$m sized particles by a Couette flow~\cite{vassileva2007fragmentation} have shown indeed that the threshold erosion is slightly lower than the one of fragmentation.} In our experiments, the fragmentation and erosion are likely {also} caused by the viscous stresses induced by the horizontal shearing from the surface flow.

\begin{figure}[h]
\centering
\includegraphics[width=1\textwidth]{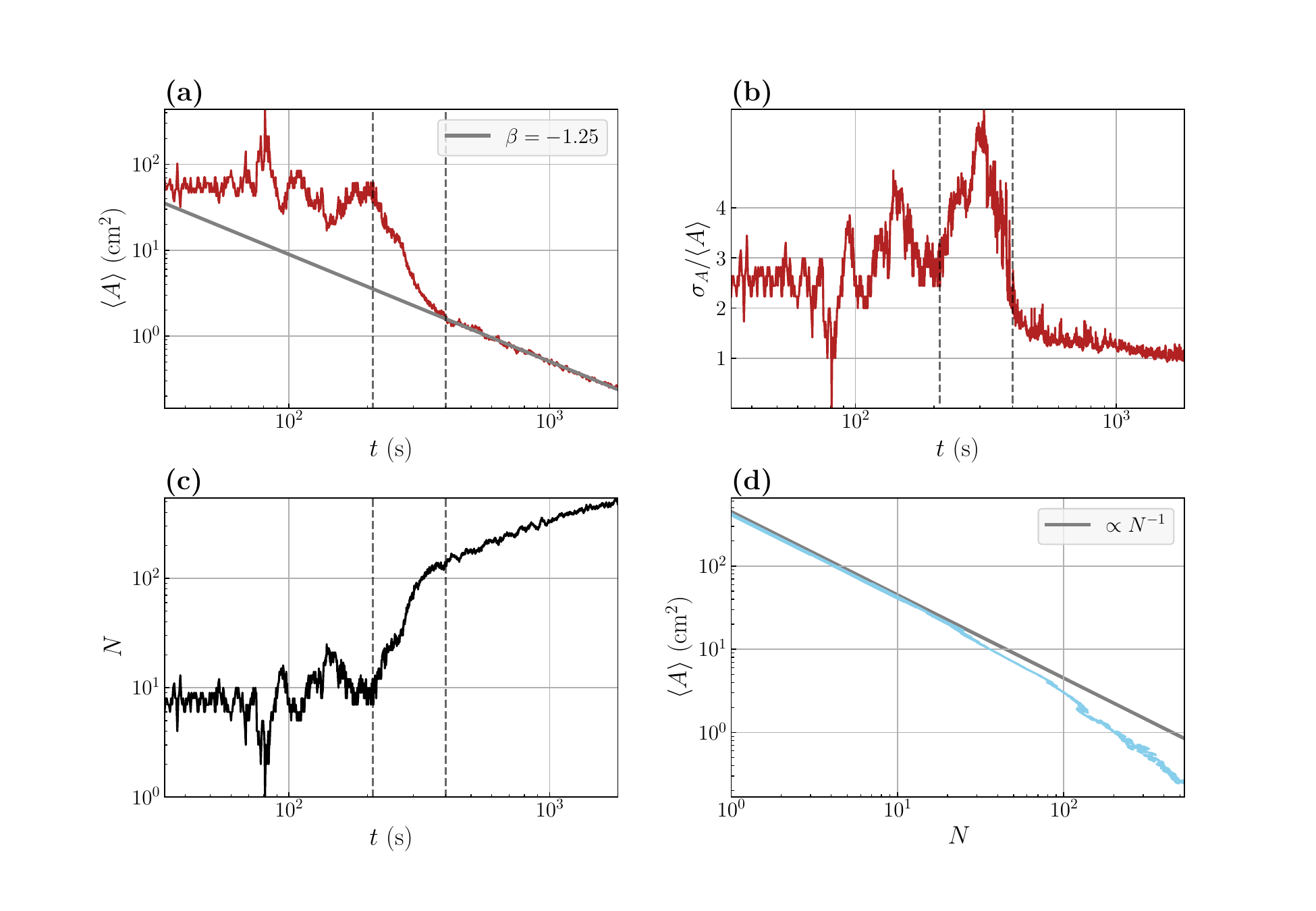}
\caption{(a) Decrease of the averaged area $\langle A \rangle$ of each graphite floes as a function of time. In a first regime, delimited by two vertical dashed lines, for $205 < t < 400$ s, $A$ drops sharply. Then, in a second regime, for $t>t^\star=400$ s, the decrease is well fitted by a power law with an exponent $\beta=-1.25$. (b) Corresponding fluctuations of the area, $\sigma_A / \langle A \rangle $, where $\sigma_A$ is the standard deviation of $A$. The fluctuation rate is maximal close to $t^\star$ and then decreases strongly in the second regime. (c) Time evolution of the number of detected floes $N$ at time $t$. We distinguish also the two regimes, with a fast increase in the first and a slower growth in the second. (d) Evolution of the average area $A$ of floes as a function of the number of floes $N$. Parameters of the experiment : $a=1.27$ mm, $f=3$ Hz and Tank 2.}  
\label{FloeEvoltime}
\end{figure}

{Let us, characterize the floe area statistics in a typical experiment in the second tank with standing waves, where the incoming waves have an amplitude $a=1.27 $ mm and a frequency $f=3$ Hz, acquired with the Basler 8 bits camera {(resolution 0.16 mm per pixel)}. The corresponding movie is available in supplemental (see Movie S5). In Fig.~\ref{FloeEvoltime} (a), we observe well a decrease of the average floe area $\langle A \rangle = (1/N) \, \sum_i^N \, A_i $ with time, due to fragmentation (breaking in smaller pieces) and also erosion (loss of grains at the edges of the floes). For the first $205$ s, the mean area $\langle A \rangle$ is nearly constant before to drop strongly by a factor $50$ during a first regime. Then, for $t$ larger than $t^\star=450$ s, in a second regime the decrease is less intense  and is well described by a power law of time $\langle A \rangle \propto t^{\beta}$ with $\beta \approx -1.25$. The fluctuations of the area quantified by the ratio of the standard deviation by the mean value $\sigma_A / \langle A \rangle $ shows in Fig.~\ref{FloeEvoltime} (b) that the heterogeneity of the distribution is maximal during the fragmentation process in the first regime and decreases for $t>t^*$ in the second regime to reach values close to $1$. Simultaneously, as illustrated in Fig.~\ref{FloeEvoltime} (c), the number of floes $N$ increases strongly in the first regime between $205$ s and $t^\star$ and slowly after. When, we plot $\langle A \rangle$ as a function of $N$, in Fig.~\ref{FloeEvoltime} (d), we observe that for small values of $N$, the mean area $\langle A \rangle$ is well proportional to $1/N$. This shows that the total area occupied by the floes $\sum_i^N A_i$ is conserved with time, meaning that the erosion process is negligible for small values of $N$. Therefore, we conclude that this first step corresponds to a fragmentation process, the number of floes increases without change of the total area. Then, for approximately $\langle A \rangle < 5$ cm$^2$ and $N >80$, the mean area decreases faster than the number of floes. {This evolution corresponds to the second regime $t>t^\star$. As the floes are not imaged on the entire tank surface and some of the detected floes leave the observation window, we cannot conclude definitively that this second regime is a combination of fragmentation and erosion. The existence of a power-law evolution is not understood yet in this second regime, but the fact that the system is open and the possible coexistence of two processes, fragmentation and erosion makes the modelization of the floe evolution challenging.}

{Moreover, the visual observation seems to show that floes at the beginning polygonal, become more and more rounded due to erosion and successive brittle fractures. To quantify this observation, a simple characterization consists in computing the circularity parameter $\kappa= \langle p \rangle^2/(4\pi \, \langle A \rangle)$, whose value is $1$ for a circle and becomes large for an elongated 2D object. In Fig.~\ref{Circularity} (a), $\kappa$ falls quickly to reach a saturation value $\kappa_\infty \approx 1.32$ for $t$ roughly larger than $t^\star=400$ s, \textit{i.e.} in the second regime. In Fig.~\ref{Circularity} (b), we show that the asymptotic value does not depend significantly on the forcing amplitude and remains above $1.25$. The rounding of the floes {could be} caused by the erosion process, but the reason of the saturation to a value slightly larger than for a circle remains unexplained.}

\begin{figure}[h!]
\centering
\includegraphics[width=1.\textwidth]{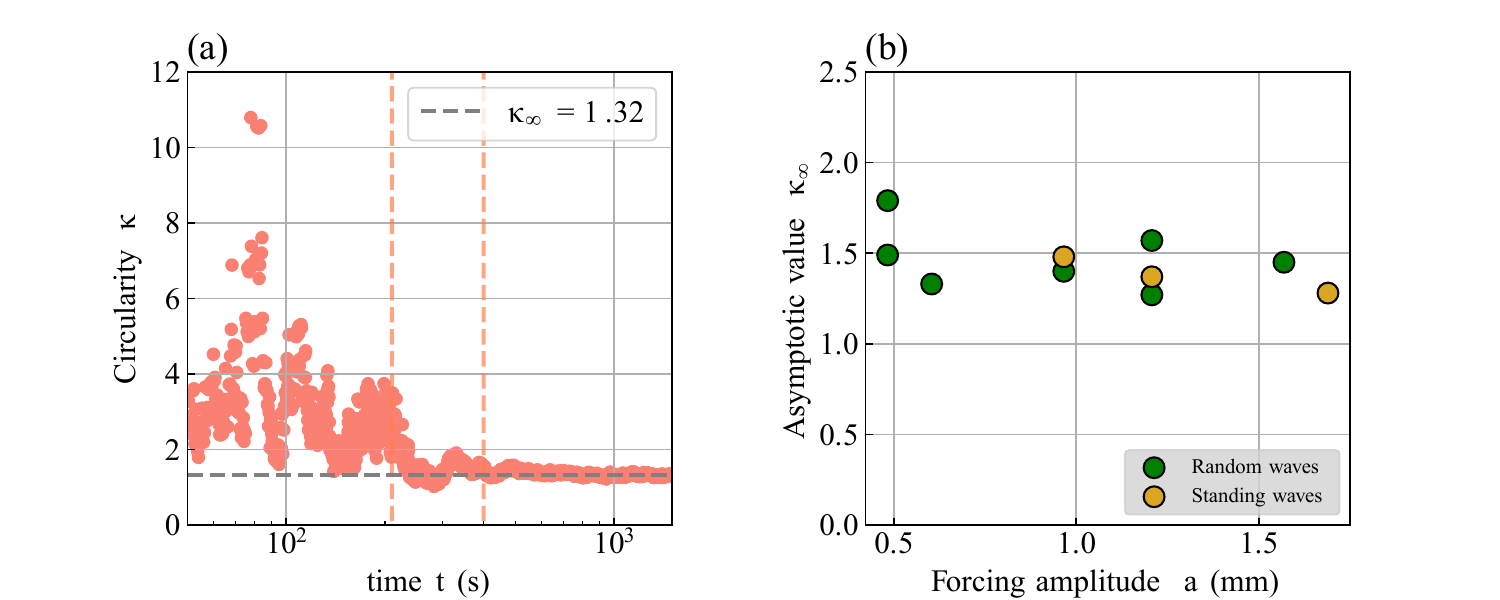}
\caption{(a) Evolution of the circularity parameter $\kappa=\dfrac{\langle p \rangle^2}{4\pi \, \langle A \rangle}$ with time. $\kappa$ decreases meaning that the floe shape becomes more circular, but without reaching the value for a disk equal to $1$. Parameters of the experiment : $a=1.27$ mm, $f=3$ Hz and Tank 2. (b) Asymptotic value $\kappa_\infty$ for various forcing amplitudes in the Tank 2. $\kappa_\infty$ does not depend significantly on the forcing.}  
\label{Circularity}
\end{figure}

\begin{figure}[h!]
\centering
\includegraphics[width=1\textwidth]{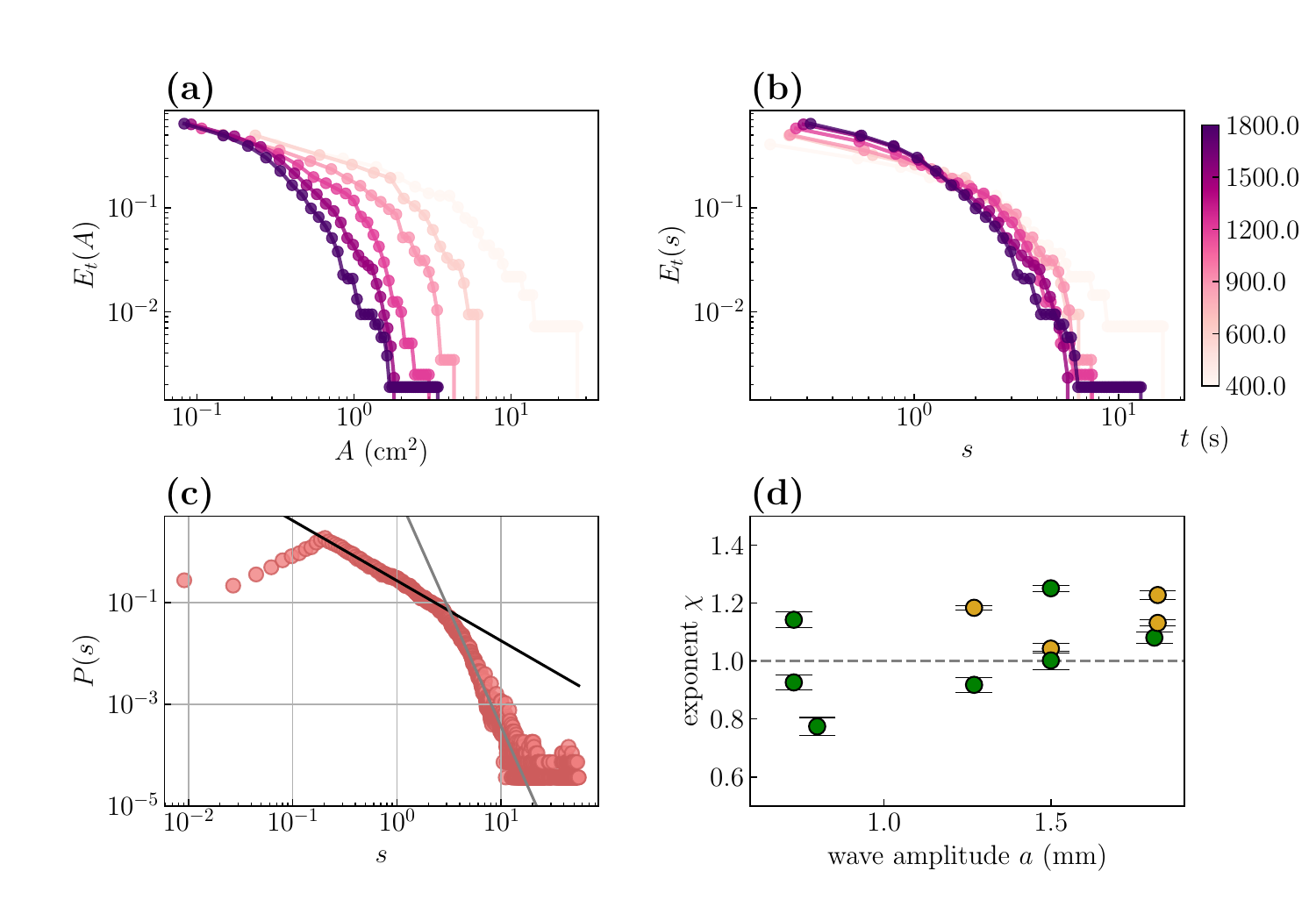}
\caption{ {(a) Complementary cumulative distribution function or exceedance function $E_t(A)$ of the area $A$ for one experiment (Parameters: $a=1.27$ mm, $f=3$ Hz and Tank 2) at the following time steps  $400$, $600$, $900$, $1200$, $1500$ and $1800$ s. (b) Complementary cumulative distribution function of the rescaled area $s=A/\langle A \rangle$ at the same time steps $t$. We note that the curves overlap, once the rescaling has been done.  (c) Probability distribution function $P(s)$ of $s$, for the values during all the duration of the experimental run. For intermediate values, $s \in [0.2,4]$, the distribution is compatible with a power law {$P(s) \propto s^{-\chi}$ with $\chi=1.2$. At larger values of $s$, the decay can be fitted by another power law with an exponent $-4.6$. (d) Values of the exponent $\chi$ as a function of the wave amplitude $a$ for the intermediate range of $s$. $\chi$ does not vary significantly with wave amplitude and remains compatible with an exponent $-1$ at first order.The error bars correspond to the uncertainties of power-law fitting, but the range of variation shows that the uncertainty on $\chi$ is greater.} }}  
\label{FloeAreadistribution}
\end{figure}

For sea ice in the field, floe size distributions have been extensively studied especially in the marginal ice zone. However, the distributions are highly variable and depend on the thickness and concentration of the sea ice influencing its mechanical properties and of the swell conditions. Moreover, this distribution evolves in time due to melting and floe collisions. When the floe population is sufficiently old to display a large variation of sizes, the floe size distributions are usually fitted as power laws with a cut-off at large scale~\cite{toyota2006characteristics,toyota2011size,Gherardi2015,alberello2019brief,Horvat2022,hwang2022multi}. Other distributions have been also observed and proposed. For example, young floes resulting from breaking at a characteristic scale are better described by a Gaussian distribution~\cite{dumas2021aerial}. Fragmentation cascade predicts log-normal distributions~\cite{villermaux2007fragmentation,montiel2022theoretical}. Statistical models of the floe population dynamic through breaking events give truncated power law distributions~\cite{Herman2010,Gherardi2015} called Gamma distributions when the cut-off is exponential. In the field, the reported exponents of the power law distributions are also very variable but typically between $-1$ and $-2$ for the floe length~\cite{stern2018reconciling}, \textit{i.e.} between  $-1$ and $-1.5$ for the floe area. Experiments in a large wave basin covered by an ice layer~\cite{Herman2018} report for the floe size distribution a combination between Gaussian and Gamma distribution, but in this study the floe population results directly from the first steps of the wave breaking as the floes cannot drift significantly.  

Here, we characterize in our experiment the graphite floe area distribution in Fig.~\ref{FloeAreadistribution} first for a run where $a=1.27$ mm, $f=3$ Hz and $t>400$ s. For one time step, the number of floes $N$, of samples is moderate. Thus, in order to present a better converged quantity, we compute the complementary cumulative distribution function or exceedance $E_t (A)$ in Fig.~\ref{FloeAreadistribution} (a) for selected values, $400$, $600$, $900$, $1200$, $1500$ and $1800$ s. $E_t (A)$ corresponds to the probability to select a floe of area larger than the value $A$. $E_t$ is related to the probability density function $P_t$ by the relation:
\begin{equation}
  {E_t(A)=\int_{A}^\infty \, P_t ({A}^{'})\, \mathrm{d}{A}^{'} \,.
}
\end{equation}

The first time value is chosen to have a sufficient statistics, \textit{i.e.} more than 100 floes. Due to the notable decrease of the mean area $\langle A \rangle$ shown in Fig.~\ref{FloeEvoltime} (a) as a function of time, the function $E_t(A)$ shifts towards the left axis with time. To reduce the effect of the time varying typical size of floes, we study now the distribution of the rescaled area, $s=A/\langle A \rangle$, as it has been suggested by Gherardi and Lagomarsino~\cite{Gherardi2015} motivated by the scaling analysis of fragmentation~\cite{cheng1988scaling}. As shown in Fig.~\ref{FloeAreadistribution} (b), the use of the rescaled area gathers the distributions obtained at various time steps, on nearly one curve.
Therefore, the evolution of the floe area distribution is mainly controlled by the time variation of $\langle A \rangle$, the first moment of the distribution, at least when $\MA{\sigma_A/\langle A \rangle }$ remains of order 1 in the second regime (Fig.~\ref{FloeEvoltime} (b) for $t>t^\star=400$ s). Then, the distribution of $s$ is stationary in good approximation. We can thus consider the data set composed of the floe area measured at all time steps to compute the probability density function $P(s)$ and benefit to the larger statistics. The result is displayed in Fig.~\ref{FloeAreadistribution} (c). At small scale, the arbitrary cut-off for minimal cluster size makes the interpretation questionable. For intermediate values on roughly one decade, \textit{i.e.} $s \in [0.2,4]$, the measured distribution is compatible with a power law {$P(s) \propto s^{-\chi}$, with $\chi = 1.2 $.} For $s>4$, the distribution is better fitted by a second power law with an exponent about $-4.6$, than by a decreasing exponential. Our data are thus not well described by the Gamma distribution formed by a power law with an exponential cut-off at large scale predicted by fragmentation models~\cite{Herman2010,Gherardi2015}. Note that in the field a probability density function made of two distinct power laws at small and large scales has been reported for relatively small ice floes (``pancakes floes")~\cite{alberello2019brief}. However, in our experiment, due to the small range of scales available and the limited statistics for large floes are not, it is difficult for now to assess the robustness of the statistics of large floes and we discuss thus the behavior in the intermediate domain. Moreover, we characterize the second regime, for $t>t^\star$, which seems to combine fragmentation and erosion of the floes, a regime not described by quantitative models in the literature. 

By varying the wave amplitude as shown in Fig.~\ref{FloeAreadistribution} (d), we do not observe a significant dependency, of the exponent $\chi$, which is roughly compatible with the value $-1$. Qualitatively, similar observations are made, on the different experiments with different experimental parameters, for the decrease of the floe size with time and the floe size distribution properties{, although the quantitative results display a significant variability. Nevertheless, $P(s)$ display for all experiments a robust power-law of exponent close to $-1$ for the range $ s \in [0.2,4]$.} Further studies are needed and we plan to investigate specifically the floe area evolution and distribution in a dedicated forthcoming work, where tracking of individual floes enable to characterize the population evolution during all the fragmentation process.

\section{Conclusion and perspectives}

In this work, we have explored the breaking and the fragmentation of a graphite particle raft due to surface gravity waves. By sprinkling a graphite powder with particles about 10 $\mu$m on water surface, a nearly homogeneous elastic monolayer is formed at the water surface under the action of attractive capillary forces. We estimate the mechanical properties of this membrane, from the wavelength of the buckling ripples made by compressing the raft. In a series of experiments, a homogeneous raft with dimensions of a few tens of centimeters in length and width is prepared on one part of the water tank. The raft is then subjected to water waves with normal incidence. For sufficiently large wave amplitudes, opening fractures occur on the raft's edge, while oblique fractures appear in its bulk, leading to the fragmentation of the raft in many fragments or floes. Because the raft breaks as a brittle material, the first graphite floes have initially polygonal shapes. Our observations reveal that the primary contribution to the mechanical response of the raft and floes is not from stretching and bending due to free-surface deformation, but from viscous stresses induced by the surface velocity field. This velocity field consists of the orbital flow associated with linear waves and the streaming velocity field resulting from nonlinear effects, which causes a long-term drift of the floes.  The streaming flows exhibit complex geometries with large-scale recirculations, influencing the breaking and transport of the floes. We estimate the breaking threshold for graphite rafts subjected to waves and estimate that the critical stress is about one tenth of the Young's modulus of the raft. Additionally, by tracking individual floes, we determine their areas. We show that due to fragmentation and erosion, the floe size decreases in time and that they become more circular with time. The floe area distributions at different time steps overlap when normalized by the average area at a given time, indicating that the time dependency of the floe size distribution is encapsulated in the mean decrease of floe size. We find that the distributions can be reasonably fitted by a power law with an exponent close to $-1$, although there is considerable variation between experiments without a clear dependency on the amplitude of the incoming waves. Further investigations are needed to provide a more quantitative description of raft fragmentation and floe drift. 

In this exploratory work, we have demonstrated that the floating rafts made with 10 $\mu$m particles serve as a suitable model system for studying the fracture and fragmentation of thin floating elastic membranes. {As evoked in the introduction, the tea scum represents another example from everyday life, of a 2D fragmentation of a floating solid. The case of sea ice constitutes the most important case of fragmentation of a floating solid, due to its implications for Earth's climate. In our laboratory experiments, using waves to break a particle raft, we observe a striking visual resemblance with the ice floes. We obtain two dimensional polygonal graphite fragment or floes, which display a large variation of scales.} However, we cannot scale down all the parameters. In particular, in our case, the raft breaks in response to the viscous stresses due to surface flows, whereas for sea ice the viscous stresses are negligible and the ice layer breaks mainly by bending. It seems thus, that the quasi-2D geometry and the fact that our rafts are brittle, explain the visual similarity. In conclusion, the large diversity of behaviors for different hydrodynamic conditions shows that particle raft made with particles of few tens of microns constitute systems with a rich physics, which have not been sufficiently studied quantitatively. By subjecting the material to length-scales larger than its thickness, we visualize mechanical response of this brittle two-dimensional elastic material, with fractures occurring at specific locations and altering the global rheology of the material. Previous studies on the breaking of capillary aggregates~\cite{vassileva2006restructuring,vassileva2007fragmentation,lagarde2020probing,xiao2020strain,to2023rifts,protiere2023particle} mostly focused on small aggregates consisting of identical particles, treating them as cohesive floating granular media rather than continuous elastic mediums. Finally, our experiments have evidenced significant streaming flows at the surface resulting likely from nonlinear effects. These surface currents are most of the time ignored in laboratory experiments. However, they are primordial to describe the drift of floating objects like pollutants or solid fragments like our graphite floes.

\begin{acknowledgments}
We acknowledge at MSC, Universit\'e Paris Cit\'e,  Yann Le Goas and Alexandre Di Palma for technical assistance, as well as Marc Durand, Drazen Zanchi, Sylvain Courrech du Pont and Eduardo Monsalve for help and discussions. We thank also Dassine Zouaoui for performing preliminary experiments. We are grateful to Christophe Gissinger, François Pétrélis and Stéphan Fauve at LPENS, PSL university, for equipment loan. We acknowledge for scientific discussions Baptiste Auvity, Antonin Eddi and Stéphane Perrard from PMMH, PSL university and also Frédéric Moisy from FAST, Paris Saclay University. {We thank Benoit Roman from PMMH, PSL university and Elsa Bayart \& Mokhtar Adda-Bedia from ENS Lyon for advice about elasticity of thin plates. Finally, we thank the anonymous referees for their insightful comments}.
\end{acknowledgments}

\appendix

\appendix
\section{Young's Modulus Measurement of the floating particle raft}
\label{Emod}

 According to Vella et al.~\cite{Vella2004} the measurement of the wavelength $\lambda_w$ of surface wrinkles due to the buckling in response to an uniaxial compression raft provides the value of the Young's modulus $E$:
\begin{equation}
E=\dfrac{3}{4\pi^4}\,\dfrac{\rho\,g (1-\nu^2)\,\lambda_w^4}{e^3}\, ,
\end{equation}
where $\rho$ is the water density, $g$ the gravity acceleration and $\nu$ the Poisson modulus here evaluated equal to $1/3$ and $e$ the raft thickness is assimilated to the particle diameter $d$. We assume indeed that our membrane is a monolayer. If we consider in a plane hexagonally close-packed identical spheres of radius $R$ (see Fig.~\ref{ridesGraphite} (a), we find $\delta=- \, \epsilon / \sqrt{3}$. The Poisson coefficient is defined as the ratio of the relative elongation  $\nu=-\dfrac{\mathrm{d} X /X}{\mathrm{d} Y /Y}$. Thus, $ \nu=-(2 \delta /2 \epsilon) \times (Y/X)=1/\sqrt{3} \times(2 R /(2R \, \sqrt{3}))=1/3$. Note, that this result differs from the finding of Vella et al.~\cite{Vella2004}, who found with the same reasoning $\nu=1/ \sqrt{3}$. In our case, $\nu=1/3$ is well below $1/2$, which corresponds to the limit value for a perfectly incompressible material. {However, the raft is in fact assimilated to a 2D material, as its thickness does not change. In this case, the incompressible limit corresponds to $\nu=1$~\cite{thorpe1992new}.} \\

By compressing a membrane made according to our protocol as depicted in Fig.~\ref{ridesGraphite} (b), we find after a statistical analysis $\lambda_w = (0.59 \pm 0.15)$ mm. From this measurement, we deduce $E = 0.8 \times 10^4$ Pa in the range $[0.25,2.0]$ $10^4$ Pa. Given the large uncertainty, this value must be taken as an order of magnitude. As pointed previously~\cite{Vella2004}, for a particle raft whose cohesion is ensured by the capillarity, the Young's modulus scales as $\gamma/d \approx 0.7\times 10^4$ Pa. The bending stiffness can be then evaluated by $B=\dfrac{E\, d^3}{12\,(1-\nu^2)}\approx 10^{-14}$ Pa m$^{3}$. This small value shows that only deformations at very small scales will contribute to the bending elasticity. 

\begin{figure}[h!]
\centering

\includegraphics[width=1\textwidth]{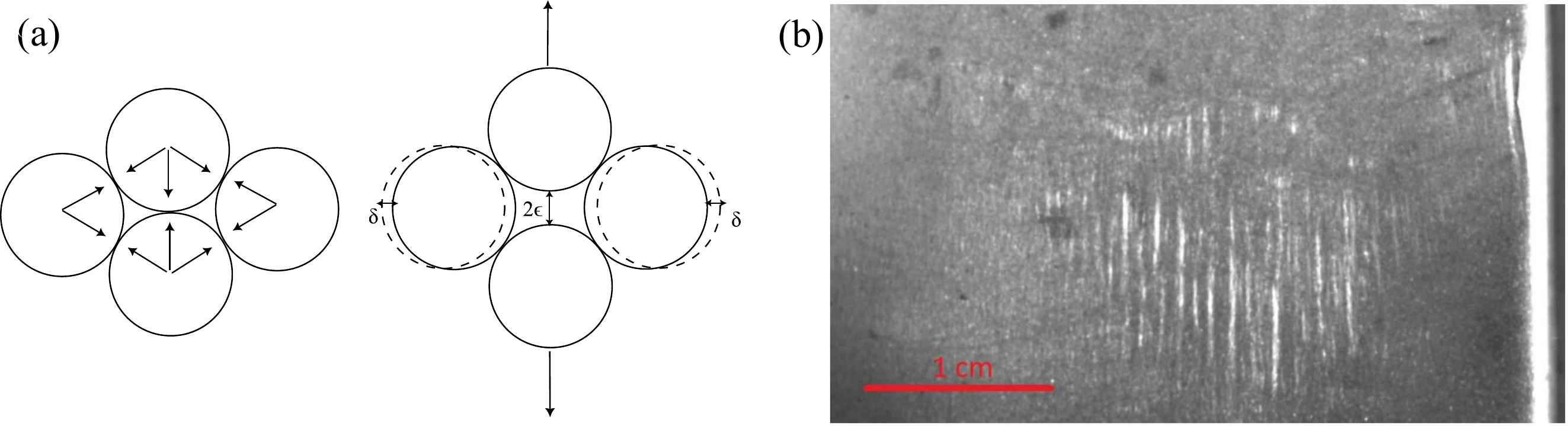}
\caption{(a) Origin of the Poisson coefficient for close packed spheres from Vella et al.~\cite{Vella2004}. (b) example of wrinkles view from the top observed by compressing a particle raft made of graphite particles ( 325 mesh, typical diameter $10$ $\mu$m)  and added with a surface density of $11.1$ g/m$^2$.}
\label{ridesGraphite}
\end{figure}

\section{Surface wave propagation}
\label{SurfaceWavespropagation}

We consider propagation of sinusoidal surface wave deforming the free-surface in presence or not of the floating membrane.
The free-surface deformation assimilated to the wave elevation reads $\eta(x,t)=a\,\cos (\omega \, t - k\,x )$, with $k$ the wave number ($k= (2\pi)/\lambda$), $\omega=2\pi\,f$ the pulsation and $a$ the wave amplitude.\\
Without membrane, assuming small deformation ($a\,k \ll 1$), small viscous dissipation and without current, the dispersion relation reads:
\begin{equation}
\omega^2=\tanh(k\,h)\,\left(g\,k+\dfrac{\gamma}{\rho}\,k^3 \right)
\label{RD1}
\end{equation}
with $h$ the fluid depth at rest, $g=9.81\,\mathrm{m}/\mathrm{s}^2$ the gravity acceleration at the Earth surface, $\gamma$ the surface tension coefficient. For pure water at $25^\circ$\,C, and $\gamma=72$\,mN/m, but without particular precautions a water surface is easily contaminated. Then, we estimate the surface tension to be about $\gamma=60$\,mN/m. $\rho$ is again the water density ($\rho=998$\,kg m$^{-3}$ at $25^\circ$ C). The phase velocity is defined by $v_\phi=\omega/k$ and the group velocity by $v_g=(\partial \omega)/(\partial k)$. If $k\,h \ll 1$, the effect of finite depth is negligible in wave propagation, \textit{i.e.} $\tanh(k\,h) \approx 1$ corresponding to the deep water regime. \\

The propagation of surface waves are also affected by the elasticity of the floating particle raft. In deep water regime, Planchette et al.~\cite{Planchette2012} demonstrated that the waves follow the dispersion relation of hydroelastic waves~\cite{Deike2013,Domino2018}:
\begin{equation}
\omega^2\,\left(1+d\,\dfrac{\rho_r}{\rho}\, k \right)=g\,k+\dfrac{\gamma_r}{\rho}\,k^3+\dfrac{B}{\rho}\,k^5 \, .
\end{equation}
Far small enough membrane, the inertia is mainly caused by the water flow below the surface, whereas the membrane elasticity brings new restoring mechanisms, stretching at intermediate wavelengths and bending at small ones. $\rho_r$ is the density of the raft, $B$ the bending modulus and $\gamma_r$ an effective surface tension which replaces the air/water surface tension, but has the same order of magnitude~\cite{Planchette2012}. Consequently, we can define and estimate characteristic lengths which separate gravity, capillary (surface tension) and bending wave regimes: \\
the gravity-capillary length $l_c=\sqrt{\dfrac{\gamma_r}{\rho\,g}}\approx 2.5 \,$ mm, 
the gravity-bending length $l_{Bg}=\left({\dfrac{B}{\rho\,g}}\right)^{1/4}\approx 32 \,$ $\mu$m,
the capillary-bending length $l_{Bc}=\sqrt{\dfrac{B}{\gamma_r}}\approx 0.4 \,$ $\mu$m.
Therefore,  for a raft with a thickness of a few tens of microns and wavelengths about few centimeters the bending and stretching of the membrane are negligible. The waves considered here resume thus to standard gravity surface waves. As the propagation medium does not change significantly at the interface between free-water and the floating raft, the waves are likely not reflected. 
 
However, the particle raft could further modify wave attenuation. The solid membrane acts indeed as a solid boundary condition, for the flow associated with the free-surface wave and changes the structure of the viscous boundary layer. In this case, the damping rate (inverse of a time) due to viscous dissipation is written as~\cite{Lamb1932}: 
\begin{equation}
\delta_w=\sqrt{2}/4\,\sqrt{\nu\,\omega}\,k \, ,
\label{deltaw}
\end{equation}
with $\nu$ the kinematic viscosity coefficient ($\nu \approx 10^{-6}$ m$^2$ s$^{-1}$ for water at ambient temperature). The spatial decay length becomes $l_d=v_g\,/ \delta$, with $v_g=\sqrt{g/(2\,k)} $. For a wave of frequency $3$ Hz, $l_d  \approx 4.8$ m. A wave emitted in $x=0$, behaves then as:  $\eta(x,t)=a\,\mathrm{e}^{-x/l_d}\cos ( \omega \,t -k\,x)$. We note also, that without particular precautions, the tap water is contaminated by atmospheric surfactants and the dissipation rate of surface waves is reasonably depicted by the Eq.~\ref{deltaw}, corresponding to the model of the inextensible film~\cite{Berhanu2022}. Therefore, we expect that the surface waves generated by the wavemaker in our experiments have a similar attenuation rate before the raft, on the raft and in presence of many floes smaller than the wavelength.

\begin{figure}[h!]
\centering
\includegraphics[width=.7\textwidth]{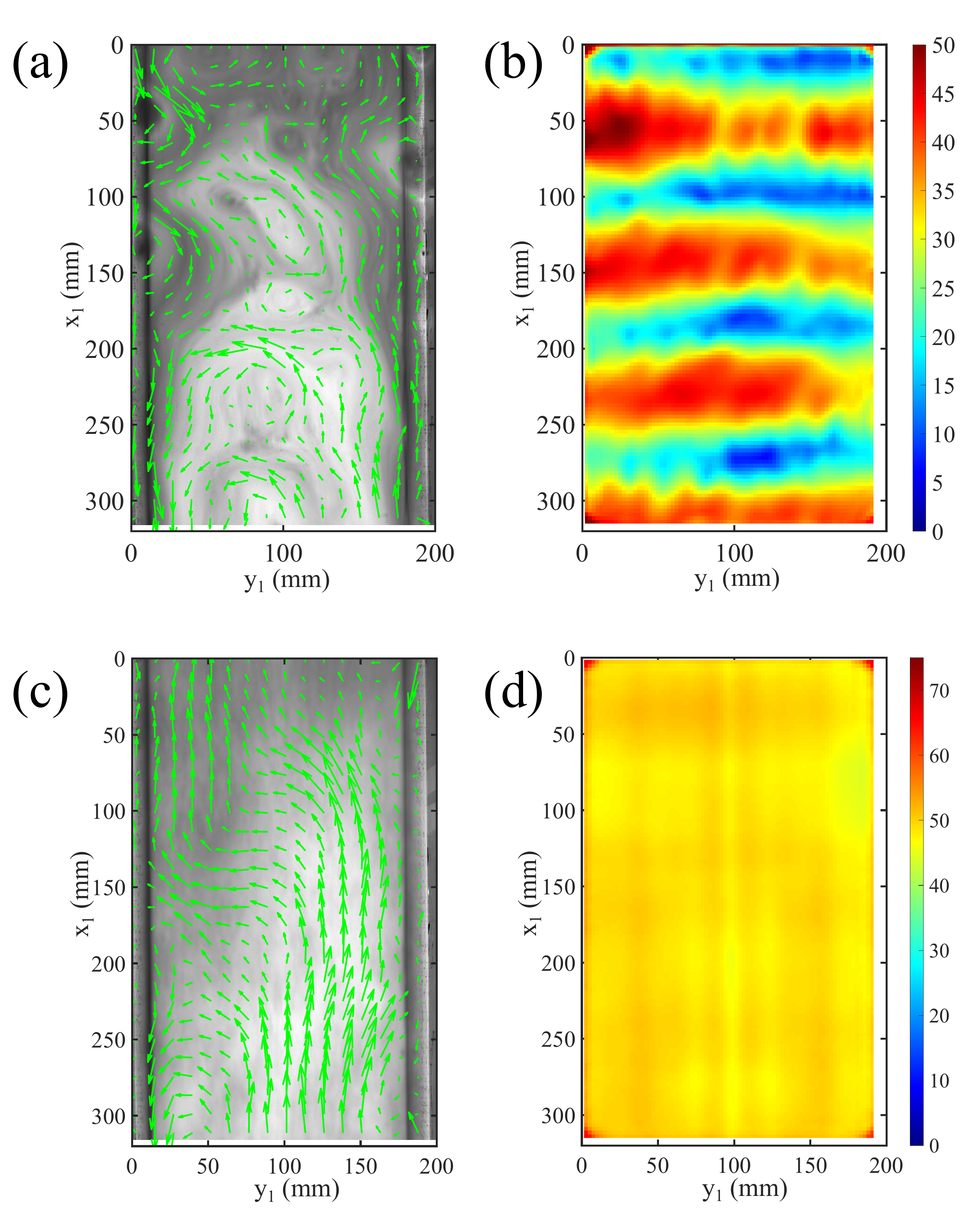} \hfill
\caption{Example of velocity field characterization using Particle Image Velocimetry algorithms. The experiments are performed in the second tank ($120\times 20$ cm) and the wavemaker is located on the top at a distance of $40$ cm from the line $x_1=0$ mm. (a) and (b) Standing waves ($f=3$ Hz and $a=1.45$ mm). (c) and (d) Random waves ($f \in [2.5,3.5]$ Hz and $a=1.45$ mm).
 (a) and (c) Time-averaged horizontal velocity field at the free-surface  $\langle \mathbf{u} \rangle_t$ corresponding to the wave-induced streaming flow plotted above the time-average of images. The top view is taken on the tank width and a portion of its length. The length of arrows is arbitrary. (b) and (d) Time-averaged norm of the velocity field $\langle ||u_x^2+u_y^2||\rangle_t$. The colormap is in mm per second. The time averaged are performed on $165$ s. In both case, an average flow arises from the oscillatory forcing by the waves.  However, for regular standing waves, the norm of the horizontal velocity is maximal  at the positions of wave nodes separated by $87.5$ mm close to half wavelength, whereas for random waves the wave field is homogeneous. }
\label{PIVfig}
\end{figure}

\section{Measurement of the surface flow by Particle Image Velocimetry}
\label{PIV}
In order to evaluate the surface flows induced by the wave motion, we measure the velocity field by applying Particle Image Velocimetry (P.I.V.) algorithms to the top-view images once the particle raft is fully disintegrated for a steady wave forcing. The very small floes creates a small scale texture on the images, which allows us to measure the displacement field with a correlation algorithm between two successive images. We use the software PIVlab~\cite{thielicke2014pivlab} to compute the velocity field on a grid of resolution of $2.5 \times 2.5 $ mm. For the experiments with the tank 2 only and from the instantaneous velocity field, we derive by time averaging the mean velocity field $\langle \mathbf{u} \rangle_t$ and also the mean average of the norm of the surface velocity field $\langle \parallel \mathbf{u}\parallel \rangle_t$. The oscillatory contribution due to the orbital velocity field linked to the standing wave cancels for the first quantity but not for the second. The relative magnitude of the orbital velocity field at the surface and of the drift flow can be separately evaluated. In Fig.~\ref{PIVfig}, we plot for regular and random waves, the time averaged wave field geometry $\langle \mathbf{u} \rangle_t$ with arrows and the $\langle || \mathbf{u} || \rangle_t$. The instantaneous velocity field mainly directed along the $x$ axis corresponds to the orbital velocity field associated to waves (not shown). After time averaging, for regular and random waves, a mean flow of complex geometry appears, with clear presence for regular waves only of two-dimensional vortices. For this example, where the wave amplitude in the field of view is $1.45$ mm, we measure $|| \langle \mathbf{u} \rangle_t || =3.24 $ mm/s for regular waves and $||  \langle  \mathbf{u}  \rangle_t || =1.79$ mm/s for random waves. By computing the time average of the horizontal velocity field, we access to the magnitude of the orbital velocity field, that is $\langle || \mathbf{u} || \rangle_t=29.29 $ mm/s for regular waves and $\langle || \mathbf{u} || \rangle_t=48.92 $ mm/s for random waves. We note, that the first value is close to the expected order of magnitude for linear waves $a\,\omega \approx 27.33$ mm/s, whereas an analog order of magnitude is more difficult to obtain for random waves.
For these wave amplitudes, the streaming velocity field is roughly ten times smaller than the orbital velocity field. We note that the Stokes drift magnitude (valid only for progressive waves) gives $U_{SD}=c_\phi\,(a\,k)^2 \approx 1.42$ mm/s, thus the correct order of magnitude, which suggests that the streaming flow generation arises from a nonlinear effect.

\clearpage

\section{Fracture by transverse stretching}
\label{StretchingMechanism}

{In this appendix, we consider the possible mechanism of raft breaking by stretching due to the adhesion of the raft on the lateral walls. In our experiments, we have seen in section~\ref{ObservationsStandingWaves}, that the presence of a nylon mesh on the walls favoring the sliding of the contact line and of the raft change qualitatively the breaking of the particle raft. Without the mesh and glass walls, in the conditions of experiments and for small wave amplitude, the contact line tends to be pinned, implying that the wave deformation depends on the $y$-coordinate, as described in Monsalve et al.~\cite{monsalve2022space}. Moreover, in the first tank of width $7.5$ cm, we observe sometimes few shear fractures near the walls and perpendicular to them (see Fig.~\ref{Shear90fracts}). This fracture pattern cannot be explained by the mechanisms described in section~\ref{Fracturepatterns} and is never observed in the second tank of width $20$ cm. We want thus evaluate in this appendix the possible mechanical effect of the walls on the breaking of the raft and how this effect depends on the tank width. We investigate thus the stretching of the raft due to its pinning on the walls.}

\begin{figure}[h!]
\centering
\includegraphics[width=.4\textwidth]{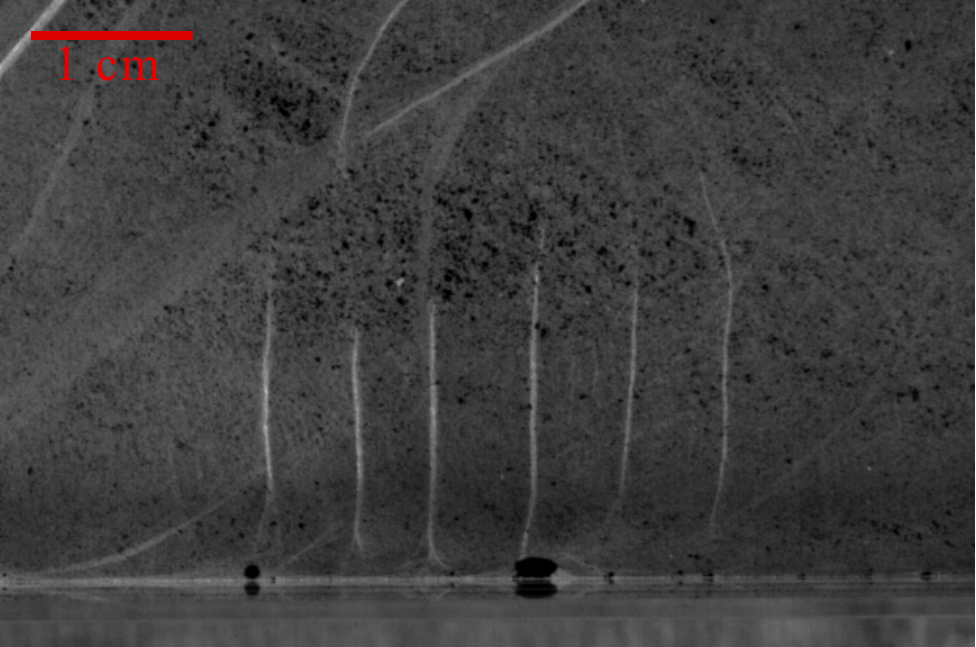}
\caption{{Example of small shearing fractures perpendicular to the walls in the first tank. Experimental conditions: $a=2.2$ mm, $\lambda=22$ cm, $V=3.75$ cm.} }
\label{Shear90fracts}
\end{figure}

{The raft is indeed not constrained along the $x$ axis, allowing it to freely responds to the wave propagation. The spatial extension of the pinned zone is of order the capillary length $l_c=\sqrt{\dfrac{\gamma_r}{\rho\,g}}$ (see Appendix~\ref{SurfaceWavespropagation}), because the deformation results from the competition between the gravity surface wave and the capillary force at the pinned contact line. Therefore, in the presence of {progressive} waves and considering weak deformation, a simple representation of the raft deformation would be the following profile:}

\begin{equation}
  { \xi(x,y,t)=a \cos(kx-\omega t)\left(1-\dfrac{\cosh(\kappa\,y)}{\cosh(\kappa\,V)}\right) }
  \label{Wavelateralprofile}
\end{equation}
 {where $a$ and $k$ are the amplitude and the wave number, $\kappa=1/l_c$ is the inverse of the capillary length and $V=W/2$ is the half-width of the tank. The static meniscus has been neglected. For a distance to the wall larger than $2\,l_c$, the surface deformation is quasi-flat. For narrow channels, of width smaller than few $l_c$, this shape can be approximated to a parabola shape. The shape of the free-surface deformation in presence of pinning of the contact line due to gravity surface wave propagation has been measured in a channel of width between $5$ and $7$ $l_c$ by Monsalve et al.~\cite{monsalve2022space}. The authors find that the lateral wave profile can be decomposed along $y$ into modes of form $\cos\left((2\,n-1) \frac{\pi/W} y \right)$ with $n=1,2,3 \dots$ The lowest order modes dominate then the decomposition for a narrow channel. We use in this appendix the surface shape given by Eq.~\ref{Wavelateralprofile}, although a more accurate shape could be given using a modal decomposition along $y$~\cite{monsalve2022space,shankar2007frequencies}.}  

{We want here to determine the stress field in the raft for a prescribed deformation given by $\xi(x,y,t)$ (Eq.~\ref{Wavelateralprofile}). For the sake of simplicity, we will work in the wave referential, so that we can remove the temporal dependency: $\xi(x,y)=a \cos(kx)\left(1-\frac{\cosh(\kappa\,y)}{\cosh(\kappa\,V)}\right)$. Because of the periodicity and the symmetries of the system, we limit our study to the domain $x\in[-\lambda/4,\lambda/4]$ and $y\in[-V,V]$. Considering that the raft thickness $e$ is about $10$ $\mu$m, the wavelength $\lambda$ about $22$ cm and the wave amplitude about $1$ mm, we make the following hypotheses:}
\begin{itemize}[label=---]
    \item {{\underline{Thin plate.} The particle raft is thin enough to consider that shear stresses are homogeneous in the thickness. The problem becomes 2D-dimensional. } }
    
\item {{\underline{Pure stretching.} Plate deformations are due solely to wave-induced stretching. The pure bending energy per surface unit scales as $\mathcal{E}_\mathrm{bend}\sim\frac{Ee^3a^2}{V^4}$ whereas the pure stretching energy per surface units scales as $\mathcal{E}_\mathrm{stretch}\sim\frac{Eea^4}{V^4}$ \cite{LandauElasticity}. Hence, $\mathcal{E}_\mathrm{bend}/\mathcal{E}_\mathrm{stretch}\sim e^2/a^2\sim 10^{-4}$.}}

\item {{ \underline{Boundary conditions for the plate.} We denote $\mathbf{u}=(u_x,u_y)$ the displacement vector. The pinning of the lateral extremities of the raft to the tank walls implies $u_y(x,y=\pm V)=0$. However during the deformation, horizontal sliding is allowed at the walls, which translates into a condition on the stress tensor components $\sigma_{xy}(x,y=\pm V)=\mu \, \sigma_{yy}(x,y=\pm V)$, where $\mu$ is a constant friction coefficient. This friction condition at the walls has been proposed to model the compression of particle rafts~\cite{aumaitre2011measurement}, but the value of the coefficient $\mu$ of order one is not known. We also have periodic boundary conditions on the x-axis extremities such that $u_x(x=-\lambda/4,y)=u_x(x=\lambda/4,y)$ and $u_y(x=-\lambda/4,y)=u_y(x=\lambda/4,y)$.}}
    
\end{itemize}

\subsection*{{Calculation of deformations and stresses}}

{First of all, we need to specify what is the deformation of the plate as the wave passes through. In a plate subjected to large deflections, supplementary deformations will arise from the out of plane deformations.
Thus, neglecting quadratic terms in the derivatives of $u_i$ (small deformation assumption), the components of the strain tensor read \cite{LandauElasticity}:}


\begin{eqnarray}
   \epsilon_{xx} &=& {\frac{\partial u_x}{\partial x} + \frac{1}{2}\left(\frac{\partial \xi}{\partial x}\right)^2 } \label{eq:strain} \\
    \epsilon_{yy} &=& \frac{\partial u_y}{\partial y}+\frac{1}{2}\left(\frac{\partial \xi}{\partial y}\right)^2 \\
    \epsilon_{xy} &=& \epsilon_{yx} = {\frac{1}{2}\left(\frac{\partial u_x}{\partial y}+\frac{\partial u_y}{\partial x}\right) +\frac{1}{2}\left(\frac{\partial \xi}{\partial x}\right) \left(\frac{\partial \xi}{\partial y}\right) } \, .
\end{eqnarray}

{The pure stretching stress tensor $\sigma$ \cite{LandauElasticity} can then be calculated injecting these three expressions in}
\begin{eqnarray}
  \sigma_{xx} &=& \frac{E}{1-\nu^2}(\epsilon_{xx}+\nu\epsilon_{yy}) \\\
   \sigma_{yy} &=& \frac{E}{1-\nu^2}(\nu\epsilon_{xx}+\epsilon_{yy})  \\
   \sigma_{xy} &=& \sigma_{yx} = \frac{E}{1+\nu}\epsilon_{xy} \label{eq:stress} \, .
\end{eqnarray}


{The minimum energy condition~\cite{LandauElasticity} for the in-plane stresses implies the following two equations:} 
\begin{equation}
   {\frac{\partial \sigma_{xx}}{\partial x}+\frac{\partial \sigma_{xy}}{\partial y}=0 }
    \label{equili2}
\end{equation}
\begin{equation}
   { \frac{\partial \sigma_{yx}}{\partial x}+\frac{\partial \sigma_{yy}}{\partial y}=0 \, . }
    \label{equili3}
\end{equation}

\subsubsection*{{Numerical simulation}}

{In order to find a solution that satisfies the boundary conditions, we solve numerically the problem. Instead of solving Eqs.~\ref{equili2} and \ref{equili3} associated to the stress field, we solve the equations associated to the displacements $u_x$ and $u_y$ to manage more easily the boundary conditions. We can find these equations injecting Eq.~\ref{eq:strain} to \ref{eq:stress} into Eq.~\ref{equili2} and \ref{equili3}~\cite{LandauElasticity}:}
\begin{eqnarray}
  { \frac{\partial^2 u_x}{\partial x^2}+\frac{1-\nu}{2}\frac{\partial^2 u_x}{\partial y^2}+\frac{1+\nu}{2}\frac{\partial^2 u_y}{\partial x \partial y} }&{=}&{ -\frac{\partial \xi}{\partial x}\frac{\partial^2 \xi}{\partial x^2}-\frac{1-\nu}{2}\frac{\partial \xi}{\partial x}\frac{\partial^2 \xi}{\partial y^2}-\frac{1+\nu}{2}\frac{\partial \xi}{\partial y}\frac{\partial^2 \xi}{\partial x \partial y} } \\
  {  \frac{\partial^2 u_y}{\partial y^2}+\frac{1-\nu}{2}\frac{\partial^2 u_y}{\partial x^2}+\frac{1+\nu}{2}\frac{\partial^2 u_x}{\partial x \partial y} } & {=}& { -\frac{\partial \xi}{\partial y}\frac{\partial^2 \xi}{\partial y^2}-\frac{1-\nu}{2}\frac{\partial \xi}{\partial y}\frac{\partial^2 \xi}{\partial x^2}-\frac{1+\nu}{2}\frac{\partial \xi}{\partial x}\frac{\partial^2 \xi}{\partial x \partial y} \, . }
\end{eqnarray}

{We use the same boundary conditions that the one specified at the beginning of this appendix. Noticing $\frac{\partial u_y}{\partial x}(x,y=\pm V)=0$ and $\frac{\partial \xi}{\partial x}(x,y=\pm V)=0$, sliding condition reads in term of displacement:}
$$ {2\mu\left\{\frac{\partial u_y}{\partial y}(x,y=\pm V)+\nu\frac{\partial u_x}{\partial x}(x,y=\pm V)+\frac{1}{2}\left(\frac{\partial\xi}{\partial y}(x,y=\pm V)\right)^2\right\}=(1-\nu)\frac{\partial u_x}{\partial y}(x,y=\pm V) \, .}$$

{We implement a spectral method (Dedalus Project~\cite{Dedalus}) using a Chebyshev discretization on axis $y$ and a Fourier discretization on axis $x$.} {We use the values of the elastic parameters found previously in Appendix~\ref{Emod}, \textit{i.e.} $E=10^4$ Pa and $\nu=1/3$. The value of the friction coefficient is arbitrary. Here, we test $\mu=0$ (free slip) and $\mu=0.5$, as in previous numerical simulations of particle raft with friction on the walls, where the values $\mu=0.5$ and $\mu=0.75$ have been tested~\cite{aumaitre2011measurement}. {We use a value of the surface tension for the raft surface $\gamma_r=60$ mN/m, which corresponds to a capillary length $l_c=2.47$ mm. In the presence of a particle raft, surface tension can be lowered, but remains of the order of $\gamma_r=60$ mN/m \cite{Planchette2012}. \\} } 

\begin{figure}[h!]
\centering
\includegraphics[width=\textwidth]{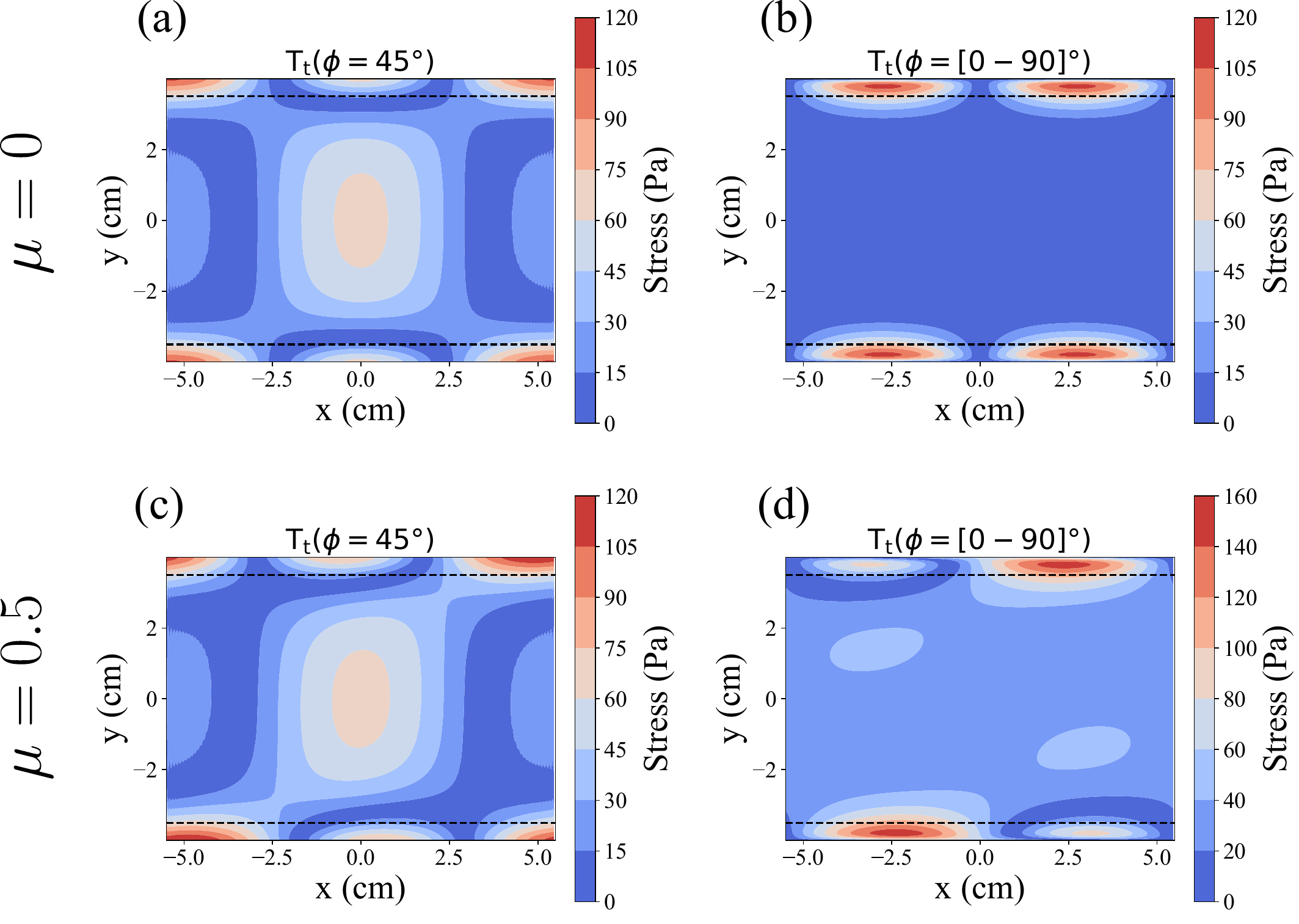}
\caption{{Tangential shearing for 45$^\circ$ fractures (a) and vertical/horizontal fractures (b) for $\mu=0$. (c) and (d) show the results for $\mu=0.5$.  The plots result from the numerical simulation. Parameters are chosen to be close from the experiments performed in the narrowest tank: $a=2.2$ mm, $\lambda=22$ cm, $V=4$ cm. We set $\nu=1/3$. Black dashed lines represent the distance $2l_c$ from the walls. 45$^\circ$ fractures seem to be favored at the antinodes of the wave close to the lateral walls. Vertical/horizontal fractures seem to be favored in between nodes and antinodes, close to the walls.}} 
\label{fig:Tt}
\end{figure}

As in Section~\ref{Bending}, we can evaluate the stress associated to a fracture which has an angle $\phi$ with the $x$ axis (see Fig.~\ref{schemafrac}):
\begin{equation}
\mathbf{T}=\mathbf{\sigma}\cdot\mathbf{n}
\quad \mathrm{with} \quad \mathbf{\sigma}=
\begin{pmatrix}
\sigma_{xx} & \sigma_{xy} \\
\sigma_{xy} & \sigma_{yy}
\end{pmatrix}
\quad \mathrm{and}\quad \mathbf{n}=
\begin{pmatrix}
\cos\phi \\
\sin\phi
\end{pmatrix} \, .
\end{equation}
Then, if we look for shear mode fractures, we should get interested in the component of $\mathbf{T}$ which is tangential to the fracture:
\begin{equation}
    T_t=||\mathbf{T}-(\mathbf{T}\cdot\mathbf{n})\cdot\mathbf{n}|| \, .
\end{equation}
The special cases $\phi=$0, 45 and 90$^\circ$ read
\begin{eqnarray}
    T_t(\phi=[0-90]^\circ) &=& |\sigma_{xy}|\\
    T_t(\phi=45^\circ) &=& \frac{|\sigma_{xx}-\sigma_{yy}|}{2} \, .
\end{eqnarray}
and are plotted in Fig.~\ref{fig:Tt}. {Tangential shearing are almost concentrated close to the wall. 45$^\circ$ fractures seem to be favored at the antinodes of the wave whereas vertical/horizontal fractures seem to be favored in between nodes and antinodes.}\\

\begin{figure}[h!]
\centering
\includegraphics[width=\textwidth]{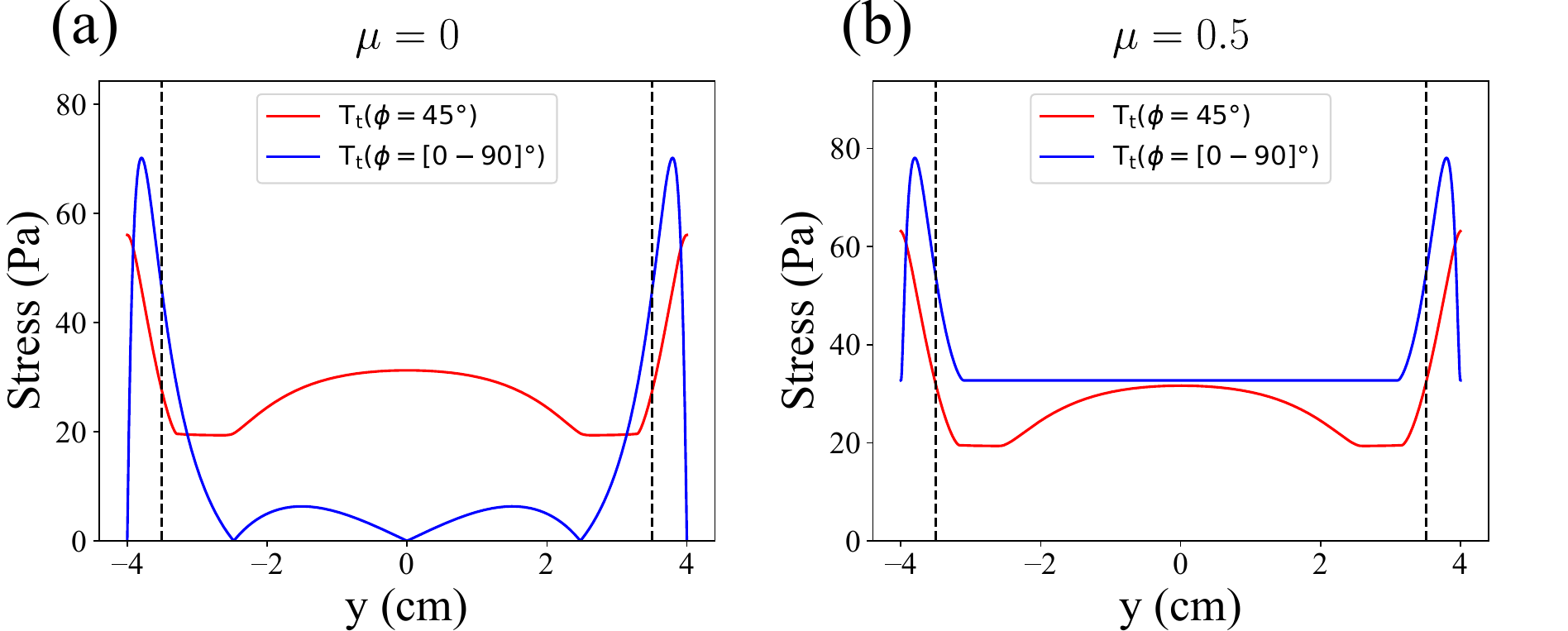}
\caption{{$x$ averaged profiles of the tangential shearing shown in Fig.~\ref{fig:Tt}  for $\mu=0$ (a) and $\mu=0.5$ (b). Black dashed lines represent the distance $2l_c$ from the walls. The values of the parameters are $a=2.2$ mm, $\lambda=22$ cm ($k=28.6$ m$^{-1}$), $V=4$ cm. We set $\nu=1/3$.}} 
\label{fig:y_profile}
\end{figure}

{$x$ averaged profiles of tangential shearing are plotted Fig~\ref{fig:y_profile} for $\mu=0$ and $\mu=0.5$. In both cases, maximal stresses are reached in the meniscus. For $\mu=0$, the stress associated with horizontal/vertical fractures goes to zero at the walls $y=\pm V$. This is because the free-slip condition imposes $\sigma_{xy}(x,y=\pm V)=0$. However for $\mu=0.5$, it is non-zero because the sliding condition imposes $\sigma_{xy}(x,y=\pm V)=\mu \, \sigma_{yy}(x,y=\pm V)$. Note also that $T_t(\phi=[0-90]^\circ)$ is higher and almost constant in the core of the particle raft when $\mu=0.5$.}\\

\begin{figure}[h!]
\centering
\includegraphics[width=.95\textwidth]{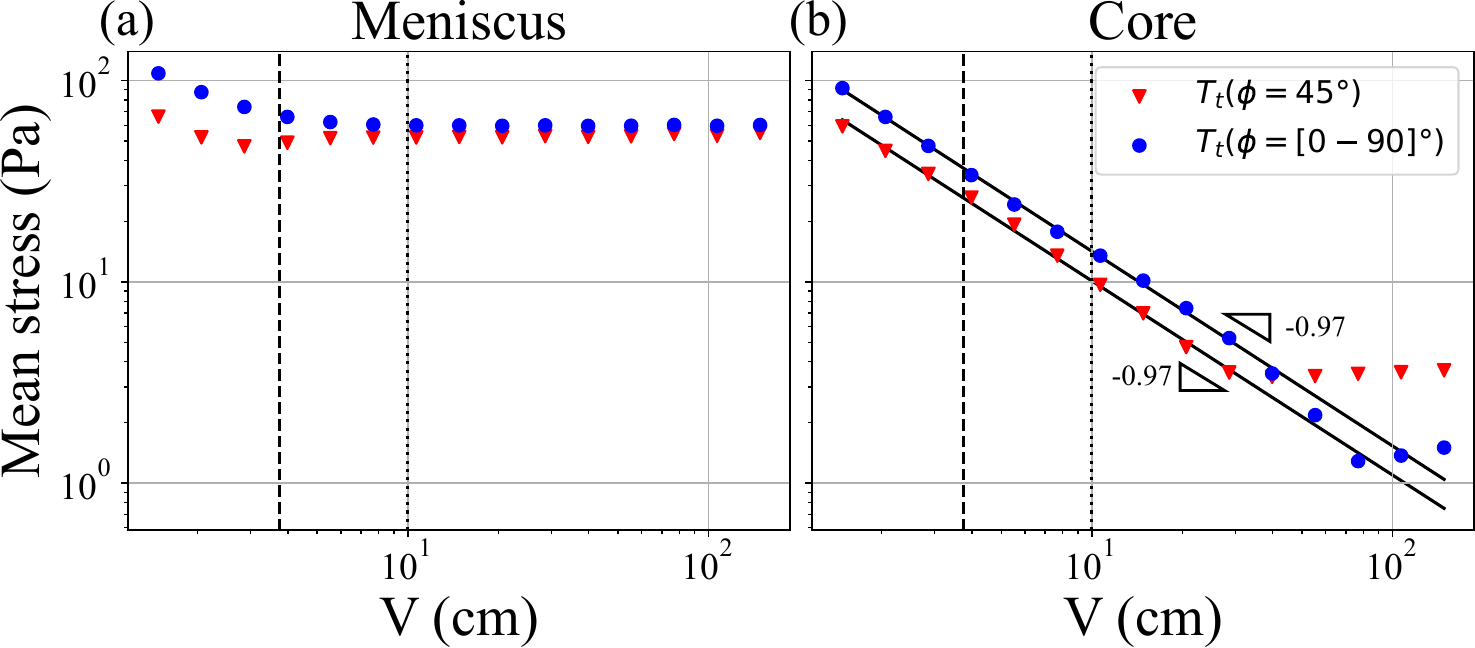}
\caption{{Tangential stresses for 45$^\circ$ fractures (red triangles) and vertical/horizontal fractures (blue circles) averaged in the meniscus ($y<-V+2l_c$ and $y>V-2l_c$) (a) and in the core ($-V+2l_c<y<V-2l_c$) (b). We vary the tank half-width $V$. The values of the parameters are $a=2.2$ mm, $\lambda=22$ cm ($k=28.6$ m$^{-1}$). We set $\nu=1/3$ and $\mu=0.5$. Dashed (resp. dotted) black line shows the half width $V$ of tank 1 (resp. tank 2).}} 
\label{Fig:scale_V}
\end{figure}

{Then, we investigate the dependence of tangential shear on the tank half-width $V$, plotting in Fig.~\ref{Fig:scale_V} the spatial average of tangential stresses for the two crack orientations considered $  T_t(\phi=45^\circ) $ and $T_t(\phi=[0-90]^\circ)$, for both the meniscus region and the core region of the particle raft. For the meniscus region, stresses first decrease as $V$ increases, then appear to stabilize at a constant value of order 60 Pa when $V>7$ cm. In fact, when the raft is large enough, its half-width $V$ no longer has any impact on the shape and size of the meniscus. In the core region however, stresses continue to decrease with the scaling $V^{-1}$ and then stabilize at values of order 1 Pa.}\\

\begin{figure}[h!]
\centering
\includegraphics[width=1\textwidth]{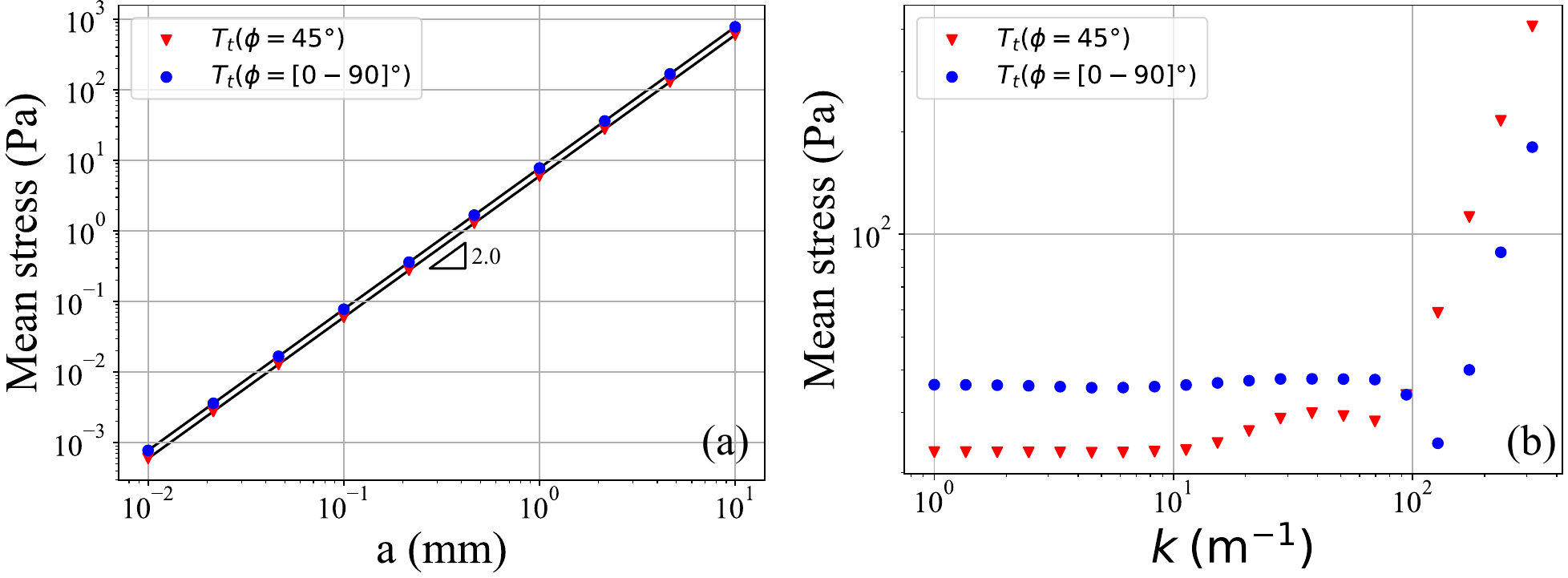}
\caption{{Mean tangential stresses for 45$^\circ$ fractures (red triangles) and vertical/horizontal fractures (blue circles). We vary the wave amplitude $a$ in (a) and the wave number $k$ in (b). When not specified in the $x$ axis, the values of the parameters are $a=2.2$ mm, $\lambda=22$ cm ($k=28.6$ m$^{-1}$), $V=4$ cm. We set $\nu=1/3$ and $\mu=0.5$.}}
\label{Fig:scale_a_k}
\end{figure}

{In Fig.~\ref{Fig:scale_a_k}, we also study the dependence of tangential shear on the wave amplitude $a$ (a) and the wavenumber $k$ (b), by plotting the spatial average of tangential stresses for the two crack orientations over the entire domain.  First, we note that stresses are well proportional to $a^2$. Second, $T_t(\phi=45^\circ)$ and $T_t(\phi=[0-90]^\circ)$ appear to be nearly constants and of order 30 Pa for $k<100$ m$^{-1}$ and then increase drastically with $k$. This corresponds to the regime where wavelength is lower than 6 cm, which is never the case in our experiments.}\\


\newpage
{According to this numerical study, the magnitude of the tangential stresses for 45$^\circ$ and 0$^\circ$/90$^\circ$ fractures is approximately of order 100 Pa for $V=W/2=4$ cm (see Fig.~\ref{fig:Tt} and Fig.~\ref{fig:y_profile}). This is two time less than the shear stress caused by the wave induced flow under the plate (Section~\ref{ViscousStresses}). Furthermore, they are mainly localised in the meniscus region of the plate whereas viscous stress is applied under the whole plate. Fig.~\ref{Fig:scale_V} also shows that the stresses due to pinning effect decrease in the core of the raft as $V$ increases. Experimentally, 45$^\circ$ cracks are observed at any transverse location of the raft, even when the raft is not pinned to the lateral walls of the tank. For all these reasons, the viscous stress is still the most likely candidate to explain the breaking of the particle rafts studied in this article, especially in the second tank of half-width $V=W/2=10$ cm.}\\

{Concerning the shear fractures oriented at $90^\circ$ observed near the walls in the first tank (see Fig.~\ref{Shear90fracts}), the numerical study performed in this appendix suggests that it is difficult to attribute definitively this fracture pattern to the stretching mechanism due to the lateral deflection of the raft. In fact, we found that 45$^\circ$ cracks should be predominant at the immediate vicinity of the walls according to Fig.~\ref{fig:Tt} and Fig.~\ref{fig:y_profile}. Nevertheless, at about a distance $l_c$ of the walls of the walls, the 90$^\circ$ become dominant. Further studies would be useful to model the stresses on a floating membrane subjected to water waves, when the sides are attached on a wall.}


\begin{thebibliography}{77}%
\makeatletter
\providecommand \@ifxundefined [1]{%
 \@ifx{#1\undefined}
}%
\providecommand \@ifnum [1]{%
 \ifnum #1\expandafter \@firstoftwo
 \else \expandafter \@secondoftwo
 \fi
}%
\providecommand \@ifx [1]{%
 \ifx #1\expandafter \@firstoftwo
 \else \expandafter \@secondoftwo
 \fi
}%
\providecommand \natexlab [1]{#1}%
\providecommand \enquote  [1]{``#1''}%
\providecommand \bibnamefont  [1]{#1}%
\providecommand \bibfnamefont [1]{#1}%
\providecommand \citenamefont [1]{#1}%
\providecommand \href@noop [0]{\@secondoftwo}%
\providecommand \href [0]{\begingroup \@sanitize@url \@href}%
\providecommand \@href[1]{\@@startlink{#1}\@@href}%
\providecommand \@@href[1]{\endgroup#1\@@endlink}%
\providecommand \@sanitize@url [0]{\catcode `\\12\catcode `\$12\catcode
  `\&12\catcode `\#12\catcode `\^12\catcode `\_12\catcode `\%12\relax}%
\providecommand \@@startlink[1]{}%
\providecommand \@@endlink[0]{}%
\providecommand \url  [0]{\begingroup\@sanitize@url \@url }%
\providecommand \@url [1]{\endgroup\@href {#1}{\urlprefix }}%
\providecommand \urlprefix  [0]{URL }%
\providecommand \Eprint [0]{\href }%
\providecommand \doibase [0]{https://doi.org/}%
\providecommand \selectlanguage [0]{\@gobble}%
\providecommand \bibinfo  [0]{\@secondoftwo}%
\providecommand \bibfield  [0]{\@secondoftwo}%
\providecommand \translation [1]{[#1]}%
\providecommand \BibitemOpen [0]{}%
\providecommand \bibitemStop [0]{}%
\providecommand \bibitemNoStop [0]{.\EOS\space}%
\providecommand \EOS [0]{\spacefactor3000\relax}%
\providecommand \BibitemShut  [1]{\csname bibitem#1\endcsname}%
\let\auto@bib@innerbib\@empty
\bibitem [{\citenamefont {Spiro}\ and\ \citenamefont
  {Jaganyl}(1993)}]{Spiro1993}%
  \BibitemOpen
  \bibfield  {author} {\bibinfo {author} {\bibfnamefont {M.}~\bibnamefont
  {Spiro}}\ and\ \bibinfo {author} {\bibfnamefont {D.}~\bibnamefont
  {Jaganyl}},\ }\bibfield  {title} {\bibinfo {title} {What causes scum on
  tea?},\ }\href {https://doi.org/10.1038/364581a0} {\bibfield  {journal}
  {\bibinfo  {journal} {Nature}\ }\textbf {\bibinfo {volume} {364}},\ \bibinfo
  {pages} {581} (\bibinfo {year} {1993})}\BibitemShut {NoStop}%
\bibitem [{\citenamefont {Enzweiler}\ and\ \citenamefont
  {de~Oliveira}(1994)}]{Enzweiler1994}%
  \BibitemOpen
  \bibfield  {author} {\bibinfo {author} {\bibfnamefont {J.}~\bibnamefont
  {Enzweiler}}\ and\ \bibinfo {author} {\bibfnamefont {M.~G.}\ \bibnamefont
  {de~Oliveira}},\ }\bibfield  {title} {\bibinfo {title} {More on tea scum.},\
  }\href {https://doi.org/10.1038/364581a0} {\bibfield  {journal} {\bibinfo
  {journal} {Nature}\ }\textbf {\bibinfo {volume} {367}},\ \bibinfo {pages}
  {602} (\bibinfo {year} {1994})}\BibitemShut {NoStop}%
\bibitem [{\citenamefont {Lehoucq}\ \emph {et~al.}(2015)\citenamefont
  {Lehoucq}, \citenamefont {Weiss}, \citenamefont {Dubrulle}, \citenamefont
  {Amon}, \citenamefont {Le~Bouil}, \citenamefont {Crassous}, \citenamefont
  {Amitrano},\ and\ \citenamefont {Graner}}]{Lehoucq2015}%
  \BibitemOpen
  \bibfield  {author} {\bibinfo {author} {\bibfnamefont {R.}~\bibnamefont
  {Lehoucq}}, \bibinfo {author} {\bibfnamefont {J.}~\bibnamefont {Weiss}},
  \bibinfo {author} {\bibfnamefont {B.}~\bibnamefont {Dubrulle}}, \bibinfo
  {author} {\bibfnamefont {A.}~\bibnamefont {Amon}}, \bibinfo {author}
  {\bibfnamefont {A.}~\bibnamefont {Le~Bouil}}, \bibinfo {author}
  {\bibfnamefont {J.}~\bibnamefont {Crassous}}, \bibinfo {author}
  {\bibfnamefont {D.}~\bibnamefont {Amitrano}},\ and\ \bibinfo {author}
  {\bibfnamefont {F.}~\bibnamefont {Graner}},\ }\bibfield  {title} {\bibinfo
  {title} {Analysis of image vs. position, scale and direction reveals pattern
  texture anisotropy},\ }\bibfield  {journal} {\bibinfo  {journal} {Frontiers
  in Physics}\ }\textbf {\bibinfo {volume} {2}},\ \href
  {https://doi.org/10.3389/fphy.2014.00084} {10.3389/fphy.2014.00084} (\bibinfo
  {year} {2015})\BibitemShut {NoStop}%
\bibitem [{\citenamefont {Tanizawa}\ \emph {et~al.}(2007)\citenamefont
  {Tanizawa}, \citenamefont {Abe},\ and\ \citenamefont
  {Yamada}}]{Tanizawa2007}%
  \BibitemOpen
  \bibfield  {author} {\bibinfo {author} {\bibfnamefont {Y.}~\bibnamefont
  {Tanizawa}}, \bibinfo {author} {\bibfnamefont {T.}~\bibnamefont {Abe}},\ and\
  \bibinfo {author} {\bibfnamefont {K.}~\bibnamefont {Yamada}},\ }\bibfield
  {title} {\bibinfo {title} {Black tea stain formed on the surface of teacups
  and pots. part 1. - study on the chemical composition and structure},\ }\href
  {https://doi.org/10.1016/j.foodchem.2006.05.068} {\bibfield  {journal}
  {\bibinfo  {journal} {Food Chemistry}\ }\textbf {\bibinfo {volume} {103}},\
  \bibinfo {pages} {1} (\bibinfo {year} {2007})}\BibitemShut {NoStop}%
\bibitem [{\citenamefont {Proti{\`e}re}(2023)}]{protiere2023particle}%
  \BibitemOpen
  \bibfield  {author} {\bibinfo {author} {\bibfnamefont {S.}~\bibnamefont
  {Proti{\`e}re}},\ }\bibfield  {title} {\bibinfo {title} {Particle rafts and
  armored droplets},\ }\href
  {https://doi.org/10.1146/annurev-fluid-030322-015150} {\bibfield  {journal}
  {\bibinfo  {journal} {Annual Review of Fluid Mechanics}\ }\textbf {\bibinfo
  {volume} {55}},\ \bibinfo {pages} {459} (\bibinfo {year} {2023})}\BibitemShut
  {NoStop}%
\bibitem [{\citenamefont {Lagarde}\ \emph {et~al.}(2019)\citenamefont
  {Lagarde}, \citenamefont {Josserand},\ and\ \citenamefont
  {Protiere}}]{lagarde2019capillary}%
  \BibitemOpen
  \bibfield  {author} {\bibinfo {author} {\bibfnamefont {A.}~\bibnamefont
  {Lagarde}}, \bibinfo {author} {\bibfnamefont {C.}~\bibnamefont {Josserand}},\
  and\ \bibinfo {author} {\bibfnamefont {S.}~\bibnamefont {Protiere}},\
  }\bibfield  {title} {\bibinfo {title} {The capillary interaction between
  pairs of granular rafts},\ }\href {https://doi.org/10.1039/C9SM00476A}
  {\bibfield  {journal} {\bibinfo  {journal} {Soft Matter}\ }\textbf {\bibinfo
  {volume} {15}},\ \bibinfo {pages} {5695} (\bibinfo {year}
  {2019})}\BibitemShut {NoStop}%
\bibitem [{\citenamefont {Chan}\ \emph {et~al.}(1981)\citenamefont {Chan},
  \citenamefont {Henry~Jr},\ and\ \citenamefont {White}}]{chan1981interaction}%
  \BibitemOpen
  \bibfield  {author} {\bibinfo {author} {\bibfnamefont {D.}~\bibnamefont
  {Chan}}, \bibinfo {author} {\bibfnamefont {J.}~\bibnamefont {Henry~Jr}},\
  and\ \bibinfo {author} {\bibfnamefont {L.}~\bibnamefont {White}},\ }\bibfield
   {title} {\bibinfo {title} {The interaction of colloidal particles collected
  at fluid interfaces},\ }\href {https://doi.org/10.1016/0021-9797(81)90092-8}
  {\bibfield  {journal} {\bibinfo  {journal} {Journal of Colloid and Interface
  Science}\ }\textbf {\bibinfo {volume} {79}},\ \bibinfo {pages} {410}
  (\bibinfo {year} {1981})}\BibitemShut {NoStop}%
\bibitem [{\citenamefont {Vella}\ and\ \citenamefont
  {Mahadevan}(2005)}]{Vella2005cheerios}%
  \BibitemOpen
  \bibfield  {author} {\bibinfo {author} {\bibfnamefont {D.}~\bibnamefont
  {Vella}}\ and\ \bibinfo {author} {\bibfnamefont {L.}~\bibnamefont
  {Mahadevan}},\ }\bibfield  {title} {\bibinfo {title} {The cheerios effect},\
  }\href {https://doi.org/10.1119/1.1898523} {\bibfield  {journal} {\bibinfo
  {journal} {American journal of physics}\ }\textbf {\bibinfo {volume} {73}},\
  \bibinfo {pages} {817} (\bibinfo {year} {2005})}\BibitemShut {NoStop}%
\bibitem [{\citenamefont {Vassileva}\ \emph {et~al.}(2005)\citenamefont
  {Vassileva}, \citenamefont {van~den Ende}, \citenamefont {Mugele},\ and\
  \citenamefont {Mellema}}]{vassileva2005capillary}%
  \BibitemOpen
  \bibfield  {author} {\bibinfo {author} {\bibfnamefont {N.~D.}\ \bibnamefont
  {Vassileva}}, \bibinfo {author} {\bibfnamefont {D.}~\bibnamefont {van~den
  Ende}}, \bibinfo {author} {\bibfnamefont {F.}~\bibnamefont {Mugele}},\ and\
  \bibinfo {author} {\bibfnamefont {J.}~\bibnamefont {Mellema}},\ }\bibfield
  {title} {\bibinfo {title} {Capillary forces between spherical particles
  floating at a liquid- liquid interface},\ }\href
  {https://doi.org/10.1021/la051186o} {\bibfield  {journal} {\bibinfo
  {journal} {Langmuir}\ }\textbf {\bibinfo {volume} {21}},\ \bibinfo {pages}
  {11190} (\bibinfo {year} {2005})}\BibitemShut {NoStop}%
\bibitem [{\citenamefont {Dalbe}\ \emph {et~al.}(2011)\citenamefont {Dalbe},
  \citenamefont {Cosic}, \citenamefont {Berhanu},\ and\ \citenamefont
  {Kudrolli}}]{dalbe2011aggregation}%
  \BibitemOpen
  \bibfield  {author} {\bibinfo {author} {\bibfnamefont {M.-J.}\ \bibnamefont
  {Dalbe}}, \bibinfo {author} {\bibfnamefont {D.}~\bibnamefont {Cosic}},
  \bibinfo {author} {\bibfnamefont {M.}~\bibnamefont {Berhanu}},\ and\ \bibinfo
  {author} {\bibfnamefont {A.}~\bibnamefont {Kudrolli}},\ }\bibfield  {title}
  {\bibinfo {title} {Aggregation of frictional particles due to capillary
  attraction},\ }\href {https://doi.org/10.1103/PhysRevE.83.051403} {\bibfield
  {journal} {\bibinfo  {journal} {Physical Review E}\ }\textbf {\bibinfo
  {volume} {83}},\ \bibinfo {pages} {051403} (\bibinfo {year}
  {2011})}\BibitemShut {NoStop}%
\bibitem [{\citenamefont {Kralchevsky}\ and\ \citenamefont
  {Nagayama}(2000)}]{kralchevsky2000capillary}%
  \BibitemOpen
  \bibfield  {author} {\bibinfo {author} {\bibfnamefont {P.~A.}\ \bibnamefont
  {Kralchevsky}}\ and\ \bibinfo {author} {\bibfnamefont {K.}~\bibnamefont
  {Nagayama}},\ }\bibfield  {title} {\bibinfo {title} {Capillary interactions
  between particles bound to interfaces, liquid films and biomembranes},\
  }\href {https://doi.org/10.1016/S0001-8686(99)00016-0} {\bibfield  {journal}
  {\bibinfo  {journal} {Advances in colloid and interface science}\ }\textbf
  {\bibinfo {volume} {85}},\ \bibinfo {pages} {145} (\bibinfo {year}
  {2000})}\BibitemShut {NoStop}%
\bibitem [{\citenamefont {Stamou}\ \emph {et~al.}(2000)\citenamefont {Stamou},
  \citenamefont {Duschl},\ and\ \citenamefont {Johannsmann}}]{stamou2000long}%
  \BibitemOpen
  \bibfield  {author} {\bibinfo {author} {\bibfnamefont {D.}~\bibnamefont
  {Stamou}}, \bibinfo {author} {\bibfnamefont {C.}~\bibnamefont {Duschl}},\
  and\ \bibinfo {author} {\bibfnamefont {D.}~\bibnamefont {Johannsmann}},\
  }\bibfield  {title} {\bibinfo {title} {Long-range attraction between
  colloidal spheres at the air-water interface: The consequence of an irregular
  meniscus},\ }\href {https://doi.org/10.1103/PhysRevE.62.5263} {\bibfield
  {journal} {\bibinfo  {journal} {Physical Review E}\ }\textbf {\bibinfo
  {volume} {62}},\ \bibinfo {pages} {5263} (\bibinfo {year}
  {2000})}\BibitemShut {NoStop}%
\bibitem [{\citenamefont {Kralchevsky}\ \emph {et~al.}(2001)\citenamefont
  {Kralchevsky}, \citenamefont {Denkov},\ and\ \citenamefont
  {Danov}}]{kralchevsky2001particles}%
  \BibitemOpen
  \bibfield  {author} {\bibinfo {author} {\bibfnamefont {P.~A.}\ \bibnamefont
  {Kralchevsky}}, \bibinfo {author} {\bibfnamefont {N.~D.}\ \bibnamefont
  {Denkov}},\ and\ \bibinfo {author} {\bibfnamefont {K.~D.}\ \bibnamefont
  {Danov}},\ }\bibfield  {title} {\bibinfo {title} {Particles with an undulated
  contact line at a fluid interface: interaction between capillary quadrupoles
  and rheology of particulate monolayers},\ }\href
  {https://doi.org/10.1021/la0109359} {\bibfield  {journal} {\bibinfo
  {journal} {Langmuir}\ }\textbf {\bibinfo {volume} {17}},\ \bibinfo {pages}
  {7694} (\bibinfo {year} {2001})}\BibitemShut {NoStop}%
\bibitem [{\citenamefont {Fournier}\ and\ \citenamefont
  {Galatola}(2002)}]{fournier2002anisotropic}%
  \BibitemOpen
  \bibfield  {author} {\bibinfo {author} {\bibfnamefont {J.-B.}\ \bibnamefont
  {Fournier}}\ and\ \bibinfo {author} {\bibfnamefont {P.}~\bibnamefont
  {Galatola}},\ }\bibfield  {title} {\bibinfo {title} {Anisotropic capillary
  interactions and jamming of colloidal particles trapped at a liquid-fluid
  interface},\ }\href {https://doi.org/10.1103/PhysRevE.65.031601} {\bibfield
  {journal} {\bibinfo  {journal} {Physical Review E}\ }\textbf {\bibinfo
  {volume} {65}},\ \bibinfo {pages} {031601} (\bibinfo {year}
  {2002})}\BibitemShut {NoStop}%
\bibitem [{\citenamefont {Botto}\ \emph {et~al.}(2012)\citenamefont {Botto},
  \citenamefont {Lewandowski}, \citenamefont {Cavallaro},\ and\ \citenamefont
  {Stebe}}]{botto2012capillary}%
  \BibitemOpen
  \bibfield  {author} {\bibinfo {author} {\bibfnamefont {L.}~\bibnamefont
  {Botto}}, \bibinfo {author} {\bibfnamefont {E.~P.}\ \bibnamefont
  {Lewandowski}}, \bibinfo {author} {\bibfnamefont {M.}~\bibnamefont
  {Cavallaro}},\ and\ \bibinfo {author} {\bibfnamefont {K.~J.}\ \bibnamefont
  {Stebe}},\ }\bibfield  {title} {\bibinfo {title} {Capillary interactions
  between anisotropic particles},\ }\href {https://doi.org/10.1039/c2sm25929j}
  {\bibfield  {journal} {\bibinfo  {journal} {Soft Matter}\ }\textbf {\bibinfo
  {volume} {8}},\ \bibinfo {pages} {9957} (\bibinfo {year} {2012})}\BibitemShut
  {NoStop}%
\bibitem [{\citenamefont {Vella}\ \emph {et~al.}(2004)\citenamefont {Vella},
  \citenamefont {Aussillous},\ and\ \citenamefont {Mahadevan}}]{Vella2004}%
  \BibitemOpen
  \bibfield  {author} {\bibinfo {author} {\bibfnamefont {D.}~\bibnamefont
  {Vella}}, \bibinfo {author} {\bibfnamefont {P.}~\bibnamefont {Aussillous}},\
  and\ \bibinfo {author} {\bibfnamefont {L.}~\bibnamefont {Mahadevan}},\
  }\bibfield  {title} {\bibinfo {title} {Elasticity of an interfacial particle
  raft},\ }\href {https://doi.org/10.1209/epl/i2004-10202-x} {\bibfield
  {journal} {\bibinfo  {journal} {Europhys. Lett.}\ }\textbf {\bibinfo {volume}
  {68}},\ \bibinfo {pages} {212} (\bibinfo {year} {2004})}\BibitemShut
  {NoStop}%
\bibitem [{\citenamefont {Vella}\ \emph {et~al.}(2006)\citenamefont {Vella},
  \citenamefont {Kim}, \citenamefont {Aussillous},\ and\ \citenamefont
  {Mahadevan}}]{Vella2006}%
  \BibitemOpen
  \bibfield  {author} {\bibinfo {author} {\bibfnamefont {D.}~\bibnamefont
  {Vella}}, \bibinfo {author} {\bibfnamefont {H.-Y.}\ \bibnamefont {Kim}},
  \bibinfo {author} {\bibfnamefont {P.}~\bibnamefont {Aussillous}},\ and\
  \bibinfo {author} {\bibfnamefont {L.}~\bibnamefont {Mahadevan}},\ }\bibfield
  {title} {\bibinfo {title} {Dynamics of surfactant-driven fracture of particle
  rafts},\ }\href {https://doi.org/10.1103/PhysRevLett.96.178301} {\bibfield
  {journal} {\bibinfo  {journal} {Phys. Rev. Lett.}\ }\textbf {\bibinfo
  {volume} {96}},\ \bibinfo {pages} {178301} (\bibinfo {year}
  {2006})}\BibitemShut {NoStop}%
\bibitem [{\citenamefont {Bandi}\ \emph {et~al.}(2011)\citenamefont {Bandi},
  \citenamefont {Tallinen},\ and\ \citenamefont {Mahadevan}}]{bandi2011shock}%
  \BibitemOpen
  \bibfield  {author} {\bibinfo {author} {\bibfnamefont {M.}~\bibnamefont
  {Bandi}}, \bibinfo {author} {\bibfnamefont {T.}~\bibnamefont {Tallinen}},\
  and\ \bibinfo {author} {\bibfnamefont {L.}~\bibnamefont {Mahadevan}},\
  }\bibfield  {title} {\bibinfo {title} {Shock-driven jamming and periodic
  fracture of particulate rafts},\ }\href
  {https://doi.org/10.1209/0295-5075/96/36008} {\bibfield  {journal} {\bibinfo
  {journal} {Europhysics Letters}\ }\textbf {\bibinfo {volume} {96}},\ \bibinfo
  {pages} {36008} (\bibinfo {year} {2011})}\BibitemShut {NoStop}%
\bibitem [{\citenamefont {Peco}\ \emph {et~al.}(2017)\citenamefont {Peco},
  \citenamefont {Chen}, \citenamefont {Liu}, \citenamefont {Bandi},
  \citenamefont {Dolbow},\ and\ \citenamefont {Fried}}]{peco2017influence}%
  \BibitemOpen
  \bibfield  {author} {\bibinfo {author} {\bibfnamefont {C.}~\bibnamefont
  {Peco}}, \bibinfo {author} {\bibfnamefont {W.}~\bibnamefont {Chen}}, \bibinfo
  {author} {\bibfnamefont {Y.}~\bibnamefont {Liu}}, \bibinfo {author}
  {\bibfnamefont {M.}~\bibnamefont {Bandi}}, \bibinfo {author} {\bibfnamefont
  {J.~E.}\ \bibnamefont {Dolbow}},\ and\ \bibinfo {author} {\bibfnamefont
  {E.}~\bibnamefont {Fried}},\ }\bibfield  {title} {\bibinfo {title} {Influence
  of surface tension in the surfactant-driven fracture of closely-packed
  particulate monolayers},\ }\href {https://doi.org/10.1039/C7SM01245D}
  {\bibfield  {journal} {\bibinfo  {journal} {Soft matter}\ }\textbf {\bibinfo
  {volume} {13}},\ \bibinfo {pages} {5832} (\bibinfo {year}
  {2017})}\BibitemShut {NoStop}%
\bibitem [{\citenamefont {Vassileva}\ \emph {et~al.}(2006)\citenamefont
  {Vassileva}, \citenamefont {van~den Ende}, \citenamefont {Mugele},\ and\
  \citenamefont {Mellema}}]{vassileva2006restructuring}%
  \BibitemOpen
  \bibfield  {author} {\bibinfo {author} {\bibfnamefont {N.~D.}\ \bibnamefont
  {Vassileva}}, \bibinfo {author} {\bibfnamefont {D.}~\bibnamefont {van~den
  Ende}}, \bibinfo {author} {\bibfnamefont {F.}~\bibnamefont {Mugele}},\ and\
  \bibinfo {author} {\bibfnamefont {J.}~\bibnamefont {Mellema}},\ }\bibfield
  {title} {\bibinfo {title} {Restructuring and break-up of two-dimensional
  aggregates in shear flow},\ }\href {https://doi.org/10.1021/la053460k}
  {\bibfield  {journal} {\bibinfo  {journal} {Langmuir}\ }\textbf {\bibinfo
  {volume} {22}},\ \bibinfo {pages} {4959} (\bibinfo {year}
  {2006})}\BibitemShut {NoStop}%
\bibitem [{\citenamefont {Vassileva}\ \emph {et~al.}(2007)\citenamefont
  {Vassileva}, \citenamefont {van~den Ende}, \citenamefont {Mugele},\ and\
  \citenamefont {Mellema}}]{vassileva2007fragmentation}%
  \BibitemOpen
  \bibfield  {author} {\bibinfo {author} {\bibfnamefont {N.~D.}\ \bibnamefont
  {Vassileva}}, \bibinfo {author} {\bibfnamefont {D.}~\bibnamefont {van~den
  Ende}}, \bibinfo {author} {\bibfnamefont {F.}~\bibnamefont {Mugele}},\ and\
  \bibinfo {author} {\bibfnamefont {J.}~\bibnamefont {Mellema}},\ }\bibfield
  {title} {\bibinfo {title} {Fragmentation and erosion of two-dimensional
  aggregates in shear flow},\ }\href {https://doi.org/10.1021/la0625087}
  {\bibfield  {journal} {\bibinfo  {journal} {Langmuir}\ }\textbf {\bibinfo
  {volume} {23}},\ \bibinfo {pages} {2352} (\bibinfo {year}
  {2007})}\BibitemShut {NoStop}%
\bibitem [{\citenamefont {Huang}\ \emph {et~al.}(2012)\citenamefont {Huang},
  \citenamefont {Brinkmann},\ and\ \citenamefont {Herminghaus}}]{huang2012wet}%
  \BibitemOpen
  \bibfield  {author} {\bibinfo {author} {\bibfnamefont {K.}~\bibnamefont
  {Huang}}, \bibinfo {author} {\bibfnamefont {M.}~\bibnamefont {Brinkmann}},\
  and\ \bibinfo {author} {\bibfnamefont {S.}~\bibnamefont {Herminghaus}},\
  }\bibfield  {title} {\bibinfo {title} {Wet granular rafts: aggregation in two
  dimensions under shear flow},\ }\href {https://doi.org/10.1039/C2SM26074C}
  {\bibfield  {journal} {\bibinfo  {journal} {Soft Matter}\ }\textbf {\bibinfo
  {volume} {8}},\ \bibinfo {pages} {11939} (\bibinfo {year}
  {2012})}\BibitemShut {NoStop}%
\bibitem [{\citenamefont {Kim}\ \emph {et~al.}(2019)\citenamefont {Kim},
  \citenamefont {Rendos}, \citenamefont {Ganesh},\ and\ \citenamefont
  {Brown}}]{kim2019failure}%
  \BibitemOpen
  \bibfield  {author} {\bibinfo {author} {\bibfnamefont {B.~L.}\ \bibnamefont
  {Kim}}, \bibinfo {author} {\bibfnamefont {A.}~\bibnamefont {Rendos}},
  \bibinfo {author} {\bibfnamefont {P.}~\bibnamefont {Ganesh}},\ and\ \bibinfo
  {author} {\bibfnamefont {K.~A.}\ \bibnamefont {Brown}},\ }\bibfield  {title}
  {\bibinfo {title} {Failure of particle-laden interfaces studied using the
  funnel method},\ }\href {https://doi.org/10.1016/j.colcom.2018.11.008}
  {\bibfield  {journal} {\bibinfo  {journal} {Colloid and Interface Science
  Communications}\ }\textbf {\bibinfo {volume} {28}},\ \bibinfo {pages} {54}
  (\bibinfo {year} {2019})}\BibitemShut {NoStop}%
\bibitem [{\citenamefont {Xiao}\ \emph {et~al.}(2020)\citenamefont {Xiao},
  \citenamefont {Ivancic},\ and\ \citenamefont {Durian}}]{xiao2020strain}%
  \BibitemOpen
  \bibfield  {author} {\bibinfo {author} {\bibfnamefont {H.}~\bibnamefont
  {Xiao}}, \bibinfo {author} {\bibfnamefont {R.~J.}\ \bibnamefont {Ivancic}},\
  and\ \bibinfo {author} {\bibfnamefont {D.~J.}\ \bibnamefont {Durian}},\
  }\bibfield  {title} {\bibinfo {title} {Strain localization and failure of
  disordered particle rafts with tunable ductility during tensile
  deformation},\ }\href {https://doi.org/10.1039/D0SM00839G} {\bibfield
  {journal} {\bibinfo  {journal} {Soft Matter}\ }\textbf {\bibinfo {volume}
  {16}},\ \bibinfo {pages} {8226} (\bibinfo {year} {2020})}\BibitemShut
  {NoStop}%
\bibitem [{\citenamefont {To}\ and\ \citenamefont {Nagel}(2023)}]{to2023rifts}%
  \BibitemOpen
  \bibfield  {author} {\bibinfo {author} {\bibfnamefont {K.-I.}\ \bibnamefont
  {To}}\ and\ \bibinfo {author} {\bibfnamefont {S.~R.}\ \bibnamefont {Nagel}},\
  }\bibfield  {title} {\bibinfo {title} {Rifts in rafts},\ }\href
  {https://doi.org/10.1039/D2SM01451C} {\bibfield  {journal} {\bibinfo
  {journal} {Soft Matter}\ }\textbf {\bibinfo {volume} {19}},\ \bibinfo {pages}
  {905} (\bibinfo {year} {2023})}\BibitemShut {NoStop}%
\bibitem [{\citenamefont {Lagarde}\ and\ \citenamefont
  {Proti{\`e}re}(2020)}]{lagarde2020probing}%
  \BibitemOpen
  \bibfield  {author} {\bibinfo {author} {\bibfnamefont {A.}~\bibnamefont
  {Lagarde}}\ and\ \bibinfo {author} {\bibfnamefont {S.}~\bibnamefont
  {Proti{\`e}re}},\ }\bibfield  {title} {\bibinfo {title} {Probing the erosion
  and cohesion of a granular raft in motion},\ }\href
  {https://doi.org/10.1103/PhysRevFluids.5.044003} {\bibfield  {journal}
  {\bibinfo  {journal} {Physical Review Fluids}\ }\textbf {\bibinfo {volume}
  {5}},\ \bibinfo {pages} {044003} (\bibinfo {year} {2020})}\BibitemShut
  {NoStop}%
\bibitem [{\citenamefont {Squire}\ \emph {et~al.}(1995)\citenamefont {Squire},
  \citenamefont {Dugan}, \citenamefont {Wadhams}, \citenamefont {Rottier},\
  and\ \citenamefont {Liu}}]{squire1995ocean}%
  \BibitemOpen
  \bibfield  {author} {\bibinfo {author} {\bibfnamefont {V.~A.}\ \bibnamefont
  {Squire}}, \bibinfo {author} {\bibfnamefont {J.~P.}\ \bibnamefont {Dugan}},
  \bibinfo {author} {\bibfnamefont {P.}~\bibnamefont {Wadhams}}, \bibinfo
  {author} {\bibfnamefont {P.~J.}\ \bibnamefont {Rottier}},\ and\ \bibinfo
  {author} {\bibfnamefont {A.~K.}\ \bibnamefont {Liu}},\ }\bibfield  {title}
  {\bibinfo {title} {Of ocean waves and sea ice},\ }\href
  {https://doi.org/10.1146/annurev.fl.27.010195.000555} {\bibfield  {journal}
  {\bibinfo  {journal} {Annual Review of Fluid Mechanics}\ }\textbf {\bibinfo
  {volume} {27}},\ \bibinfo {pages} {115} (\bibinfo {year} {1995})}\BibitemShut
  {NoStop}%
\bibitem [{\citenamefont {Dumont}\ \emph {et~al.}(2011)\citenamefont {Dumont},
  \citenamefont {Kohout},\ and\ \citenamefont {Bertino}}]{dumont2011wave}%
  \BibitemOpen
  \bibfield  {author} {\bibinfo {author} {\bibfnamefont {D.}~\bibnamefont
  {Dumont}}, \bibinfo {author} {\bibfnamefont {A.}~\bibnamefont {Kohout}},\
  and\ \bibinfo {author} {\bibfnamefont {L.}~\bibnamefont {Bertino}},\
  }\bibfield  {title} {\bibinfo {title} {A wave-based model for the marginal
  ice zone including a floe breaking parameterization},\ }\bibfield  {journal}
  {\bibinfo  {journal} {Journal of Geophysical Research: Oceans}\ }\textbf
  {\bibinfo {volume} {116}},\ \href {https://doi.org/10.1029/2010JC006682}
  {10.1029/2010JC006682} (\bibinfo {year} {2011})\BibitemShut {NoStop}%
\bibitem [{\citenamefont {Feltham}(2008)}]{feltham2008sea}%
  \BibitemOpen
  \bibfield  {author} {\bibinfo {author} {\bibfnamefont {D.~L.}\ \bibnamefont
  {Feltham}},\ }\bibfield  {title} {\bibinfo {title} {Sea ice rheology},\
  }\href {https://doi.org/10.1146/annurev.fluid.40.111406.102151} {\bibfield
  {journal} {\bibinfo  {journal} {Annu. Rev. Fluid Mech.}\ }\textbf {\bibinfo
  {volume} {40}},\ \bibinfo {pages} {91} (\bibinfo {year} {2008})}\BibitemShut
  {NoStop}%
\bibitem [{\citenamefont {Herman}(2022)}]{herman2022granular}%
  \BibitemOpen
  \bibfield  {author} {\bibinfo {author} {\bibfnamefont {A.}~\bibnamefont
  {Herman}},\ }\bibfield  {title} {\bibinfo {title} {Granular effects in sea
  ice rheology in the marginal ice zone},\ }\href
  {https://doi.org/10.1098/rsta.2021.0260} {\bibfield  {journal} {\bibinfo
  {journal} {Philosophical Transactions of the Royal Society A}\ }\textbf
  {\bibinfo {volume} {380}},\ \bibinfo {pages} {20210260} (\bibinfo {year}
  {2022})}\BibitemShut {NoStop}%
\bibitem [{\citenamefont {Squire}(2020)}]{squire2020ocean}%
  \BibitemOpen
  \bibfield  {author} {\bibinfo {author} {\bibfnamefont {V.~A.}\ \bibnamefont
  {Squire}},\ }\bibfield  {title} {\bibinfo {title} {Ocean wave interactions
  with sea ice: a reappraisal},\ }\href
  {https://doi.org/10.1146/annurev.fl.27.010195.000555} {\bibfield  {journal}
  {\bibinfo  {journal} {Annual Review of Fluid Mechanics}\ }\textbf {\bibinfo
  {volume} {52}},\ \bibinfo {pages} {37} (\bibinfo {year} {2020})}\BibitemShut
  {NoStop}%
\bibitem [{\citenamefont {Horvat}(2022)}]{Horvat2022}%
  \BibitemOpen
  \bibfield  {author} {\bibinfo {author} {\bibfnamefont {C.}~\bibnamefont
  {Horvat}},\ }\bibfield  {title} {\bibinfo {title} {Floes, the marginal ice
  zone and coupled wave-sea-ice feedbacks},\ }\href
  {https://doi.org/10.1098/rsta.2021.0252} {\bibfield  {journal} {\bibinfo
  {journal} {Philosophical Transactions of the Royal Society A}\ }\textbf
  {\bibinfo {volume} {380}},\ \bibinfo {pages} {20210252} (\bibinfo {year}
  {2022})}\BibitemShut {NoStop}%
\bibitem [{\citenamefont {Brenner}\ \emph {et~al.}(2023)\citenamefont
  {Brenner}, \citenamefont {Horvat}, \citenamefont {Hall}, \citenamefont
  {Lo~Piccolo}, \citenamefont {Fox-Kemper}, \citenamefont {Labbé},\ and\
  \citenamefont {Dansereau}}]{Brenner2023}%
  \BibitemOpen
  \bibfield  {author} {\bibinfo {author} {\bibfnamefont {S.}~\bibnamefont
  {Brenner}}, \bibinfo {author} {\bibfnamefont {C.}~\bibnamefont {Horvat}},
  \bibinfo {author} {\bibfnamefont {P.}~\bibnamefont {Hall}}, \bibinfo {author}
  {\bibfnamefont {A.}~\bibnamefont {Lo~Piccolo}}, \bibinfo {author}
  {\bibfnamefont {B.}~\bibnamefont {Fox-Kemper}}, \bibinfo {author}
  {\bibfnamefont {S.}~\bibnamefont {Labbé}},\ and\ \bibinfo {author}
  {\bibfnamefont {V.}~\bibnamefont {Dansereau}},\ }\bibfield  {title} {\bibinfo
  {title} {Scale-dependent air-sea exchange in the polar oceans: Floe-floe and
  floe-flow coupling in the generation of ice-ocean boundary layer
  turbulence},\ }\href {https://doi.org/https://doi.org/10.1029/2023GL105703}
  {\bibfield  {journal} {\bibinfo  {journal} {Geophysical Research Letters}\
  }\textbf {\bibinfo {volume} {50}},\ \bibinfo {pages} {e2023GL105703}
  (\bibinfo {year} {2023})},\ \bibinfo {note} {e2023GL105703 2023GL105703},\
  \Eprint
  {https://arxiv.org/abs/https://agupubs.onlinelibrary.wiley.com/doi/pdf/10.1029/2023GL105703}
  {https://agupubs.onlinelibrary.wiley.com/doi/pdf/10.1029/2023GL105703}
  \BibitemShut {NoStop}%
\bibitem [{\citenamefont {Herman}(2010)}]{Herman2010}%
  \BibitemOpen
  \bibfield  {author} {\bibinfo {author} {\bibfnamefont {A.}~\bibnamefont
  {Herman}},\ }\bibfield  {title} {\bibinfo {title} {Sea-ice floe-size
  distribution in the context of spontaneous scaling emergence in stochastic
  systems},\ }\href {https://doi.org/10.1103/PhysRevE.81.066123} {\bibfield
  {journal} {\bibinfo  {journal} {Physical Review E}\ }\textbf {\bibinfo
  {volume} {81}},\ \bibinfo {pages} {066123} (\bibinfo {year}
  {2010})}\BibitemShut {NoStop}%
\bibitem [{\citenamefont {Gherardi}\ and\ \citenamefont
  {Lagomarsino}(2015)}]{Gherardi2015}%
  \BibitemOpen
  \bibfield  {author} {\bibinfo {author} {\bibfnamefont {M.}~\bibnamefont
  {Gherardi}}\ and\ \bibinfo {author} {\bibfnamefont {M.~C.}\ \bibnamefont
  {Lagomarsino}},\ }\bibfield  {title} {\bibinfo {title} {Characterizing the
  size and shape of sea ice floes},\ }\href {https://doi.org/10.1038/srep10226}
  {\bibfield  {journal} {\bibinfo  {journal} {Scientific reports}\ }\textbf
  {\bibinfo {volume} {5}},\ \bibinfo {pages} {1} (\bibinfo {year}
  {2015})}\BibitemShut {NoStop}%
\bibitem [{\citenamefont {Horvat}\ and\ \citenamefont
  {Tziperman}(2017)}]{Horvat2017}%
  \BibitemOpen
  \bibfield  {author} {\bibinfo {author} {\bibfnamefont {C.}~\bibnamefont
  {Horvat}}\ and\ \bibinfo {author} {\bibfnamefont {E.}~\bibnamefont
  {Tziperman}},\ }\bibfield  {title} {\bibinfo {title} {The evolution of
  scaling laws in the sea ice floe size distribution},\ }\href
  {https://doi.org/10.1002/2016JC012573} {\bibfield  {journal} {\bibinfo
  {journal} {Journal of Geophysical Research: Oceans}\ }\textbf {\bibinfo
  {volume} {122}},\ \bibinfo {pages} {7630} (\bibinfo {year}
  {2017})}\BibitemShut {NoStop}%
\bibitem [{\citenamefont {Montiel}\ and\ \citenamefont
  {Mokus}(2022)}]{montiel2022theoretical}%
  \BibitemOpen
  \bibfield  {author} {\bibinfo {author} {\bibfnamefont {F.}~\bibnamefont
  {Montiel}}\ and\ \bibinfo {author} {\bibfnamefont {N.}~\bibnamefont
  {Mokus}},\ }\bibfield  {title} {\bibinfo {title} {Theoretical framework for
  the emergent floe size distribution in the marginal ice zone: the case for
  log-normality},\ }\href {https://doi.org/10.1098/rsta.2021.0257} {\bibfield
  {journal} {\bibinfo  {journal} {Philosophical Transactions of the Royal
  Society A}\ }\textbf {\bibinfo {volume} {380}},\ \bibinfo {pages} {20210257}
  (\bibinfo {year} {2022})}\BibitemShut {NoStop}%
\bibitem [{\citenamefont {Dumas-Lefebvre}\ and\ \citenamefont
  {Dumont}(2021{\natexlab{a}})}]{Dumas2021}%
  \BibitemOpen
  \bibfield  {author} {\bibinfo {author} {\bibfnamefont {E.}~\bibnamefont
  {Dumas-Lefebvre}}\ and\ \bibinfo {author} {\bibfnamefont {D.}~\bibnamefont
  {Dumont}},\ }\bibfield  {title} {\bibinfo {title} {Aerial observations of sea
  ice break-up by ship waves},\ }\href {https://doi.org/10.5194/tc-17-827-2023}
  {\bibfield  {journal} {\bibinfo  {journal} {The Cryosphere Discussions}\ ,\
  \bibinfo {pages} {1}} (\bibinfo {year} {2021}{\natexlab{a}})}\BibitemShut
  {NoStop}%
\bibitem [{\citenamefont {Herman}\ \emph {et~al.}(2018)\citenamefont {Herman},
  \citenamefont {Evers},\ and\ \citenamefont {Reimer}}]{Herman2018}%
  \BibitemOpen
  \bibfield  {author} {\bibinfo {author} {\bibfnamefont {A.}~\bibnamefont
  {Herman}}, \bibinfo {author} {\bibfnamefont {K.-U.}\ \bibnamefont {Evers}},\
  and\ \bibinfo {author} {\bibfnamefont {N.}~\bibnamefont {Reimer}},\
  }\bibfield  {title} {\bibinfo {title} {Floe-size distributions in laboratory
  ice broken by waves},\ }\href {https://doi.org/10.5194/tc-12-685-2018}
  {\bibfield  {journal} {\bibinfo  {journal} {The Cryosphere}\ }\textbf
  {\bibinfo {volume} {12}},\ \bibinfo {pages} {685} (\bibinfo {year}
  {2018})}\BibitemShut {NoStop}%
\bibitem [{\citenamefont {Garcia}\ \emph {et~al.}(2008)\citenamefont {Garcia},
  \citenamefont {Osborne},\ and\ \citenamefont {Subashi}}]{Garcia2008}%
  \BibitemOpen
  \bibfield  {author} {\bibinfo {author} {\bibfnamefont {R.}~\bibnamefont
  {Garcia}}, \bibinfo {author} {\bibfnamefont {K.}~\bibnamefont {Osborne}},\
  and\ \bibinfo {author} {\bibfnamefont {E.}~\bibnamefont {Subashi}},\
  }\bibfield  {title} {\bibinfo {title} {Validity of the "sharp-kink
  approximation" for water and other fluids},\ }\href
  {https://doi.org/10.1021/jp712181m} {\bibfield  {journal} {\bibinfo
  {journal} {J. Phys. Chem. B}\ }\textbf {\bibinfo {volume} {112}},\ \bibinfo
  {pages} {8114} (\bibinfo {year} {2008})}\BibitemShut {NoStop}%
\bibitem [{\citenamefont {Kozbial}\ \emph {et~al.}(2017)\citenamefont
  {Kozbial}, \citenamefont {Trouba}, \citenamefont {Liu},\ and\ \citenamefont
  {Li}}]{Kozbial2018}%
  \BibitemOpen
  \bibfield  {author} {\bibinfo {author} {\bibfnamefont {A.}~\bibnamefont
  {Kozbial}}, \bibinfo {author} {\bibfnamefont {C.}~\bibnamefont {Trouba}},
  \bibinfo {author} {\bibfnamefont {H.}~\bibnamefont {Liu}},\ and\ \bibinfo
  {author} {\bibfnamefont {L.}~\bibnamefont {Li}},\ }\bibfield  {title}
  {\bibinfo {title} {Characterization of the intrinsic water wettability of
  graphite using contact angle measurements: Effect of defects on static and
  dynamic contact angles},\ }\href
  {https://doi.org/10.1021/acs.langmuir.6b04193} {\bibfield  {journal}
  {\bibinfo  {journal} {Langmuir}\ }\textbf {\bibinfo {volume} {33}},\ \bibinfo
  {pages} {959} (\bibinfo {year} {2017})}\BibitemShut {NoStop}%
\bibitem [{\citenamefont {Lehle}\ \emph {et~al.}(2008)\citenamefont {Lehle},
  \citenamefont {Noruzifar},\ and\ \citenamefont
  {Oettel}}]{lehle2008ellipsoidal}%
  \BibitemOpen
  \bibfield  {author} {\bibinfo {author} {\bibfnamefont {H.}~\bibnamefont
  {Lehle}}, \bibinfo {author} {\bibfnamefont {E.}~\bibnamefont {Noruzifar}},\
  and\ \bibinfo {author} {\bibfnamefont {M.}~\bibnamefont {Oettel}},\
  }\bibfield  {title} {\bibinfo {title} {Ellipsoidal particles at fluid
  interfaces},\ }\href {https://doi.org/10.1140/epje/i2007-10314-1} {\bibfield
  {journal} {\bibinfo  {journal} {The European Physical Journal E}\ }\textbf
  {\bibinfo {volume} {26}},\ \bibinfo {pages} {151} (\bibinfo {year}
  {2008})}\BibitemShut {NoStop}%
\bibitem [{\citenamefont {Villermaux}(2007)}]{villermaux2007fragmentation}%
  \BibitemOpen
  \bibfield  {author} {\bibinfo {author} {\bibfnamefont {E.}~\bibnamefont
  {Villermaux}},\ }\bibfield  {title} {\bibinfo {title} {Fragmentation},\
  }\href {https://doi.org/10.1146/annurev.fluid.39.050905.110214} {\bibfield
  {journal} {\bibinfo  {journal} {Annu. Rev. Fluid Mech.}\ }\textbf {\bibinfo
  {volume} {39}},\ \bibinfo {pages} {419} (\bibinfo {year} {2007})}\BibitemShut
  {NoStop}%
\bibitem [{\citenamefont {Schwarz}\ and\ \citenamefont
  {Weeks}(1977)}]{Schwarz1977}%
  \BibitemOpen
  \bibfield  {author} {\bibinfo {author} {\bibfnamefont {J.}~\bibnamefont
  {Schwarz}}\ and\ \bibinfo {author} {\bibfnamefont {W.}~\bibnamefont
  {Weeks}},\ }\bibfield  {title} {\bibinfo {title} {Engineering properties of
  sea ice},\ }\href {https://doi.org/10.3189/S0022143000029476} {\bibfield
  {journal} {\bibinfo  {journal} {Journal of Glaciology}\ }\textbf {\bibinfo
  {volume} {19}},\ \bibinfo {pages} {499} (\bibinfo {year} {1977})}\BibitemShut
  {NoStop}%
\bibitem [{\citenamefont {Mellor}(1986)}]{Mellor1986}%
  \BibitemOpen
  \bibfield  {author} {\bibinfo {author} {\bibfnamefont {M.}~\bibnamefont
  {Mellor}},\ }\href@noop {} {\emph {\bibinfo {title} {Mechanical behavior of
  sea ice}}}\ (\bibinfo  {publisher} {Springer},\ \bibinfo {year}
  {1986})\BibitemShut {NoStop}%
\bibitem [{\citenamefont {Timco}\ and\ \citenamefont
  {Weeks}(2010)}]{Timco2010}%
  \BibitemOpen
  \bibfield  {author} {\bibinfo {author} {\bibfnamefont {G.}~\bibnamefont
  {Timco}}\ and\ \bibinfo {author} {\bibfnamefont {W.}~\bibnamefont {Weeks}},\
  }\bibfield  {title} {\bibinfo {title} {A review of the engineering properties
  of sea ice},\ }\href {https://doi.org/10.1016/j.coldregions.2009.10.003}
  {\bibfield  {journal} {\bibinfo  {journal} {Cold regions science and
  technology}\ }\textbf {\bibinfo {volume} {60}},\ \bibinfo {pages} {107}
  (\bibinfo {year} {2010})}\BibitemShut {NoStop}%
\bibitem [{\citenamefont {Heorton}\ \emph {et~al.}(2018)\citenamefont
  {Heorton}, \citenamefont {Feltham},\ and\ \citenamefont
  {Tsamados}}]{Heorton2018}%
  \BibitemOpen
  \bibfield  {author} {\bibinfo {author} {\bibfnamefont {H.}~\bibnamefont
  {Heorton}}, \bibinfo {author} {\bibfnamefont {D.}~\bibnamefont {Feltham}},\
  and\ \bibinfo {author} {\bibfnamefont {M.}~\bibnamefont {Tsamados}},\
  }\bibfield  {title} {\bibinfo {title} {Stress and deformation characteristics
  of sea ice in a high resolution, anisotropic sea ice model},\ }\href
  {https://doi.org/10.1098/rsta.2017.0349} {\bibfield  {journal} {\bibinfo
  {journal} {Philosophical Transactions A.}\ }\textbf {\bibinfo {volume}
  {376}},\ \bibinfo {pages} {20170349} (\bibinfo {year} {2018})}\BibitemShut
  {NoStop}%
\bibitem [{\citenamefont {McGoldrick}(1971)}]{mcgoldrick1971sensitive}%
  \BibitemOpen
  \bibfield  {author} {\bibinfo {author} {\bibfnamefont {L.}~\bibnamefont
  {McGoldrick}},\ }\bibfield  {title} {\bibinfo {title} {A sensitive linear
  capacitance-to-voltage converter, with applications to surface wave
  measurements},\ }\href {https://doi.org/10.1063/1.1685094} {\bibfield
  {journal} {\bibinfo  {journal} {Review of Scientific Instruments}\ }\textbf
  {\bibinfo {volume} {42}},\ \bibinfo {pages} {359} (\bibinfo {year}
  {1971})}\BibitemShut {NoStop}%
\bibitem [{\citenamefont {Kim}\ \emph {et~al.}(2016)\citenamefont {Kim},
  \citenamefont {Moon},\ and\ \citenamefont {Kim}}]{kim2016dynamics}%
  \BibitemOpen
  \bibfield  {author} {\bibinfo {author} {\bibfnamefont {J.}~\bibnamefont
  {Kim}}, \bibinfo {author} {\bibfnamefont {M.-W.}\ \bibnamefont {Moon}},\ and\
  \bibinfo {author} {\bibfnamefont {H.-Y.}\ \bibnamefont {Kim}},\ }\bibfield
  {title} {\bibinfo {title} {Dynamics of hemiwicking},\ }\href
  {https://doi.org/10.1017/jfm.2016.386} {\bibfield  {journal} {\bibinfo
  {journal} {Journal of Fluid Mechanics}\ }\textbf {\bibinfo {volume} {800}},\
  \bibinfo {pages} {57} (\bibinfo {year} {2016})}\BibitemShut {NoStop}%
\bibitem [{\citenamefont {Monsalve}\ \emph {et~al.}(2022)\citenamefont
  {Monsalve}, \citenamefont {Maurel}, \citenamefont {Pagneux},\ and\
  \citenamefont {Petitjeans}}]{monsalve2022space}%
  \BibitemOpen
  \bibfield  {author} {\bibinfo {author} {\bibfnamefont {E.}~\bibnamefont
  {Monsalve}}, \bibinfo {author} {\bibfnamefont {A.}~\bibnamefont {Maurel}},
  \bibinfo {author} {\bibfnamefont {V.}~\bibnamefont {Pagneux}},\ and\ \bibinfo
  {author} {\bibfnamefont {P.}~\bibnamefont {Petitjeans}},\ }\bibfield  {title}
  {\bibinfo {title} {Space-time-resolved measurements of the effect of pinned
  contact line on the dispersion relation of water waves},\ }\href
  {https://doi.org/10.1103/PhysRevFluids.7.014802} {\bibfield  {journal}
  {\bibinfo  {journal} {Physical Review Fluids}\ }\textbf {\bibinfo {volume}
  {7}},\ \bibinfo {pages} {014802} (\bibinfo {year} {2022})}\BibitemShut
  {NoStop}%
\bibitem [{\citenamefont {van~den Bremer}\ and\ \citenamefont
  {Breivik}(2018)}]{vandenBremer2018}%
  \BibitemOpen
  \bibfield  {author} {\bibinfo {author} {\bibfnamefont {T.~S.}\ \bibnamefont
  {van~den Bremer}}\ and\ \bibinfo {author} {\bibfnamefont {{\O}.}~\bibnamefont
  {Breivik}},\ }\bibfield  {title} {\bibinfo {title} {Stokes drift},\ }\href
  {https://doi.org/10.1098/rsta.2017.0104} {\bibfield  {journal} {\bibinfo
  {journal} {Philosophical Transactions of the Royal Society A: Mathematical,
  Physical and Engineering Sciences}\ }\textbf {\bibinfo {volume} {376}},\
  \bibinfo {pages} {20170104} (\bibinfo {year} {2018})}\BibitemShut {NoStop}%
\bibitem [{\citenamefont {von Kameke}\ \emph {et~al.}(2011)\citenamefont {von
  Kameke}, \citenamefont {Huhn}, \citenamefont {Fernandez-Garcia},
  \citenamefont {Munuzuri},\ and\ \citenamefont
  {Perez-Munuzuri}}]{vonKameke2011}%
  \BibitemOpen
  \bibfield  {author} {\bibinfo {author} {\bibfnamefont {A.}~\bibnamefont {von
  Kameke}}, \bibinfo {author} {\bibfnamefont {F.}~\bibnamefont {Huhn}},
  \bibinfo {author} {\bibfnamefont {G.}~\bibnamefont {Fernandez-Garcia}},
  \bibinfo {author} {\bibfnamefont {A.~P.}\ \bibnamefont {Munuzuri}},\ and\
  \bibinfo {author} {\bibfnamefont {V.}~\bibnamefont {Perez-Munuzuri}},\
  }\bibfield  {title} {\bibinfo {title} {Double cascade turbulence and
  richardson dispersion in a horizontal fluid flow induced by faraday waves},\
  }\href {https://doi.org/10.1103/PhysRevLett.107.074502} {\bibfield  {journal}
  {\bibinfo  {journal} {Phys. Rev. Lett.}\ }\textbf {\bibinfo {volume} {107}},\
  \bibinfo {pages} {074502} (\bibinfo {year} {2011})}\BibitemShut {NoStop}%
\bibitem [{\citenamefont {Francois}\ \emph {et~al.}(2013)\citenamefont
  {Francois}, \citenamefont {Xia}, \citenamefont {Punzmann},\ and\
  \citenamefont {Shats}}]{Francois2013}%
  \BibitemOpen
  \bibfield  {author} {\bibinfo {author} {\bibfnamefont {N.}~\bibnamefont
  {Francois}}, \bibinfo {author} {\bibfnamefont {H.}~\bibnamefont {Xia}},
  \bibinfo {author} {\bibfnamefont {H.}~\bibnamefont {Punzmann}},\ and\
  \bibinfo {author} {\bibfnamefont {M.}~\bibnamefont {Shats}},\ }\bibfield
  {title} {\bibinfo {title} {Inverse energy cascade and emergence of large
  coherent vortices in turbulence driven by faraday waves},\ }\href
  {https://doi.org/10.1103/PhysRevLett.110.194501} {\bibfield  {journal}
  {\bibinfo  {journal} {Phys. Rev. Lett.}\ }\textbf {\bibinfo {volume} {110}},\
  \bibinfo {pages} {194501} (\bibinfo {year} {2013})}\BibitemShut {NoStop}%
\bibitem [{\citenamefont {P{\'e}rinet}\ \emph {et~al.}(2017)\citenamefont
  {P{\'e}rinet}, \citenamefont {Guti{\'e}rrez}, \citenamefont {Urra},
  \citenamefont {Mujica},\ and\ \citenamefont
  {Gordillo}}]{perinet2017streaming}%
  \BibitemOpen
  \bibfield  {author} {\bibinfo {author} {\bibfnamefont {N.}~\bibnamefont
  {P{\'e}rinet}}, \bibinfo {author} {\bibfnamefont {P.}~\bibnamefont
  {Guti{\'e}rrez}}, \bibinfo {author} {\bibfnamefont {H.}~\bibnamefont {Urra}},
  \bibinfo {author} {\bibfnamefont {N.}~\bibnamefont {Mujica}},\ and\ \bibinfo
  {author} {\bibfnamefont {L.}~\bibnamefont {Gordillo}},\ }\bibfield  {title}
  {\bibinfo {title} {Streaming patterns in faraday waves},\ }\href
  {https://doi.org/10.1017/jfm.2017.166} {\bibfield  {journal} {\bibinfo
  {journal} {Journal of Fluid Mechanics}\ }\textbf {\bibinfo {volume} {819}},\
  \bibinfo {pages} {285} (\bibinfo {year} {2017})}\BibitemShut {NoStop}%
\bibitem [{\citenamefont {Punzmann}\ \emph {et~al.}(2014)\citenamefont
  {Punzmann}, \citenamefont {Francois}, \citenamefont {Xia}, \citenamefont
  {Falkovich},\ and\ \citenamefont {Shats}}]{punzmann2014generation}%
  \BibitemOpen
  \bibfield  {author} {\bibinfo {author} {\bibfnamefont {H.}~\bibnamefont
  {Punzmann}}, \bibinfo {author} {\bibfnamefont {N.}~\bibnamefont {Francois}},
  \bibinfo {author} {\bibfnamefont {H.}~\bibnamefont {Xia}}, \bibinfo {author}
  {\bibfnamefont {G.}~\bibnamefont {Falkovich}},\ and\ \bibinfo {author}
  {\bibfnamefont {M.}~\bibnamefont {Shats}},\ }\bibfield  {title} {\bibinfo
  {title} {Generation and reversal of surface flows by propagating waves},\
  }\href {https://doi.org/10.1038/nphys3041} {\bibfield  {journal} {\bibinfo
  {journal} {Nature Physics}\ }\textbf {\bibinfo {volume} {10}},\ \bibinfo
  {pages} {658} (\bibinfo {year} {2014})}\BibitemShut {NoStop}%
\bibitem [{\citenamefont {Planchette}\ \emph {et~al.}(2012)\citenamefont
  {Planchette}, \citenamefont {Lorenceau},\ and\ \citenamefont
  {Biance}}]{Planchette2012}%
  \BibitemOpen
  \bibfield  {author} {\bibinfo {author} {\bibfnamefont {C.}~\bibnamefont
  {Planchette}}, \bibinfo {author} {\bibfnamefont {E.}~\bibnamefont
  {Lorenceau}},\ and\ \bibinfo {author} {\bibfnamefont {A.}~\bibnamefont
  {Biance}},\ }\bibfield  {title} {\bibinfo {title} {Surface wave on a particle
  raft},\ }\href {https://doi.org/10.1039/C2SM06859A} {\bibfield  {journal}
  {\bibinfo  {journal} {SoftMatter}\ }\textbf {\bibinfo {volume} {8}},\
  \bibinfo {pages} {2444} (\bibinfo {year} {2012})}\BibitemShut {NoStop}%
\bibitem [{\citenamefont {Voermans}\ \emph {et~al.}(2020)\citenamefont
  {Voermans}, \citenamefont {Rabault}, \citenamefont {Filchuk}, \citenamefont
  {Ryzhov}, \citenamefont {Heil}, \citenamefont {Marchenko}, \citenamefont
  {Collins~III}, \citenamefont {Dabboor}, \citenamefont {Sutherland},\ and\
  \citenamefont {Babanin}}]{voermans2020experimental}%
  \BibitemOpen
  \bibfield  {author} {\bibinfo {author} {\bibfnamefont {J.~J.}\ \bibnamefont
  {Voermans}}, \bibinfo {author} {\bibfnamefont {J.}~\bibnamefont {Rabault}},
  \bibinfo {author} {\bibfnamefont {K.}~\bibnamefont {Filchuk}}, \bibinfo
  {author} {\bibfnamefont {I.}~\bibnamefont {Ryzhov}}, \bibinfo {author}
  {\bibfnamefont {P.}~\bibnamefont {Heil}}, \bibinfo {author} {\bibfnamefont
  {A.}~\bibnamefont {Marchenko}}, \bibinfo {author} {\bibfnamefont {C.~O.}\
  \bibnamefont {Collins~III}}, \bibinfo {author} {\bibfnamefont
  {M.}~\bibnamefont {Dabboor}}, \bibinfo {author} {\bibfnamefont
  {G.}~\bibnamefont {Sutherland}},\ and\ \bibinfo {author} {\bibfnamefont
  {A.~V.}\ \bibnamefont {Babanin}},\ }\bibfield  {title} {\bibinfo {title}
  {Experimental evidence for a universal threshold characterizing wave-induced
  sea ice break-up},\ }\href {https://doi.org/10.5194/tc-14-4265-2020}
  {\bibfield  {journal} {\bibinfo  {journal} {The Cryosphere}\ }\textbf
  {\bibinfo {volume} {14}},\ \bibinfo {pages} {4265} (\bibinfo {year}
  {2020})}\BibitemShut {NoStop}%
\bibitem [{\citenamefont {Landau}\ and\ \citenamefont
  {Lifshitz}(1959)}]{LandauElasticity}%
  \BibitemOpen
  \bibfield  {author} {\bibinfo {author} {\bibfnamefont {L.~D.}\ \bibnamefont
  {Landau}}\ and\ \bibinfo {author} {\bibfnamefont {E.~M.}\ \bibnamefont
  {Lifshitz}},\ }\href@noop {} {\emph {\bibinfo {title} {Theory of elasticity,
  Course of Theoretical Physics}}},\ Vol.~\bibinfo {volume} {7}\ (\bibinfo
  {publisher} {Pergamon},\ \bibinfo {year} {1959})\BibitemShut {NoStop}%
\bibitem [{\citenamefont {Oswald}(2009)}]{OswaldBook}%
  \BibitemOpen
  \bibfield  {author} {\bibinfo {author} {\bibfnamefont {P.}~\bibnamefont
  {Oswald}},\ }\href@noop {} {\emph {\bibinfo {title} {Rheophysics: The
  Deformation and Flow of Matter}}}\ (\bibinfo  {publisher} {Cambridge
  University Press},\ \bibinfo {year} {2009})\BibitemShut {NoStop}%
\bibitem [{\citenamefont {Allan}\ \emph {et~al.}(2021)\citenamefont {Allan},
  \citenamefont {Caswell}, \citenamefont {Keim}, \citenamefont {van~der Wel},\
  and\ \citenamefont {Verweij}}]{trackPy}%
  \BibitemOpen
  \bibfield  {author} {\bibinfo {author} {\bibfnamefont {D.~B.}\ \bibnamefont
  {Allan}}, \bibinfo {author} {\bibfnamefont {T.}~\bibnamefont {Caswell}},
  \bibinfo {author} {\bibfnamefont {N.~C.}\ \bibnamefont {Keim}}, \bibinfo
  {author} {\bibfnamefont {C.~M.}\ \bibnamefont {van~der Wel}},\ and\ \bibinfo
  {author} {\bibfnamefont {R.~W.}\ \bibnamefont {Verweij}},\ }\bibfield
  {title} {\bibinfo {title} {soft-matter/trackpy: Trackpy v0. 5.0},\
  }\href@noop {} {\bibfield  {journal} {\bibinfo  {journal} {Gen{\`e}ve:
  Zenodo}\ } (\bibinfo {year} {2021})}\BibitemShut {NoStop}%
\bibitem [{\citenamefont {Van~der Walt}\ \emph {et~al.}(2014)\citenamefont
  {Van~der Walt}, \citenamefont {Sch{\"o}nberger}, \citenamefont
  {Nunez-Iglesias}, \citenamefont {Boulogne}, \citenamefont {Warner},
  \citenamefont {Yager}, \citenamefont {Gouillart},\ and\ \citenamefont
  {Yu}}]{van2014scikit}%
  \BibitemOpen
  \bibfield  {author} {\bibinfo {author} {\bibfnamefont {S.}~\bibnamefont
  {Van~der Walt}}, \bibinfo {author} {\bibfnamefont {J.~L.}\ \bibnamefont
  {Sch{\"o}nberger}}, \bibinfo {author} {\bibfnamefont {J.}~\bibnamefont
  {Nunez-Iglesias}}, \bibinfo {author} {\bibfnamefont {F.}~\bibnamefont
  {Boulogne}}, \bibinfo {author} {\bibfnamefont {J.~D.}\ \bibnamefont
  {Warner}}, \bibinfo {author} {\bibfnamefont {N.}~\bibnamefont {Yager}},
  \bibinfo {author} {\bibfnamefont {E.}~\bibnamefont {Gouillart}},\ and\
  \bibinfo {author} {\bibfnamefont {T.}~\bibnamefont {Yu}},\ }\bibfield
  {title} {\bibinfo {title} {scikit-image: image processing in python},\ }\href
  {https://doi.org/10.7717/peerj.453} {\bibfield  {journal} {\bibinfo
  {journal} {PeerJ}\ }\textbf {\bibinfo {volume} {2}},\ \bibinfo {pages} {e453}
  (\bibinfo {year} {2014})}\BibitemShut {NoStop}%
\bibitem [{\citenamefont {Toyota}\ \emph {et~al.}(2006)\citenamefont {Toyota},
  \citenamefont {Takatsuji},\ and\ \citenamefont
  {Nakayama}}]{toyota2006characteristics}%
  \BibitemOpen
  \bibfield  {author} {\bibinfo {author} {\bibfnamefont {T.}~\bibnamefont
  {Toyota}}, \bibinfo {author} {\bibfnamefont {S.}~\bibnamefont {Takatsuji}},\
  and\ \bibinfo {author} {\bibfnamefont {M.}~\bibnamefont {Nakayama}},\
  }\bibfield  {title} {\bibinfo {title} {Characteristics of sea ice floe size
  distribution in the seasonal ice zone},\ }\href
  {https://doi.org/10.1029/2005GL024556} {\bibfield  {journal} {\bibinfo
  {journal} {Geophysical research letters}\ }\textbf {\bibinfo {volume} {33}},\
  \bibinfo {pages} {L02616} (\bibinfo {year} {2006})}\BibitemShut {NoStop}%
\bibitem [{\citenamefont {Toyota}\ \emph {et~al.}(2011)\citenamefont {Toyota},
  \citenamefont {Haas},\ and\ \citenamefont {Tamura}}]{toyota2011size}%
  \BibitemOpen
  \bibfield  {author} {\bibinfo {author} {\bibfnamefont {T.}~\bibnamefont
  {Toyota}}, \bibinfo {author} {\bibfnamefont {C.}~\bibnamefont {Haas}},\ and\
  \bibinfo {author} {\bibfnamefont {T.}~\bibnamefont {Tamura}},\ }\bibfield
  {title} {\bibinfo {title} {Size distribution and shape properties of
  relatively small sea-ice floes in the antarctic marginal ice zone in late
  winter},\ }\href {https://doi.org/10.1016/j.dsr2.2010.10.034} {\bibfield
  {journal} {\bibinfo  {journal} {Deep Sea Research Part II: Topical Studies in
  Oceanography}\ }\textbf {\bibinfo {volume} {58}},\ \bibinfo {pages} {1182}
  (\bibinfo {year} {2011})}\BibitemShut {NoStop}%
\bibitem [{\citenamefont {Alberello}\ \emph {et~al.}(2019)\citenamefont
  {Alberello}, \citenamefont {Onorato}, \citenamefont {Bennetts}, \citenamefont
  {Vichi}, \citenamefont {Eayrs}, \citenamefont {MacHutchon},\ and\
  \citenamefont {Toffoli}}]{alberello2019brief}%
  \BibitemOpen
  \bibfield  {author} {\bibinfo {author} {\bibfnamefont {A.}~\bibnamefont
  {Alberello}}, \bibinfo {author} {\bibfnamefont {M.}~\bibnamefont {Onorato}},
  \bibinfo {author} {\bibfnamefont {L.}~\bibnamefont {Bennetts}}, \bibinfo
  {author} {\bibfnamefont {M.}~\bibnamefont {Vichi}}, \bibinfo {author}
  {\bibfnamefont {C.}~\bibnamefont {Eayrs}}, \bibinfo {author} {\bibfnamefont
  {K.}~\bibnamefont {MacHutchon}},\ and\ \bibinfo {author} {\bibfnamefont
  {A.}~\bibnamefont {Toffoli}},\ }\bibfield  {title} {\bibinfo {title} {Brief
  communication: Pancake ice floe size distribution during the winter expansion
  of the antarctic marginal ice zone},\ }\href
  {https://doi.org/10.5194/tc-13-41-2019} {\bibfield  {journal} {\bibinfo
  {journal} {The Cryosphere}\ }\textbf {\bibinfo {volume} {13}},\ \bibinfo
  {pages} {41} (\bibinfo {year} {2019})}\BibitemShut {NoStop}%
\bibitem [{\citenamefont {Hwang}\ and\ \citenamefont
  {Wang}(2022)}]{hwang2022multi}%
  \BibitemOpen
  \bibfield  {author} {\bibinfo {author} {\bibfnamefont {B.}~\bibnamefont
  {Hwang}}\ and\ \bibinfo {author} {\bibfnamefont {Y.}~\bibnamefont {Wang}},\
  }\bibfield  {title} {\bibinfo {title} {Multi-scale satellite observations of
  arctic sea ice: new insight into the life cycle of the floe size
  distribution},\ }\href {https://doi.org/10.1098/rsta.2021.0259} {\bibfield
  {journal} {\bibinfo  {journal} {Philosophical Transactions of the Royal
  Society A}\ }\textbf {\bibinfo {volume} {380}},\ \bibinfo {pages} {20210259}
  (\bibinfo {year} {2022})}\BibitemShut {NoStop}%
\bibitem [{\citenamefont {Dumas-Lefebvre}\ and\ \citenamefont
  {Dumont}(2021{\natexlab{b}})}]{dumas2021aerial}%
  \BibitemOpen
  \bibfield  {author} {\bibinfo {author} {\bibfnamefont {E.}~\bibnamefont
  {Dumas-Lefebvre}}\ and\ \bibinfo {author} {\bibfnamefont {D.}~\bibnamefont
  {Dumont}},\ }\bibfield  {title} {\bibinfo {title} {Aerial observations of sea
  ice break-up by ship waves},\ }\href {https://doi.org/10.5194/tc-2021-328}
  {\bibfield  {journal} {\bibinfo  {journal} {The Cryosphere Discussions}\ ,\
  \bibinfo {pages} {1}} (\bibinfo {year} {2021}{\natexlab{b}})}\BibitemShut
  {NoStop}%
\bibitem [{\citenamefont {Stern}\ \emph {et~al.}(2018)\citenamefont {Stern},
  \citenamefont {Schweiger}, \citenamefont {Zhang},\ and\ \citenamefont
  {Steele}}]{stern2018reconciling}%
  \BibitemOpen
  \bibfield  {author} {\bibinfo {author} {\bibfnamefont {H.~L.}\ \bibnamefont
  {Stern}}, \bibinfo {author} {\bibfnamefont {A.~J.}\ \bibnamefont
  {Schweiger}}, \bibinfo {author} {\bibfnamefont {J.}~\bibnamefont {Zhang}},\
  and\ \bibinfo {author} {\bibfnamefont {M.}~\bibnamefont {Steele}},\
  }\bibfield  {title} {\bibinfo {title} {On reconciling disparate studies of
  the sea-ice floe size distribution},\ }\href
  {https://doi.org/10.1525/elementa.304} {\bibfield  {journal} {\bibinfo
  {journal} {Elementa: Science of the Anthropocene}\ }\textbf {\bibinfo
  {volume} {6}},\ \bibinfo {pages} {1} (\bibinfo {year} {2018})}\BibitemShut
  {NoStop}%
\bibitem [{\citenamefont {Cheng}\ and\ \citenamefont
  {Redner}(1988)}]{cheng1988scaling}%
  \BibitemOpen
  \bibfield  {author} {\bibinfo {author} {\bibfnamefont {Z.}~\bibnamefont
  {Cheng}}\ and\ \bibinfo {author} {\bibfnamefont {S.}~\bibnamefont {Redner}},\
  }\bibfield  {title} {\bibinfo {title} {Scaling theory of fragmentation},\
  }\href {https://doi.org/10.1103/PhysRevLett.60.2450} {\bibfield  {journal}
  {\bibinfo  {journal} {Physical review letters}\ }\textbf {\bibinfo {volume}
  {60}},\ \bibinfo {pages} {2450} (\bibinfo {year} {1988})}\BibitemShut
  {NoStop}%
\bibitem [{\citenamefont {Thorpe}\ and\ \citenamefont
  {Jasiuk}(1992)}]{thorpe1992new}%
  \BibitemOpen
  \bibfield  {author} {\bibinfo {author} {\bibfnamefont {M.}~\bibnamefont
  {Thorpe}}\ and\ \bibinfo {author} {\bibfnamefont {I.}~\bibnamefont
  {Jasiuk}},\ }\bibfield  {title} {\bibinfo {title} {New results in the theory
  of elasticity for two-dimensional composites},\ }\href
  {https://doi.org/10.1098/rspa.1992.0124} {\bibfield  {journal} {\bibinfo
  {journal} {Proceedings of the Royal Society of London. Series A: Mathematical
  and Physical Sciences}\ }\textbf {\bibinfo {volume} {438}},\ \bibinfo {pages}
  {531} (\bibinfo {year} {1992})}\BibitemShut {NoStop}%
\bibitem [{\citenamefont {Deike}\ \emph {et~al.}(2013)\citenamefont {Deike},
  \citenamefont {Bacri},\ and\ \citenamefont {Falcon}}]{Deike2013}%
  \BibitemOpen
  \bibfield  {author} {\bibinfo {author} {\bibfnamefont {L.}~\bibnamefont
  {Deike}}, \bibinfo {author} {\bibfnamefont {J.-C.}\ \bibnamefont {Bacri}},\
  and\ \bibinfo {author} {\bibfnamefont {E.}~\bibnamefont {Falcon}},\
  }\bibfield  {title} {\bibinfo {title} {Nonlinear waves on the surface of a
  fluid covered by an elastic sheet},\ }\href
  {https://doi.org/10.1017/jfm.2013.379} {\bibfield  {journal} {\bibinfo
  {journal} {Journal of Fluid Mechanics}\ }\textbf {\bibinfo {volume} {733}},\
  \bibinfo {pages} {394} (\bibinfo {year} {2013})}\BibitemShut {NoStop}%
\bibitem [{\citenamefont {Domino}\ \emph {et~al.}(2018)\citenamefont {Domino},
  \citenamefont {Fermigier}, \citenamefont {Fort},\ and\ \citenamefont
  {Eddi}}]{Domino2018}%
  \BibitemOpen
  \bibfield  {author} {\bibinfo {author} {\bibfnamefont {L.}~\bibnamefont
  {Domino}}, \bibinfo {author} {\bibfnamefont {M.}~\bibnamefont {Fermigier}},
  \bibinfo {author} {\bibfnamefont {E.}~\bibnamefont {Fort}},\ and\ \bibinfo
  {author} {\bibfnamefont {A.}~\bibnamefont {Eddi}},\ }\bibfield  {title}
  {\bibinfo {title} {Dispersion-free control of hydroelastic waves down to
  sub-wavelength scale},\ }\href {https://doi.org/10.1209/0295-5075/121/14001}
  {\bibfield  {journal} {\bibinfo  {journal} {Europhysics Letters}\ }\textbf
  {\bibinfo {volume} {121}},\ \bibinfo {pages} {14001} (\bibinfo {year}
  {2018})}\BibitemShut {NoStop}%
\bibitem [{\citenamefont {Lamb}(1932)}]{Lamb1932}%
  \BibitemOpen
  \bibfield  {author} {\bibinfo {author} {\bibfnamefont {H.}~\bibnamefont
  {Lamb}},\ }\href@noop {} {\emph {\bibinfo {title} {Hydrodynamics}}}\
  (\bibinfo  {publisher} {Springer-Verlag, Berlin},\ \bibinfo {year}
  {1932})\BibitemShut {NoStop}%
\bibitem [{\citenamefont {Berhanu}(2022)}]{Berhanu2022}%
  \BibitemOpen
  \bibfield  {author} {\bibinfo {author} {\bibfnamefont {M.}~\bibnamefont
  {Berhanu}},\ }\bibfield  {title} {\bibinfo {title} {Impact of the dissipation
  on the nonlinear interactions and turbulence of gravity-capillary waves},\
  }\href {https://doi.org/10.3390/fluids7040137} {\bibfield  {journal}
  {\bibinfo  {journal} {Fluids}\ }\textbf {\bibinfo {volume} {7}},\ \bibinfo
  {pages} {137} (\bibinfo {year} {2022})}\BibitemShut {NoStop}%
\bibitem [{\citenamefont {Thielicke}\ and\ \citenamefont
  {Stamhuis}(2014)}]{thielicke2014pivlab}%
  \BibitemOpen
  \bibfield  {author} {\bibinfo {author} {\bibfnamefont {W.}~\bibnamefont
  {Thielicke}}\ and\ \bibinfo {author} {\bibfnamefont {E.}~\bibnamefont
  {Stamhuis}},\ }\bibfield  {title} {\bibinfo {title} {Pivlab--towards
  user-friendly, affordable and accurate digital particle image velocimetry in
  matlab},\ }\href {https://doi.org/10.5334/jors.bl} {\bibfield  {journal}
  {\bibinfo  {journal} {Journal of open research software}\ }\textbf {\bibinfo
  {volume} {2}},\ \bibinfo {pages} {e30} (\bibinfo {year} {2014})}\BibitemShut
  {NoStop}%
\bibitem [{\citenamefont {Shankar}(2007)}]{shankar2007frequencies}%
  \BibitemOpen
  \bibfield  {author} {\bibinfo {author} {\bibfnamefont {P.}~\bibnamefont
  {Shankar}},\ }\bibfield  {title} {\bibinfo {title} {Frequencies of
  gravity--capillary waves on highly curved interfaces with edge constraints},\
  }\href {https://doi.org/10.1016/j.fluiddyn.2006.12.002} {\bibfield  {journal}
  {\bibinfo  {journal} {Fluid dynamics research}\ }\textbf {\bibinfo {volume}
  {39}},\ \bibinfo {pages} {457} (\bibinfo {year} {2007})}\BibitemShut
  {NoStop}%
\bibitem [{\citenamefont {Aumaitre}\ \emph {et~al.}(2011)\citenamefont
  {Aumaitre}, \citenamefont {Vella},\ and\ \citenamefont
  {Cicuta}}]{aumaitre2011measurement}%
  \BibitemOpen
  \bibfield  {author} {\bibinfo {author} {\bibfnamefont {E.}~\bibnamefont
  {Aumaitre}}, \bibinfo {author} {\bibfnamefont {D.}~\bibnamefont {Vella}},\
  and\ \bibinfo {author} {\bibfnamefont {P.}~\bibnamefont {Cicuta}},\
  }\bibfield  {title} {\bibinfo {title} {On the measurement of the surface
  pressure in langmuir films with finite shear elasticity},\ }\href
  {https://doi.org/10.1039/C0SM01213K} {\bibfield  {journal} {\bibinfo
  {journal} {Soft Matter}\ }\textbf {\bibinfo {volume} {7}},\ \bibinfo {pages}
  {2530} (\bibinfo {year} {2011})}\BibitemShut {NoStop}%
\bibitem [{\citenamefont {{Burns}}\ \emph {et~al.}(2020)\citenamefont
  {{Burns}}, \citenamefont {{Vasil}}, \citenamefont {{Oishi}}, \citenamefont
  {{Lecoanet}},\ and\ \citenamefont {{Brown}}}]{Dedalus}%
  \BibitemOpen
  \bibfield  {author} {\bibinfo {author} {\bibfnamefont {K.~J.}\ \bibnamefont
  {{Burns}}}, \bibinfo {author} {\bibfnamefont {G.~M.}\ \bibnamefont
  {{Vasil}}}, \bibinfo {author} {\bibfnamefont {J.~S.}\ \bibnamefont
  {{Oishi}}}, \bibinfo {author} {\bibfnamefont {D.}~\bibnamefont
  {{Lecoanet}}},\ and\ \bibinfo {author} {\bibfnamefont {B.~P.}\ \bibnamefont
  {{Brown}}},\ }\bibfield  {title} {\bibinfo {title} {{Dedalus: A flexible
  framework for numerical simulations with spectral methods}},\ }\href
  {https://doi.org/10.1103/PhysRevResearch.2.023068} {\bibfield  {journal}
  {\bibinfo  {journal} {Physical Review Research}\ }\textbf {\bibinfo {volume}
  {2}},\ \bibinfo {eid} {023068} (\bibinfo {year} {2020})},\ \Eprint
  {https://arxiv.org/abs/1905.10388} {arXiv:1905.10388 [astro-ph.IM]}
  \BibitemShut {NoStop}%
\end{thebibliography}
\end{document}